\documentclass[iop,numberedappendix,appendixfloats]{emulateapj}

\usepackage{amssymb}
\usepackage{amsmath}
\usepackage{float}
\usepackage[font=footnotesize,caption=false]{subfig}
\usepackage{booktabs}
\usepackage{ulem}
\usepackage[usenames,dvipsnames]{color}
\usepackage{appendix}

\usepackage{xspace}
\usepackage{graphicx}
\usepackage[colorlinks,breaklinks, citecolor=blue,linkcolor=blue,urlcolor=blue]{hyperref}
\usepackage{apjfonts}
\usepackage{etoolbox}

\usepackage{xcolor}

\def\pasa{PASA}

\def\halpha{H$\alpha$  }
\def\halphans{H$\alpha$}

\def\hbetans{H$\beta$}
\def\NII{[N\hspace{.03cm}II]}
\def\SII{[S\hspace{.03cm}II]}
\def\OI{[O\hspace{.03cm}I]}

\def\OIII{[O\hspace{.03cm}III]}

\def\Msun{\hbox{M$_{\odot}$}}
\def\sfrunits{\Msun\ yr$^{-1}$}

\def\kms{km s$^{-1}$}

\def\sfrha{SFR$_{\mathrm{H}\alpha}$}
\def\sfrsed{SFR$_{\mathrm{SED}}$}

\def\vrot{$v_\mathrm{rot}$}
\def\vobs{$v_\mathrm{obs}$}

\def\zkmos{$z_\mathrm{KMOS}$}
\def\kmostd{KMOS$^\mathrm{3D}$~}
\def\kmostdns{KMOS$^\mathrm{3D}$}
\newcommand{\spark}{\small{SPARK}\normalsize{}}
\newcommand\nar{New Astronomy Reviews} 

\def\lmstar{$\mathrm{\log(M_{\star})}$}
\def\dms{$\mathrm{\Delta MS}$}
\def\uvrest{$(U-V)_{\rm rest}$}
\def\dmre{$\mathrm{\Delta MR_{\rm e}}$}
\makeatletter
\patchcmd{\NAT@citex}
    {\@citea\NAT@hyper@{\NAT@nmfmt{\NAT@nm}\NAT@date}}
    {\@citea\NAT@nmfmt{\NAT@nm}\NAT@hyper@{\NAT@date}}
    {}
    {}
\patchcmd{\NAT@citex}
    {\@citea\NAT@hyper@{%
         \NAT@nmfmt{\NAT@nm}%
         \hyper@natlinkbreak{\NAT@aysep\NAT@spacechar}{\@citeb\@extra@b@citeb}%
         \NAT@date}}
    {\@citea\NAT@nmfmt{\NAT@nm}%
     \NAT@aysep\NAT@spacechar%
     \NAT@hyper@{\NAT@date}}
    {}
    {}
\patchcmd{\NAT@citex}
    {\@citea\NAT@hyper@{%
         \NAT@nmfmt{\NAT@nm}%
         \hyper@natlinkbreak
         {\NAT@spacechar\NAT@@open\if*#1*\else#1\NAT@spacechar\fi}%
         {\@citeb\@extra@b@citeb}%
         \NAT@date}}
    {\@citea\NAT@nmfmt{\NAT@nm}%
        \NAT@spacechar\NAT@@open\if*#1*\else#1\NAT@spacechar\fi%
        \NAT@hyper@{\NAT@date}}
    {}
    {}
\makeatother

\def\totalobs{739}

\defcitealias{2015ApJ...799..209W}{W15}

\shorttitle{The KMOS$^\mathrm{3D}$ survey}
\shortauthors{Wisnioski et al.}

\begin{document}
\title{The KMOS$^\mathrm{3D}$ Survey: data release and final survey paper\altaffilmark{$\dagger$}}
\author{
%The KMOS$^\mathrm{3D}$ Team
%1st tier
E. Wisnioski\altaffilmark{1,2,3}\altaffilmark{$\ddagger$}, N.M. F\"orster Schreiber\altaffilmark{1}\altaffilmark{$\ddagger\ddagger$},  
M. Fossati\altaffilmark{1,4,5}, J. T. Mendel\altaffilmark{1,2,3}, D. Wilman\altaffilmark{1,4}, \\
%2nd tier
R. Genzel\altaffilmark{1,6},
R. Bender\altaffilmark{1,4},
S. Wuyts\altaffilmark{7},
R. L. Davies\altaffilmark{1},  
H. \"Ubler\altaffilmark{1},\\
%3rd tier
K. Bandara\altaffilmark{1},
A. Beifiori\altaffilmark{1,4}, 
S. Belli\altaffilmark{1},
G. Brammer\altaffilmark{8}, 
J. Chan\altaffilmark{1,4,9}, 
R. I. Davies\altaffilmark{1},  
M. Fabricius\altaffilmark{1},
A. Galametz\altaffilmark{10},  
P. Lang\altaffilmark{1,11}, 
D. Lutz\altaffilmark{1}, 
E. J. Nelson\altaffilmark{1,12}, 
I. Momcheva\altaffilmark{13},
S. Price\altaffilmark{1}.
D. Rosario\altaffilmark{5},
R. Saglia\altaffilmark{1,4}, 
S. Seitz\altaffilmark{1,4},
T. Shimizu\altaffilmark{1}.
L.J. Tacconi\altaffilmark{1},
K. Tadaki,\altaffilmark{14},
P. G. van Dokkum\altaffilmark{15}  
E. Wuyts\altaffilmark{1}
}

%et al.\\
\affil{
\altaffilmark{1}{Max-Planck-Institut f\"{u}r extraterrestrische Physik (MPE), Giessenbachstr. 1, D-85748 Garching, Germany}\\
\altaffilmark{2}{Research School of Astronomy and Astrophysics, Australian National University, Canberra, ACT 2611, Australia}\\
\altaffilmark{3}{ARC Centre of Excellence for All Sky Astrophysics in 3 Dimensions (ASTRO 3D)}\\
\altaffilmark{4}{Universit\"ats-Sternwarte, Ludwig-Maximilians-Universit\"at, Scheinerstrasse 1, D-81679 M\"unchen, Germany}\\
\altaffilmark{5}{Institute for Computational Cosmology and Centre for Extragalactic Astronomy, Department of Physics, Durham University, South Road, Durham DH1 3LE, UK}\\
\altaffilmark{6}{Departments of Physics \& Astronomy, University of California, Berkeley, CA 94720, USA}\\
\altaffilmark{7}{Department of Physics, University of Bath, Claverton Down, Bath, BA2 7AY, UK}\\
\altaffilmark{8}{Cosmic Dawn Center, Niels Bohr Institute, University of Copenhagen, Juliane Maries Vej 30, DK-2100 Copenhagen, Denmark}\\
\altaffilmark{9}{Department of Physics and Astronomy, University of California, Riverside, CA 92521, USA}\\
\altaffilmark{10}{Department of Astronomy, University of Geneva, 1205, Versoix, Switzerland}\\
\altaffilmark{11}{Max-Planck-Institut für Astronomie, Königstuhl 17, D-69117 Heidelberg, Germany}\\
\altaffilmark{12}{Harvard-Smithsonian Center for Astrophysics, Cambridge, USA}\\
\altaffilmark{13}{Space Telescope Science Institute, 3700 San Martin Drive, Baltimore, MD 21218, USA}\\
\altaffilmark{14}{National Astronomical Observatory of Japan, Mitaka, Tokyo, Japan}\\
\altaffilmark{15}{Department of Astronomy, Yale University, New Haven, CT 06511, USA}
}

\altaffiltext{$\ddagger$}{\href{mailto:emily.wisnioski@anu.edu.au}{emily.wisnioski@anu.edu.au}}
{
\altaffiltext{$\ddagger\ddagger$}{\href{mailto:forster@mpe.mpg.de}{forster@mpe.mpg.de}}
}
\altaffiltext{$\dagger$}{Based on observations obtained at the Very Large Telescope (VLT) of the European Southern Observatory (ESO), Paranal, Chile (ESO program IDS 092A-0091, 093.A-0079, 093.A-0079, 094.A-0217, 095.A-0047, 096.A-0025, 097.A-0028, 098.A-0045,  099.A-0013, 0100.A-0039, and 0101.A-0022)}
\altaffiltext{$\mathsection$}{{\url{http://www.mpe.mpg.de/ir/KMOS3D}}}
%%%%%%%%%%%%%%%%%%%%%%%%%%%%%%%%%%%%%%%%%%%%%%%%
\begin{abstract}
%%%%%%%%%%%%%%%%%%%%%%%%%%%%%%%%%%%%%%%%%%%%%%%%
We present the completed \kmostd survey $-$ an integral field spectroscopic survey of \totalobs, $\log(M_{\star}/M_{\odot})>9$, galaxies at $0.6<z<2.7$ using the K-band Multi Object Spectrograph (KMOS) at the Very Large Telescope (VLT).  \kmostd provides a population-wide census of kinematics, star formation, outflows, and nebular gas conditions both on and off the star-forming galaxy main sequence through the spatially resolved and integrated properties of \halphans, \NII, and \SII~emission lines. 
We detect \halpha emission for 91\% of galaxies on the main sequence of star-formation and 79\% overall. The depth of the survey has allowed us to detect galaxies with star-formation rates below 1 \sfrunits, as well as to resolve {81}\% of detected galaxies with $\geq3$ resolution elements along the kinematic major axis. The detection fraction of \halpha is a strong function of both color and offset from the main sequence, with the detected and non-detected samples exhibiting different SED shapes. Comparison of \halpha and UV+IR star formation rates (SFRs) reveal that dust attenuation corrections may be underestimated by 0.5 dex at the highest masses ($\log(M_{\star}/M_{\odot})>10.5$). We confirm our first year results of a high rotation dominated fraction {(monotonic velocity gradient and \vrot/$\sigma_0 > \sqrt{3.36}$)} of 77\% for the full \kmostd sample. The rotation-dominated fraction is a function of both stellar mass and redshift with the strongest evolution measured {over the redshift range of the survey} for galaxies with $\log(M_{\star}/M_{\odot})<10.5$. 
With this paper we include a final data release of all \totalobs observed objects$^\mathsection$. \\
 \end{abstract}


%\clearpage

%%%%%%%%%%%%%%%%%%%%%%%%%%%%%%%%%%%%%%%%%%%%%%%%
\section{Introduction}
\setcounter{footnote}{0}
%%%%%%%%%%%%%%%%%%%%%%%%%%%%%%%%%%%%%%%%%%%%%%%%

Near-infrared (near-IR) integral field unit (IFU) spectrographs are very powerful at exploring galaxy evolution around the peak epoch of cosmic star formation activity $\sim 6 - 11$ billion years ago.  By spatially and spectrally resolving
rest-optical nebular emission lines such as H$\alpha$, H$\beta$, [\ion{N}{2}], [\ion{S}{2}], [\ion{O}{3}], and [\ion{O}{2}] of $z \sim 1 - 3$ galaxies, near-IR IFUs enable the full two-dimensional (2D) mapping of the kinematics, star formation, and physical conditions of the interstellar medium on sub-galactic scales. Over the past 15 years, results on these properties, alongside population censuses from multiwavelength lookback surveys and high-resolution broad-band optical and near-IR imaging, lent empirical support to the equilibrium growth scenario in which galaxy evolution is regulated by the balance between fairly continuous gas accretion from the cosmic web and minor mergers, internal dynamical processes such as disk instabilities, and galactic-scale outflows \citep[e.g.,][]{2005MNRAS.363....2K,2009ApJ...703..785D,2013ApJ...772..119L,2015MNRAS.450.2327Z}. This scenario naturally explains the tightness of the stellar mass versus star formation rate ($M_{\star}$$-$SFR) ``main sequence'' (MS) of star-forming galaxies (SFGs) and cold molecular gas scaling relationships that are observed out to at least $z \sim 4$ \citep[e.g.,][]{2007ApJ...660L..43N,2007AA...468...33E,2007ApJ...670..156D,2011ApJ...739L..40R,2013ApJ...768...74T,2018ApJ...853..179T,2014ApJ...795..104W,2017ApJ...837..150S}.

Observations with the first generation of sensitive near-IR IFUs on 8-10\,m-class telescopes such as SINFONI at the Very Large Telescope \citep[VLT;][]{2003SPIE.4841.1548E,2004Msngr.117...17B}, OSIRIS at Keck~II \citep{2006NewAR..50..362L}, and NIFS at Gemini North \citep{2003SPIE.4841.1581M}, were key in uncovering the importance of internal processes in the early growth of massive galaxies.  Studies with these instruments first compellingly showed from direct 2D kinematic evidence that rotating, yet turbulent disks, are common among massive SFGs despite their often irregular and clumpy appearance in the rest-UV \citep[e.g.,][] {2006Natur.442..786G,2006ApJ...645.1062F,2009ApJ...706.1364F,2008ApJ...682..231S,2008Natur.455..775S,2009ApJ...697.2057L,2009ApJ...699..421W,Jones:2010uf,2011MNRAS.417.2601W,2011arXiv1109.5952M,2011AA...528A..88G,2012AA...539A..92E,2012ApJ...760..130S}. They also uncovered the launching sites and role of galactic winds powered by star formation and active galactic nuclei (AGN) through detection of their telltale high-velocity signature, in both typical MS SFGs and more extreme quasars, submillimeter, and radio galaxies \citep[e.g.,][]{2008AA...491..407N,Shapiro:2009sj,2011ApJ...733..101G,2012ApJ...752..111N,2012ApJ...761...43N,2012AA...537L...8C,2014ApJ...787...38F,2015ApJ...799...82C}. Collectively, work with these single-IFU instruments assembled data of a couple hundred $z \sim 1 - 3$ galaxies under typical near-IR seeing conditions of $\sim 0\farcs 6$, corresponding to $\rm \sim 4.5 - 5~kpc$ for unlensed sources, and $\sim 150$ galaxies at the higher $\sim 0\farcs 1 - 0\farcs 2$ or $\rm 1 - 2~kpc$ resolution achieved with the aid of adaptive optics (AO) systems \citep[e.g.,][and references therein]{2013PASA...30...56G,2018ApJS..238...21F}.

The efficient multi-IFU \textit{K-band Multi-Object Spectrograph} (KMOS) at the VLT \citep{2004SPIE.5492.1179S} enabled the expansion of near-IR IFU surveys to much larger, homogeneous, and more complete samples. KMOS features 24 individual IFUs with each a $2\farcs 8 \times 2\farcs 8$ field of view, deployable over a patrol field of 7 arcmin diameter.  It operates in seeing-limited mode and, with its pixel scale of $0\farcs 2$ and spectral resolution of $R = \lambda/\Delta\lambda \sim 4000$, is particularly sensitive to spatially extended line emission. In the six years since KMOS was commissioned, samples of altogether $> 2000$ have been obtained through various programs, putting results on the properties of SFGs from previous surveys on a more robust statistical footing and allowing more systematic studies into new regimes of galaxy parameter space
\citep[e.g.,][]{2014ApJ...796....7G,2015ApJ...799..209W,2015ApJ...804L...4M,2016MNRAS.457.1888S,2016MNRAS.460..103T,2016MNRAS.456.1195H,2017MNRAS.467.1965H,2017ApJ...846..120B,2017ApJ...838...14M,2017ApJ...850..203P,2017MNRAS.471.1280T,2018AA...613A..72G}.

With this new opportunity, we carried out \kmostdns, a comprehensive 75-night survey of H$\alpha$$+$[\ion{N}{2}]$+$[\ion{S}{2}] emission, leveraging KMOS multiplexing to map the kinematics, star formation, gas outflows and metallicities of \totalobs~galaxies at $z \sim 0.6 - 2.7$. The overarching goal of the survey was to provide a robust census of resolved properties across the entire massive galaxy population and to track consistently the evolution thereof from the peak in cosmic SFR activity to well into the ``winding down'' epochs.  To this aim, the cornerstones of the survey strategy were
{
(1) a homogeneous coverage in redshift and galaxy stellar mass, and a wide span in SFR and colours,
}
(2) the use of the same spectral diagnostics across the entire redshift range, and (3) deep integrations to map faint, extended line emission and ensure high quality data of individual galaxies. The targets were drawn from the {\it Hubble Space Telescope\/} ({\it HST\/}) ``3D-HST'' Treasury Survey source catalog \citep{2014ApJS..214...24S,2016ApJS..225...27M}, providing a well-characterized parent sample with source detection and accurate redshifts based on rest-frame optical properties, largely reducing the bias towards blue, rest-UV bright galaxies of optical spectroscopy, which becomes especially severe at $z \ga 1.5$. 3D-HST overlaps with the CANDELS survey fields \citep{2011ApJS..197...35G,2011ApJS..197...36K} that were imaged with {\it HST\/} in the near-IR and optical, and which benefit from extensive coverage from the X-ray to far-IR radio regimes. The selection criteria were solely based on (i) stellar mass and $K$-band (rest-frame optical) magnitude cuts, (ii) reliability of the redshift, and (iii) the emission lines of interest falling in near-IR atmospheric windows and away from bright sky lines.  By avoiding selection on colors or properties sensitive to star formation or AGN activity, and by covering 5 Gyrs of cosmic time, \kmostdns\ is optimally suited for population censuses and evolutionary studies.  By emphasizing sensitive observations, the survey successfully probed line emission in parts of the galaxy population that had been unexplored by previous near-IR IFU surveys.

With this design, the main science drivers of \kmostdns\ included the dynamics, angular momentum, and structure of galaxies, galactic outflows, chemical enrichment, and quenching of star formation activity.  Key science results addressing these goals based on subsets of the targets were published in the course of the 5-year period of the observing campaigns, summarized here.
\kmostdns: \\
\hphantom{}\hspace{0.8em}$\bullet$
robustly confirmed the majority ($\ga 70\%$) of rotating disks among $z \sim 1 - 3$
SFGs with greater turbulence observed through elevated disk velocity dispersions
\citep{2015ApJ...799..209W,2019ApJ...880...48U}; \\
\hphantom{}\hspace{0.8em}$\bullet$
showed that the angular momentum distribution of high-$z$ SFGs reflects that of
their host dark matter halos
\citep{2016ApJ...826..214B}; \\
\hphantom{}\hspace{0.8em}$\bullet$
revealed that high-$z$ disks become increasingly baryon-dominated out
to $z \sim 2.5$, based on the baryon-to-dynamical mass fractions, the zero point
of the stellar and baryonic Tully-Fisher relations, and the
shape of the outer rotation curves out to $3 - 4$ times the effective radius
\citep{2016ApJ...831..149W,2017ApJ...840...92L,2017ApJ...842..121U,2017Natur.543..397G}; \\
\hphantom{}\hspace{0.8em}$\bullet$
established the trends with stellar mass and SFR of the incidence, strength,
velocity, electron density, and mass ejection rate of ionized gas outflows,
and the high duty cycle $> 50\%$ of nuclear, AGN-driven winds at
$\log(M_{\star}/{\rm M_{\odot}}) \ga 11$
\citep{2014ApJ...796....7G,2019ApJ...875...21F}; \\
\hphantom{}\hspace{0.8em}$\bullet$
provided new constraints on metallicity scaling relations and evidence in support
of typically flat gas-phase oxygen abundance gradients among high-$z$ SFGs
\citep{2014ApJ...789L..40W,2016ApJ...827...74W}; \\
\hphantom{}\hspace{0.8em}$\bullet$
shed new light on dense core formation and quenching, by unveiling star-forming
disks, gas outflows, and signs of rejuvenation events in compact SFGs and massive
sub-MS galaxies
\citep{2017ApJ...841L...6B,2018ApJ...855...97W}.

With this paper, we present the complete \kmostdns\ sample of \totalobs~galaxies, and the accompanying data release.  This release includes reduced data cubes and key galaxy properties including redshifts, SFRs, $M_*$, colors, and \halpha fluxes. 
We describe the survey design and the global properties of the sample in Section~\ref{sec.selection}. 
We present the full observational and data reduction procedures in Section~\ref{sec.obs} and \ref{sec.datareduction} respectively, and the associated products in Section~\ref{sec.results} and~\ref{sec.resolved}.
In Section~\ref{sec.sfrs}, we take advantage of the spectral and spatial resolution afforded by the KMOS data to derive SFR estimates from H$\alpha$ without contamination by neighbouring [\ion{N}{2}] line emission and underlying broad emission from outflowing gas, and examine relationships with other SFR indicators.
We revisit the kinematic properties and classification of $z \sim 0.7 - 2.7$ galaxies with the complete \kmostd sample and data sets in Section~\ref{sec.discuss}. We assume a $\Lambda$CDM cosmology with $H_0 = 70$ km s$^{-1}$ Mpc$^{-1}$, $\Omega_m = 0.3$, and $\Omega_\Lambda = 0.7$. For this cosmology, $1''$ corresponds to $\sim7.8$ kpc at $z = 0.9$, $\sim8.2$ kpc at $z = 2.3$. We adopt a \cite{2003PASP..115..763C} initial mass function.

%----------------------------------------------------------------------
\section{Sample selection}
\label{sec.selection}
%----------------------------------------------------------------------
\begin{figure*}[!ht]
%\figurenum{1}
\begin{center}
\includegraphics[scale=0.65,trim=80 20 0 30, clip=1,angle=0]{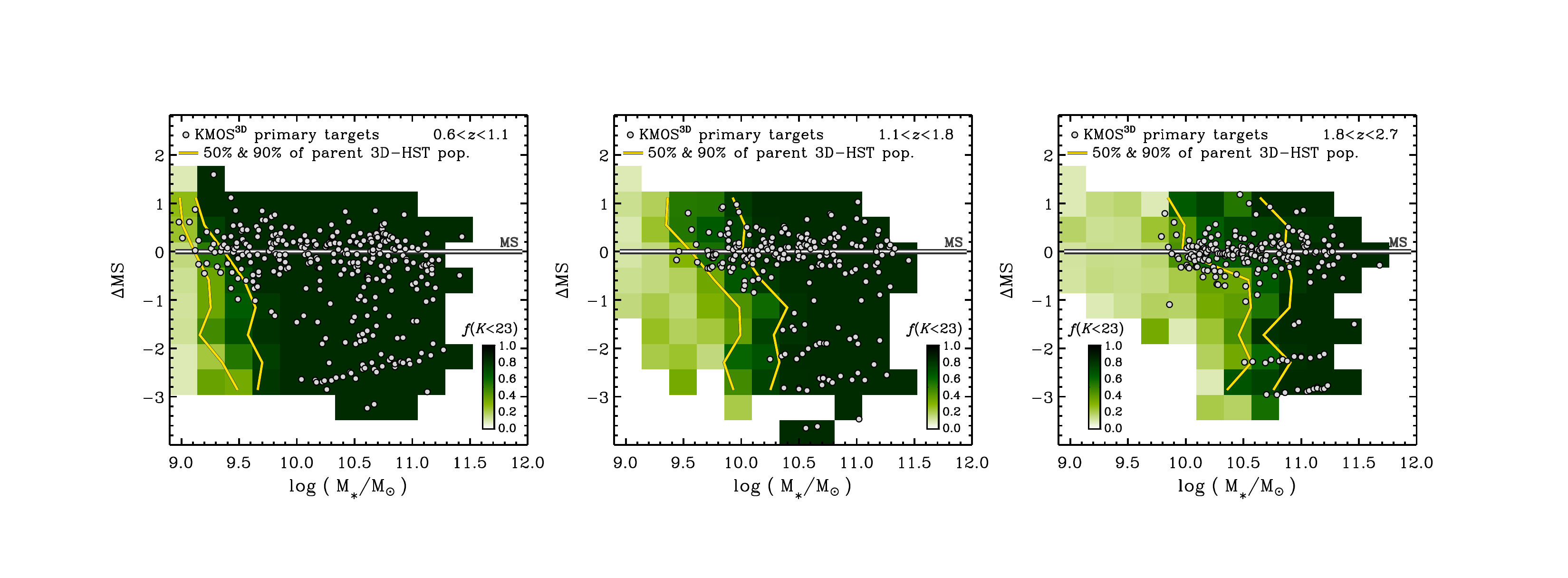}
\end{center}
\vspace{-0.9cm}
\renewcommand\baselinestretch{0.5}
\caption{
\small
%{\color{blue}
Effect of the $K$-band cut applied in selecting the \kmostd targets.
The fraction of objects at $K < 23~{\rm mag}$ relative to the number of
galaxies in the parent 3D-HST sample at $\rm \log(M_{\star}/M_{\odot}) > 9.0$
and $\rm F160W < 25.1~mag$ (the photometric 90\% completeness) is shown
in colors across the $\rm \log(M_{\star})$ versus $\rm \Delta MS$
plane.  From left to right, the panels correspond to the redshift slices
spanned by the \kmostd targets observed in the $YJ$, $H$, and $K$ band,
respectively, as labeled in the top right corners.  Low to high fractions
are represented with increasingly dark colors according to the color bars
in each panel.  Cells are not colored if they contain $< 0.05\%$
of the total number of objects in the parent sample, unless they include a
source observed in \kmostdns.  The \kmostd ``primary'' targets selected for
observations (excluding the serendipitous detections discussed in Section~\ref{sec.results}),
are overplotted as white-filled circles, and the horizontal line indicates
the MS (by definition $\rm \Delta MS = 0$) for reference.
The yellow lines mark the locii for fractions of 50\% and 90\%.
Imposing the $K < 23~{\rm mag}$ brightness cut leaves the $z \sim 1$ subset
highly complete down to $\rm \log(M_{\star}/M_{\odot}) \sim 9.5$.  The high
completeness mass range is reduced for increasing redshift intervals, and
secondarily for lower $\rm \Delta MS$ at fixed redshift.
The \kmostd targets cover the stellar mass ranges in each redshift interval
where $\ga 50\%$ of the parent 3D-HST population is brighter than $K = 23~{\rm mag}$.
%}  %  End color blue
\label{fig.K23effect}
}
\vspace{2.0ex}
\end{figure*}

All \kmostd targets were drawn from the 3D-HST grism Treasury Survey \citep{brammer:2012:04,2014ApJS..214...24S,2016ApJS..225...27M}. The 3D-HST survey observed five extragalactic fields (COSMOS, GOODS-S, GOODS-N, UDS, AEGIS) with \textit{HST} WFC3/G141 grism providing spectra with resolution of $R\sim130$ over $\lambda = 1.1-1.7$ $\mu$m. Grism redshifts derived from emission line and continuum fitting are used for selection when a prior spectroscopic redshift is unavailable. 
Targets for \kmostd were selected to be within the COSMOS, GOODS-S, and UDS fields visible from the VLT, and in the range $0.7<z<2.7$ for which the main emission lines of interest fall within near-IR atmospheric windows.  More specifically, observations through the KMOS $YJ$, $H$, and $K-$band filters cover H$\alpha$ for sources at $0.7<z<1.1$, $1.2<z<1.8$, and $1.9<z<2.7$, respectively (hereafter referred to as $z\sim1$, $z\sim1.5$, and $z\sim2$).
The selection of galaxies with a prior grism or spectroscopic redshift provides a high targeting accuracy, increasing the probability that \halpha emission falls in the observed band. {In the final sample,  36\% of targeted galaxies had a prior spectroscopic redshift. The remaining targeted galaxies were selected based on grism redshifts.} Detection fractions and redshift accuracy are discussed further in Section~\ref{sec.detectionfns}. 

The \kmostd targets have a fairly homogeneous set of spectral energy distributions (SEDs) from the extensive 
multi-wavelength coverage from X-ray to far-IR and radio available in all fields
(e.g.\ \citealt{2008ApJS..179..124U, lutz:2011aa,2011ApJS..195...10X,2012ApJS..201...30C,2013AA...553A.132M,2014ApJS..214...24S}).
The CANDELS survey contributes high-resolution WFC3 near-IR and ACS optical imaging for all the targets \citep{2011ApJS..197...35G,2011ApJS..197...36K}.
Global galaxy properties such as stellar mass, SFRs, and correction for global dust extinction are derived following \cite{2011ApJ...738..106W}. In brief, the optical to 8\,$\mu$m SEDs are fitted with \citet{2003MNRAS.344.1000B} models assuming solar metallicity, the \cite{2000ApJ...533..682C} reddening law, and either constant or exponentially declining SFRs. Star-formation rates are determined from the same SED fits or, for objects observed and detected in at least one of the mid- to far-IR (24$\mu$m to 160$\mu$m) bands with the \textit{Spitzer}/MIPS and \textit{Herschel}/PACS instruments, from rest-UV+IR luminosities through the Herschel-calibrated ladder of SFR indicators of \cite{2011ApJ...738..106W}. Resolved information of stellar populations, dust extinction and stellar mass maps derived from high resolution (FWHM$\sim0.15-0.20"$) four-band imaging ($VIJH$) in UDS and COSMOS and seven-band imaging ($BVizYJH$) in GOODS-S complement the kinematics, star formation and nebular emission data derived from KMOS for a combined view of resolved gas and stellar profiles of individual galaxies \citep{2012ApJ...753..114W,2013ApJ...779..135W,2013ApJ...763L..16N,2014ApJ...788...11L}. A complementary environment catalog in the same fields is also available from \cite{2017ApJ...835..153F}.

All \kmostd targets were selected to have a $K$-band magnitude $<23$ AB and stellar mass $>10^{9}$ \Msun. No cut involving SFR and/or colors was applied in the target selection to avoid an explicit bias towards the most actively star-forming and/or bluest galaxies. Galaxies with a grism redshift satisfied the additional criteria of having a grism quality $Q_z\le1$, a grism covering fraction $f_\mathrm{cover}>0$, a contamination integrated over the spectrum $f_\mathrm{int~contam}<1$, a fraction of flagged pixels $f_\mathrm{flagged}\le1$, and a star flag $\ne1$. These quality flags are approximately equivalent to selecting on the 3D-HST public data release v4.1.5, `use\_grism' flag. More details of the grism quality flags are found in \cite{2016ApJS..225...27M}. 
{The requirement of a sufficiently accurate redshift (i.e. a grism
redshift $z_{\rm gr}$ or spectroscopic redshift $z_{\rm sp}$) does
not appreciably alter the distribution in stellar mass, MS offset,
rest-frame $UVJ$ colors, and offset relative to the mass-size
relation of SFGs compared to a $0.7 < z < 2.7$ and
$\rm \log(M_{\star}/M_{\odot}) > 9$ sample from the 3D-HST catalog
up to the 90\% photometric completeness at $\rm F160W = 25.1~mag$
(Skelton et al. 2014) and at $K < 23~{\rm mag}$.
Specifically, and within this $z$ interval and with these $\rm M_{\star}$
and magnitude cuts, Kolmogorov-Smirnov (KS) tests show that the subset
of the 3D-HST parent population with a $z_{\rm gr}$ or $z_{\rm sp}$
does not differ significantly from the full population including also
sources with photometric redshift ($z_{\rm phot}$ only; this holds
also for the three redshift bins separately.
In contrast, the subset with only $z_{\rm sp}$ does show significantly
different distributions ($p$ values $< 0.05$), in particular with higher
levels of star formation activity, and bluer colors especially at $z > 1.3$.
We stress that the grism redshifts rely on rest-optical spectral features
including {\it both} the continuum and emission lines (Momcheva et al. 2016)
and we do not preferentially select targets with line emission detected
in the grism data.
Thus, the target selection including objects with a $z_{\rm gr}$ largely
reduced the biases towards bluer, more star-forming galaxies compared
to one considering only objects with a $z_{\rm sp}$ over the redshift
and mass range of interest.}

The $K$-band cut culls objects brighter than the 3D-HST 90\% F160W photometric completeness,
and the impact is mainly to reduce the stellar mass range of the selected targets at increasing
redshifts.  It also tends to remove a larger proportion of sub-MS objects.
Following the approach of \citet{2009ApJ...701.1765M}, marginalizing over all other galaxy parameters,
$\rm F160W = 25.1~mag$ corresponds to 90\% mass completeness at $\rm \log(M_{\star}/M_{\odot}) \sim 8.9$,
$\sim 9.5$, and $\sim 10.1$ for the $z \sim 1$, $\sim 1.5$, and $\sim 2$ intervals, respectively.
Adding the $K < 23~{\rm mag}$ criterion, the 90\% mass completeness limits become
$\rm \log(M_{\star}/M_{\odot}) \sim 9.6$, $\sim 10.2$, and $\sim 10.6$.
When distinguishing galaxies based on star formation activity using the $UVJ$ color criteria of
\citet{2011ApJ...735...86W}, the mass completeness limits are lower by $\rm \sim 0.1~dex$ for the
star-forming subset, and higher by $\rm \sim 0.1~dex$ for the quiescent one.
To further illustrate the effects of the imposed $K$-band cut across the $\log(M_{\star})$ vs.\ $\rm \Delta MS$ plane,
Figure~\ref{fig.K23effect} shows the fraction of objects with $K < 23.0~{\rm mag}$ among all galaxies at
$\rm \log(M_{\star}/M_{\odot}) > 9$ and $\rm F160W < 25.1~mag$ from the 3D-HST parent population.
The different panels correspond to the different redshift slices, and the \kmostd targets are overplotted.
The galaxies targeted in \kmostd probe regions in $\log(M_{\star})$ vs.\ $\rm \Delta MS$ where $\ga 50\%$
of the parent 3D-HST population is brighter than $K = 23~{\rm mag}$.

%OH sky avoidance
Strong spectral features from the Earth's atmosphere (e.g.\ OH sky lines, molecular features), prevalent at NIR wavelengths, can contaminate resolved emission-line observations. To reduce overlap of \kmostd emission profiles with atmospheric features, models of the sky emission and absorption features were taken into account during target selection. For each galaxy a probability of being in a clear spectral region was computed (hereafter ``visibility'') based on the Cerro Pachon site sky background and transmission spectra\footnote{The models used are for an airmass of 1.5 and a water vapor column of 4.3: cp\_skybg\_zm\_43\_15\_ph.dat, cptrans\_zm\_43\_15.dat. Available at \url{http://www.gemini.edu/sciops/telescopes-and-sites/observing-condition-constraints/ir-background-spectra}, and \url{http://www.gemini.edu/sciops/telescopes-and-sites/observing-condition-constraints/ir-transmission-spectra} }. The sky emission spectrum was inverted and multiplied with the atmospheric transmission spectrum. For each galaxy a visibility was computed, such that a visibility of 1 corresponds to a redshift with \halpha most likely clean of atmospheric effects and a visibility of 0 corresponds to a redshift with \halpha in a wavelength region with 100\% atmospheric absorption or on the brightest sky-line in that waveband. The visibility was weighted by the probability distribution function (PDF) representative of the redshift confidence. For spectroscopic redshifts the PDF was a Gaussian with a $\sigma=400$ \kms. For a grism redshift the 3D-HST PDF was convolved with a  $\sigma=1000$ \kms~Gaussian representative of the typical grism spectral resolution \citep{2016ApJS..225...27M}. Targets for KMOS were selected to have both an \halpha and \NII$\lambda6584$ visibility $\ge0.5$.  The sky line emission and low transmission avoidance criteria removes $\sim70$\% of possible targets in the full redshift range. Finally, the 3D-HST 2D spectrum for each galaxy was visually inspected by multiple team members.  Over the three fields, 142 galaxies were removed due to low-S/N in the grism or low grism coverage on the basis that their grism spectrum would not significantly improve the photometric redshift estimate ($z_\mathrm{grism} \approx z_\mathrm{phot}$). 

The final targets for observation from the resulting source list were based on the positions and density of galaxies on the sky relative to the field of view of KMOS, the availability and positioning of the 24 IFU arms, and moon  distance and illumination.

The version of the 3D-HST selection catalog changed over the course of the survey. During ESO period 92 the targets were selected from the 3D-HST v2.1 catalog. From ESO period 93 (April 2014) to 96 (September 2015) the 3D-HST v4.0 catalog was used, and from ESO period 97 (April 2016) onward targets were selected from the publicly released 3D-HST v4.1.5 catalog. The adoption of the v4.0 catalog was a result of improved imaging mosaics and thus photometry produced by the 3D-HST team. However, a shallower depth was used to extract grism redshifts for v4, $F140W<23$, than the $F140W<24$ limit of the previous catalog. As a result, when using the v4.0 catalog for target selection, additional targets to fill pointings were drawn from the v2.1 catalog with $23<F140W<24$ and fulfilling all other \kmostd criteria. The adoption of the publicly available v4.1.5 was a result of updated photometric catalogs, extraction of grism redshifts to $F140W<26$, and new quality flags. {For the data release presented here we use v4.1.5 for galaxy $ID$, $RA$, $DEC$. The redshift used to originally select the galaxies from the respective catalog is given as $z_\mathrm{best,orig}$ throughout the paper and as ``Z\_TARGETED" in the accompanying catalog.}
The targeted ID is given by ``ID\_TARGETED" in the released catalog. Galaxies selected from v2.1 have the prefix COS3, GS3, or U3 in their ``ID\_TARGETED"  for the COSMOS, GOODS-S and UDS fields respectively. Galaxies selected from v4 and v4.1.5 have the prefix COS4, GS4, or U4.\footnote{All but one galaxy, U3\_10584 selected from v2.1, has a v4.1.5 ID in \cite{2014ApJS..214...24S}. U3\_10584 at $z_\mathrm{KMOS}=2.246$, is included in this release with a numeric $ID$ 99999, with no counterpart in the \cite{2014ApJS..214...24S} catalog. Galaxy properties for this galaxy given in the \kmostd release are derived from the photometry in the v2.1 3D-HST catalog.}

%----------------------------------------------------------------------
\section{Observations}
\label{sec.obs}
%----------------------------------------------------------------------
%Seeing statistics summary, description of percent of data rejected for bad seeing, etc., observing time summary
% DEFINE NUMBERS
\def\uniquekarma{X}
\def\obstimeYJ{5.0}
\def\obstimeH{8.5}
\def\obstimeK{8.7}
\def\moonfactYJ{1.7}
\def\moonfactH{1.4}
\def\moonfactK{1.0}

\begin{figure}[!tb]
\begin{center}
\includegraphics[ scale=0.9, trim=0.5cm 16.5cm 5cm 0cm, clip]{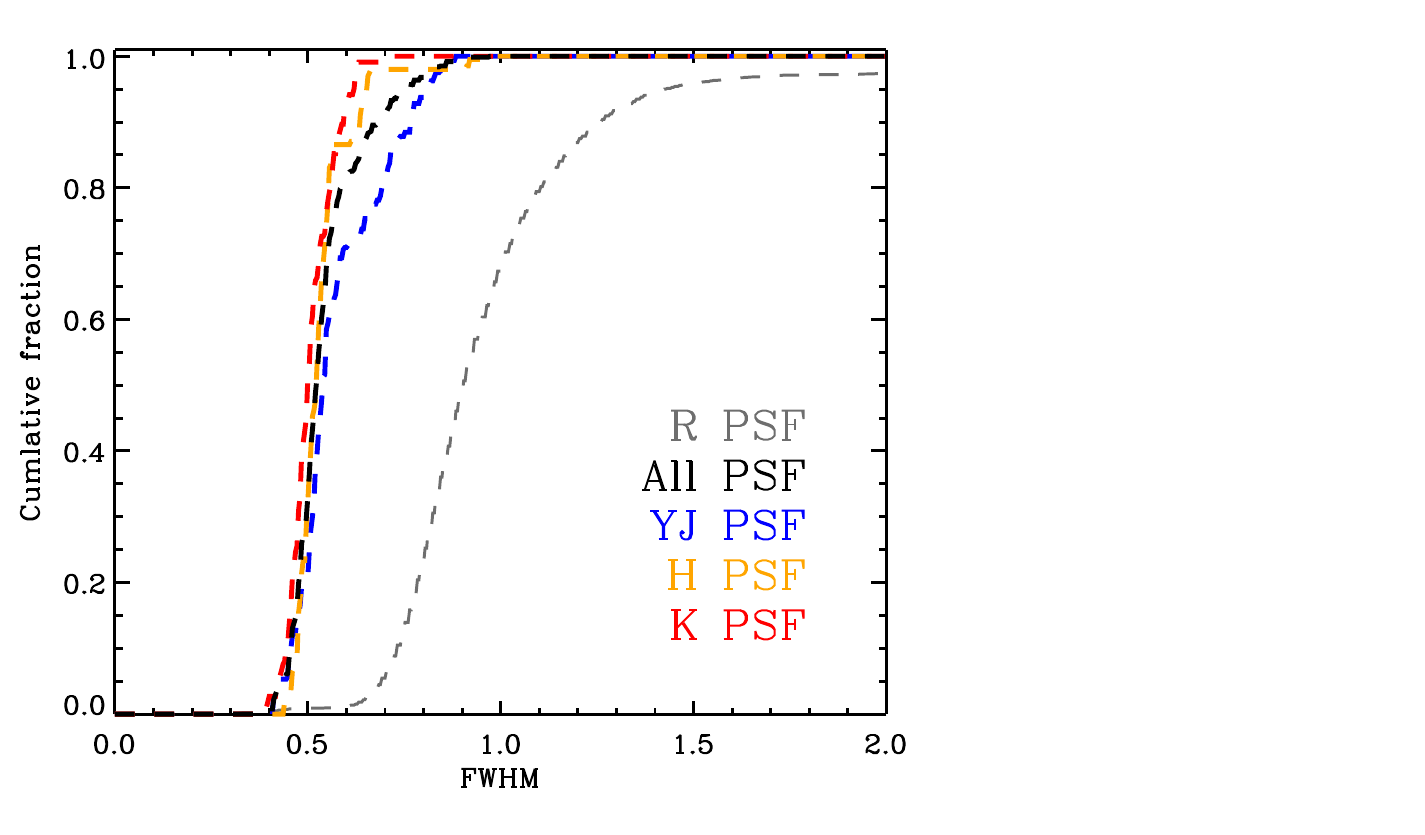}
\caption{  Cumulative distributions of seeing and image quality for all \kmostd data. The dark gray dashed line shows the distribution of airmass corrected seeing measured from the guide probe in $R$-band for all 300 second KMOS science frames. The black dotted line shows the distribution of FWHM measurements from flattened images of PSF stars observed simultaneously, in the same wave band, and same length of time as the galaxy observations. The PSF images and associated fits used here are described in detail in Section~\ref{sec.datareduction}. The blue, orange, and red distributions correspond to the PSF FWHM measurements separated into $YJ$, $H$, and $K$-band observations respectively.}
\label{fig.psffwhm}
\end{center}
\end{figure}

%General observing & seeing
%DIMM = "Differential Image Motion Monitor "
Observations with KMOS took place in Visitor Mode over 75 guaranteed time nights between October 2013 and April 2018 (ESO period 92 $-$ 101). Data were collected in excellent seeing conditions, with 70\% of the individual data frames taken in sub-arcsec seeing as measured in the $R$-band by the guide probe and corrected for airmass\footnote{ESO HIERARCH fits keyword TEL.IA.FWHM}. The image quality corresponding to the combined data cubes is higher, with 70\% of the data having a PSF FWHM $\leq0.55$ arcsec when measured in the $YJ$, $H$, or $K$ wavebands (as expected for redder wavebands). The distributions of seeing measured from the DIMM for individual frames and the PSF FWHM of the data measured from PSF stars (discussed in Section~\ref{sec.datareduction}) are shown in Figure~\ref{fig.psffwhm}. 

\begin{figure}[!tb]
\begin{center}
\includegraphics[ scale=0.38,angle=90, trim=0cm 0cm 0cm 1.0cm, clip]{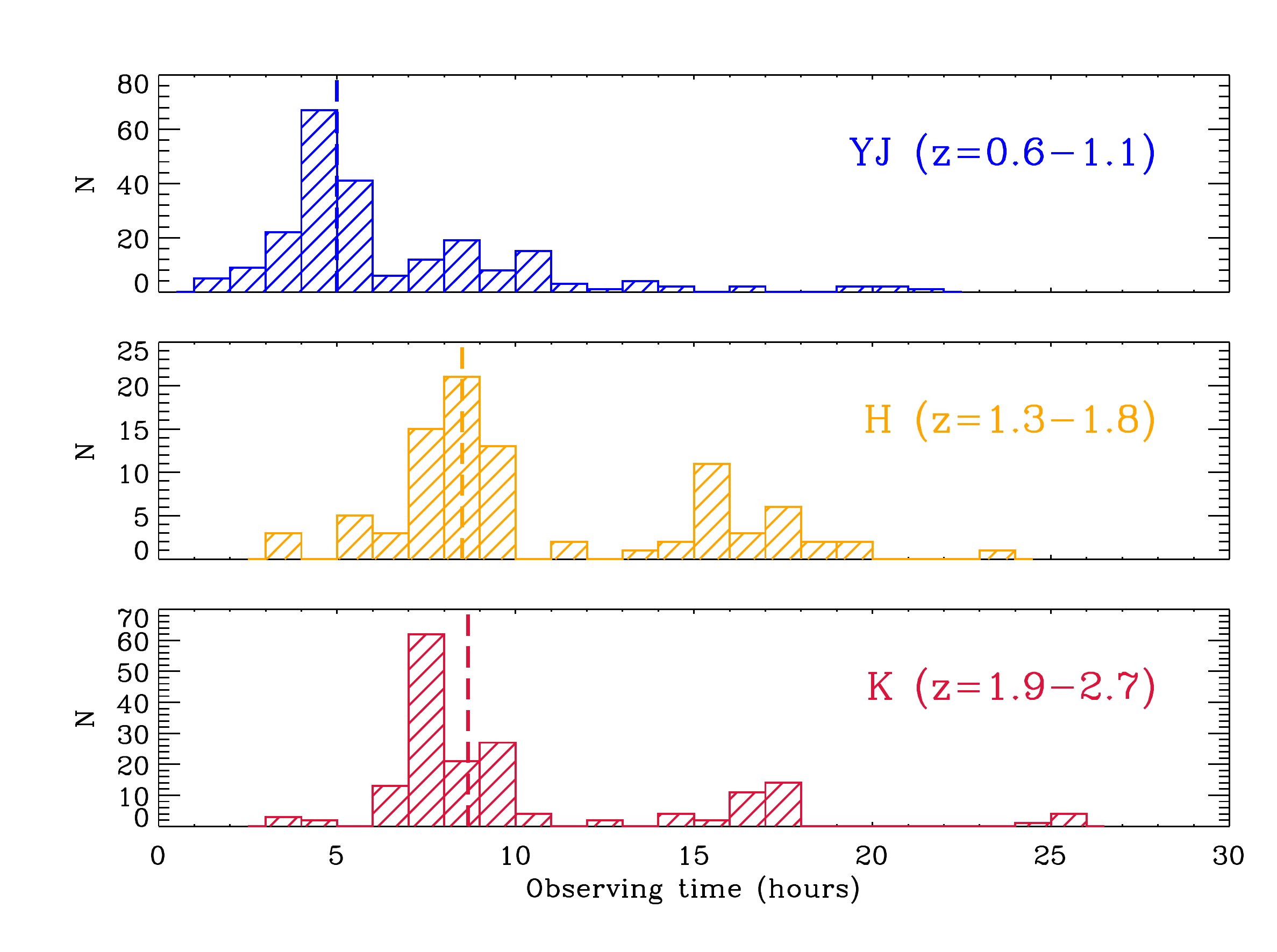}
\caption{{Observing time histograms for \kmostd galaxies targeted in the $YJ$ (top), $H$ (middle), and $K$-bands (bottom). The median observing times for galaxies in the redshift slices $z\sim1$, $z\sim1.5$, and $z\sim2$, after bad frames were removed, are \obstimeYJ, \obstimeH, and \obstimeK~(shown by the dashed vertical lines). A fraction of targets were observed in multiple pointings leading to the double peaked distributions or extended tails in the observing time histograms.}}
\label{fig.obstime}
\end{center}
\end{figure}

%definition of pointing and observing time statistics
Observational setups were prepared with the KMOS Arm Allocator (KARMA; \citealt{2008SPIE.7019E..27W}). Hereafter an individual KARMA setup, or 24 arm allocation, will be referred to as a ``pointing''.  
Individual galaxies were commonly observed in multiple KARMA pointings to obtain higher signal to noise (S/N). Targeted galaxies with the longest observing times ($>14$ hours) are objects fulfilling a key science area of the survey {reliant on detection of low surface brightness features} such as galactic scale winds \citep{2014ApJ...796....7G, 2019ApJ...875...21F} {or reliant on low-levels of star formation} such as outer rotation curves \citep{2017ApJ...840...92L,2017Natur.543..397G} and line emission detections of $UVJ$ passive galaxies \citep{2017ApJ...841L...6B}. Median on-source observing times for $z\sim0.9$, $z\sim1.4$ , and $z\sim2.3$ galaxies are \obstimeYJ, \obstimeH, and \obstimeK~hours respectively. A histogram of observing times per band is given in Figure~\ref{fig.obstime}.% as well as the MS color-coded by observing time. 

%pointing set-ups and observing strategy
Each pointing was observed for a series of 300s (ESO period 92$-$101) using a standard object(O)-sky(S) dither pattern (e.g.\ OSOOSO), where sky exposures were offset to a clear sky position. Additional subpixel/pixel shifts were included in the object-sky dithering to reduce the impact of bad pixels. Exposure maps are created for each combined datacube that trace the gradient in depth from the object location to the edge of the cube as a result of dithering. Three IFUs, one in each spectrograph sub-system of KMOS, were allocated to a  ``PSF star'' during science observations. The stars are used to monitor variations in the seeing and photometric conditions between the observed frames and in each of the three detectors. They were selected to have typical magnitudes of $16<m_\mathrm{F140W}<18$. 

%Effect of moon on background of science data?
Observations were taken in the full range of bright to dark time with $YJ$ observations prioritised in dark time and $K$ observations prioritised in bright time. Pointings were occasionally observed as close as 15 degrees from the moon. For each IFU in each 300s  O-S frame we measure the background level and error on the background (the standard deviation of the background levels). We find no difference on the average background level after a simple O-S subtraction with moon distance or illumination in any waveband. However, for $YJ$, $H$, $K$ observations we measure a \moonfactYJ, \moonfactH, and \moonfactK~factor increase in the error on the background below a moon distance of $30$ degrees, respectively. High moon illumination did not result in increased error on the background, however high moon illumination typically corresponded to observations with larger moon distances.

Calibrations are taken at then end of each night following standard ESO procedures. They are run in each waveband for which science observations were taken. These included darks (for identifying hot pixels), lamp flats, and arcs. No sky flats were taken during observation runs as they were determined to cause persistence on the detectors when observed in evening twilight. Standard stars for telluric transmission and flux calibration were typically observed at the start and end of the night as well as between pointings as discussed in Section~\ref{sec.fluxcal}.

%----------------------------------------------------------------------
\section{Data reduction}
\label{sec.datareduction}
%----------------------------------------------------------------------
% DEFINE NUMBERS
\def\zpYJstddev{0.19}
\def\zpHstddev{0.39}
\def\zpKstddev{0.26}
\def\resYJ{3515}
\def\resH{3975}
\def\resK{3860}
\def\nzrotangle{6}
\def\nzrotanglenum{39}
\def\magYJstddev{0.13}
\def\magHstddev{0.18}
\def\magKstddev{0.19}
\def\magYJmedian{ -0.09}
\def\magHmedian{-0.07}
\def\magKmedian{-0.04}
The data were reduced with the SPARK software version 1.3.5 \citep{2013AA...558A..56D} and custom PYTHON and IDL scripts. The workflow for the data reduction is described below. 

%%%%%%%%%%%%%%%%%%%%%%%%%%%%
\subsection{Detector-level corrections}
\label{detector-level corrections}
%%%%%%%%%%%%%%%%%%%%%%%%%%%%

% DETECTOR CORRECTIONS
%{\color{Gray}Detector Corrections:}\\
{
A number of processing steps were performed on the individual detector images before subsequent processing by SPARK. These include corrections for the read-out channel-dependent bias level, alternating column noise (ACN), and picture frame noise effects described by \cite{2015PASP..127.1144R}.  For each science exposure we first removed a channel-dependent bias level using reference pixels around the perimeter of each of the three KMOS HAWAII-2RG detectors.  This bias removal included a correction for ACN, which was also estimated from the reference pixel arrays. In a subset of KMOS exposures---particularly those with large negative or positive median reference pixel values---we found that the bias- and ACN-corrected frames showed significant spatial non-uniformity around their perimeters.  This appears to be a manifestation of the so-called "picture frame" noise discussed by \cite{2013SPIE.8860E..05R} and \cite{2015PASP..127.1144R}, and in KMOS appears to be due to drifts in the bias voltage (Elizabeth George, priv. comm.).  Offsets of up to $\pm$2 counts are present in 10-15\% of all exposures taken with detectors 1 and 2, and in nearly 30\% of exposures taken with detector 3; in the most extreme cases offsets of up to $\pm$10 counts are observed. In order to correct for these effects we used a set of $\sim$7500 dark frames (per detector) to estimate the correlation between the median reference pixel value and each pixel in the science array.  We then estimated and removed the residual picture frame noise based on the reference pixel values in each science exposure.}

%%%%%%%%%%%%%%%%%%%%%%%%%%%%
\subsection{Sky subtraction and heliocentric correction}
%%%%%%%%%%%%%%%%%%%%%%%%%%%%
% ZAP SKY SUBTRACTION
%{\color{Gray}Zap Sky Subtraction:}\\
{
The corrected cubes were reconstructed using standard SPARK routines, including a frame-by-frame correction for the wavelength solution based on cross-correlation with a reference OH spectrum.  Sky subtraction was then performed external to SPARK in two steps: first a simple O-S subtraction based on the adjacent sky cubes, followed by a removal of sky line residuals using a modified version of the ZAP Principle Component Analysis (PCA) sky subtraction code \citep{2016MNRAS.458.3210S}.  Due to the relatively small size of individual KMOS IFUs ($14\times14$ spaxels), principle components (PCs) for ZAP were estimated from a subsample of sky exposures taken over the lifetime of \kmostd observations.  The final set of reference spectra consisted of $\sim 5 \times 10^{5}$ individual spectra in each band with approximately uniform distributions in elevation angle and time of year.  We found that PCs measured in this way were better able to account for variations in the relative strength of different ro-vibrational OH transitions and spectrograph line spread function compared to PCs constructed from individual observing runs or observing semesters {(Mendel et al. \textit{in prep.})\footnote{For more details contact trevor.mendel@anu.edu.au}}.}

\begin{figure}[t]
\begin{center}
\includegraphics[ scale=0.37, angle=90, trim=0cm 0cm 0cm 0.5cm, clip]{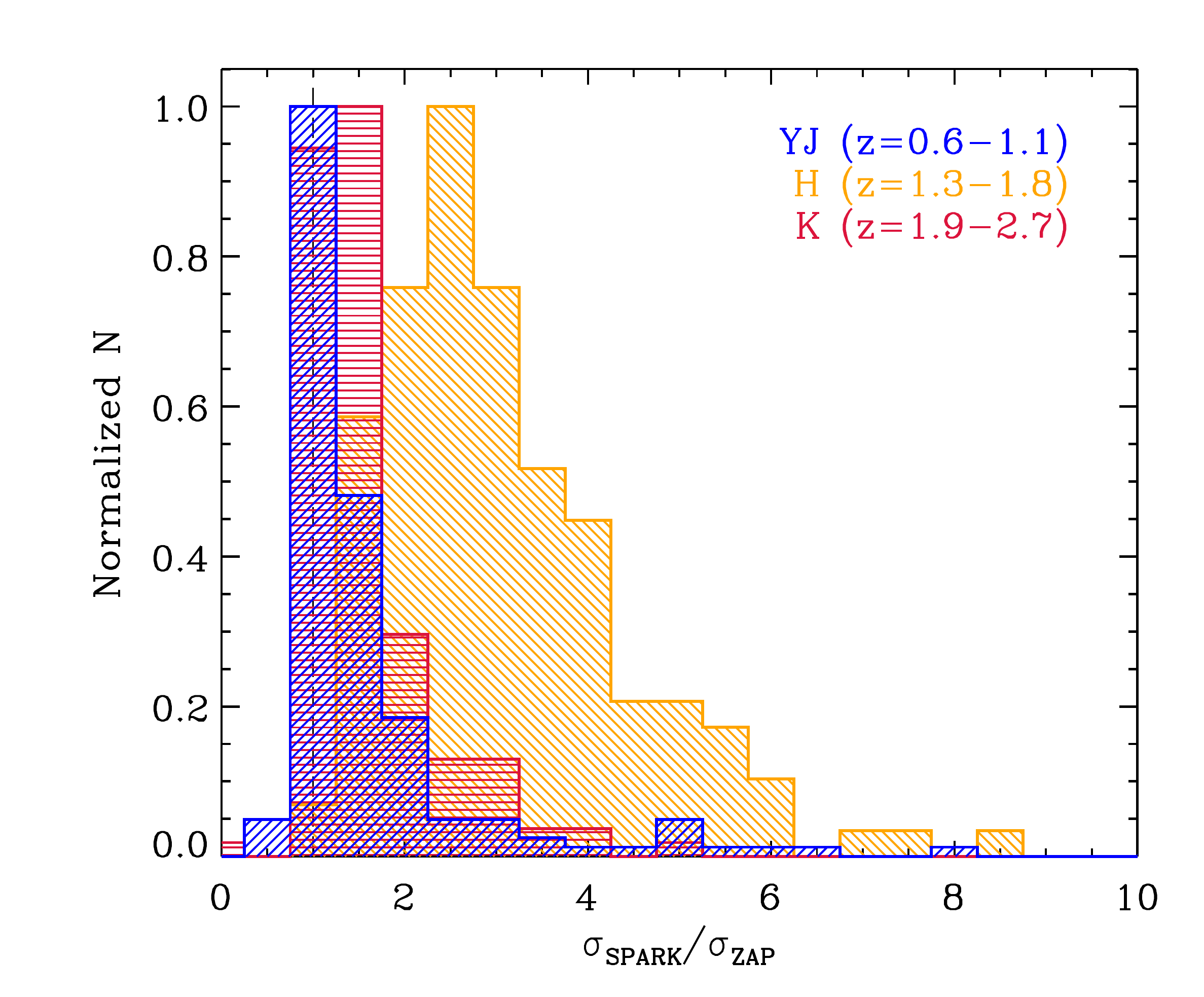}
\caption{Improvement over standard sky subtraction practices as a result of using principle component analysis (PCA) techniques. For each galaxy, the standard deviation of the spectrum, $\sigma$, is calculated over $\pm$7500 \kms~around the detected or expected location of \halpha when using either standard sky subtraction techniques implemented in SPARK,  $\sigma_\mathrm{SPARK}$ or PCA techniques used for the \kmostd cubes,  $\sigma_\mathrm{ZAP}$. The histograms show the ratio of the standard deviation of the spectra from both techniques split by observing band.}%{\color{BrickRed} should I make these cumulative distributions?}}
\label{fig.zap}
\end{center}
\end{figure}

The application of PCA sky subtraction to the KMOS data provides a noticeable reduction to both the residual OH lines and molecular features such as the O2 feature at $1.26-1.28\mu$m in the $YJ$ band. Figure~\ref{fig.zap} shows the ratio of the standard deviation for extracted galaxy spectra with the standard sky-subtraction routine implemented in SPARK to the PCA sky-subtraction routine implemented in the \kmostd data. The standard deviation is calculated over $\pm$7500 \kms~around the detected or expected location of \halpha from a spectrum extracted by summing spaxels in a $2"\times2"$ window centered on the cube. A reduction in the standard deviation of the galaxy spectra is measured in all bands. The improvement is most dramatic in the $H$-band (orange histogram) where sky emission lines are both stronger and more closely packed than in the $YJ$ (blue) and $K$-bands (red).

% HELIOCENTRIC CORRECTION
A heliocentric correction is applied to all data frames before they are combined. The corrections range from -30 km s$^{-1}$ to +30 km s$^{-1}$.
The correction is especially important for the observations of the same objects in different semesters. Uncorrected data can lead to inaccurate redshifts and to inflated integrated velocity dispersions, particularly for narrow emission lines that are near the instrument resolution limits.  

% CALIBRATIONS
%{\color{Gray}Calibrations:}\\
%%%%%%%%%%%%%%%%%%%%%%%%%%%%
\subsection{Illumination correction}
%%%%%%%%%%%%%%%%%%%%%%%%%%%%
A rotator angle dependent illumination correction is done per object-sky pair using the internal flat with the closest rotator angle. The matching angle, of the six available (30, 90, 150, 210, 270, 330), provides the best illumination correction with residual non-uniformity of $\sim\pm$3\% in IFU 1 and 23, which show the strongest gradients, and smaller elsewhere.

%%%%%%%%%%%%%%%%%%%%%%%%%%%%
\subsection{Flux calibration}
\label{sec.fluxcal}
%%%%%%%%%%%%%%%%%%%%%%%%%%%%
%FLUX CALIBRATION
%{\color{Gray}Flux calibration:}\\
Observations of A0, B, and G stars were taken before and after observing a KMOS pointing when the conditions allowed. The observed `standard' stars were selected from the Hipparcos Catalog \citep{1997AA...323L..49P} with known IR magnitudes \citep{2003yCat.2246....0C}. The star observations are used to apply both a telluric transmission correction and flux calibration to all individual science frames. The standard KMOS observing procedure was followed such that a single standard star is observed in three IFUs (one per detector). Observed stars were chosen to be at a similar airmass as the science data. 

Photometric zero points are calculated in the AB system using custom IDL routines. The observations of standard stars are collapsed to a 1D spectrum. The mean counts within a predefined wavelength range, matched to the central wavelengths of the 2MASS $J$, $H$, and $K$ filters, are used to derive the zero point. A model Moffat function is fit to the stars to correct for the small fraction of flux lost outside of the IFU {(typically 1-3\%)}. The zero point in each band is stable with standard deviations over the full survey equal to \zpYJstddev, \zpHstddev, and \zpKstddev~mags for $YJ$, $H$, and $K$ bands respectively. In the cases where the zero point is more than 2-$\sigma$ deviant from the mean, the standard star cubes are visually inspected. The few deviant zero points can be attributed to pointing errors or conditions with $>60$\% humidity.  Therefore, deviant zeropoints are replaced by the mean of all the zero points measured in the same band and detector over the duration of the survey. The zeropoints do not correlate with other recorded observing conditions such as airmass or seeing.

Many pointings were observed continuously for multiple hours during which conditions changed. Before applying a telluric and flux calibration we account for variations in the observing conditions of each frame over time. To make this correction the total flux in the PSF stars of each frame are compared to the median flux of the same stars in the three science frames observed closest in time to the standard star observations. 

A telluric transmission spectrum is created by dividing the standard star spectrum by a blackbody function of the effective temperature of the standard star and removing the intrinsic stellar absorption features.  Each spaxel is then divided by the telluric spectrum observed in the same detector.

\begin{figure}[!t]
\begin{center}
\includegraphics[ scale=0.75,angle=90, trim=10cm 5cm 0cm 0.7cm, clip]{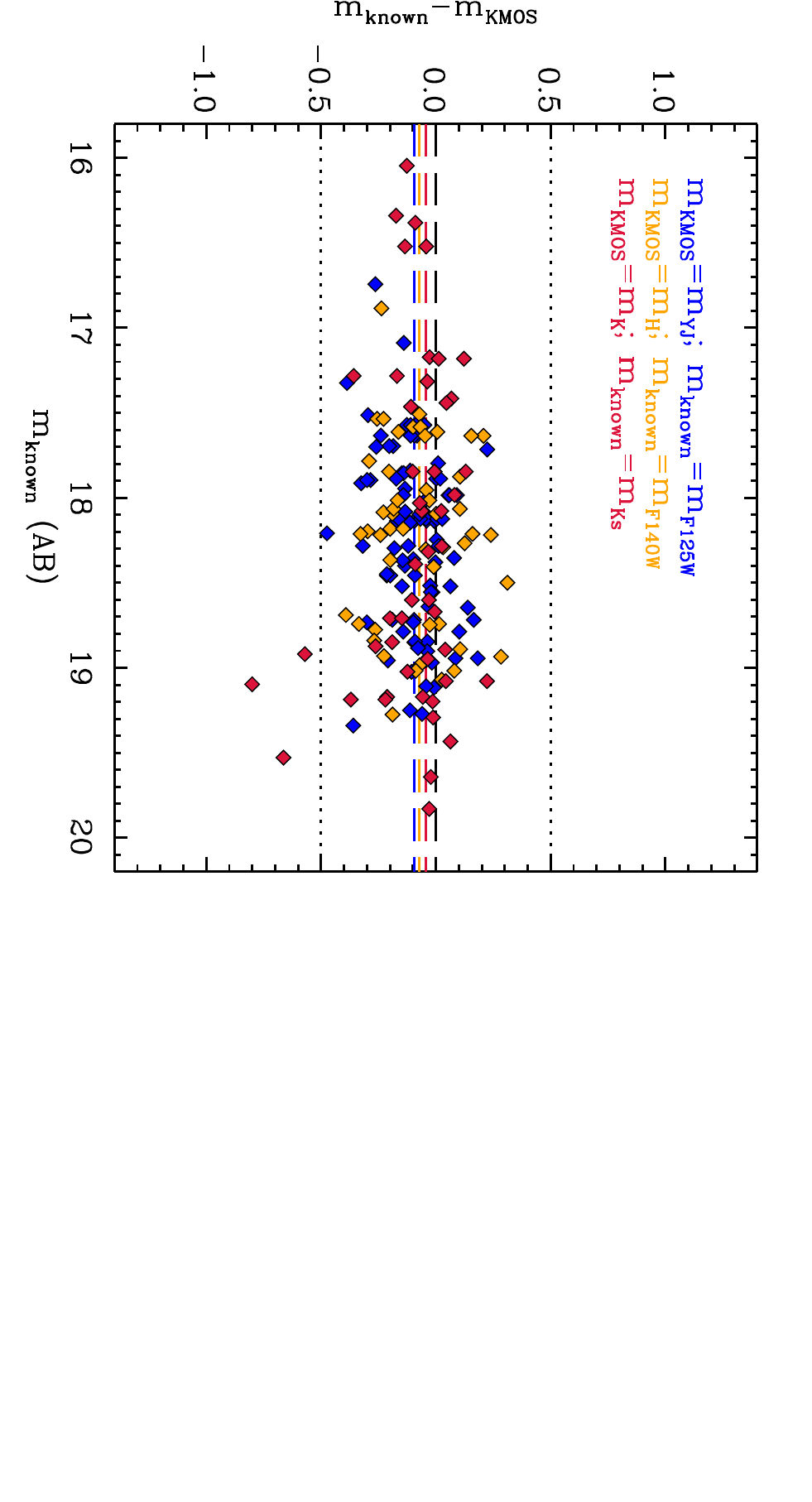}
\caption{Comparison of KMOS PSF star magnitudes  $YJ$ (blue), $H$ (orange), and $K$-band (red) to known HST (F125W, F140W) or ULTRAVISTA magnitudes (Ks). Diamonds show measurements from final combined images of the PSF stars with observation times shown in Figure~\ref{fig.obstime}. The difference between measured and known magnitude is shown as a function of known magnitude. The horizontal black dashed line shows an exact match. The dashed color lines show the median value $m_\mathrm{known}-m_\mathrm{KMOS}$ for the combined star cubes. The dotted lines show a half magnitude difference. The median and standard deviations are $\magYJmedian\pm\magYJstddev$, $\magHmedian\pm\magHstddev$,$\magKmedian\pm\magKstddev$ respectively for $YJ$, $H$, and $K$ bands.}
\label{fig.fcstars}
\end{center}
\end{figure}

An airmass correction is applied to account for the difference in elevation between observations of the standard star and each science frame. 
The zero point is then applied to derive the absolute flux scale for each science frame. Figure~\ref{fig.fcstars} shows the comparison of magnitudes derived from individual KMOS exposures of the PSF stars and known magnitudes from HST ($F125W$, $F140W$) and ULTRAVISTA ($Ks$) in similar wavebands to the KMOS $YJ$, $H$, and $K$ filters. No color correction has been applied to match the different filters. We compare the observed and known magnitudes for each of the three PSF stars in individual frames as well as the final combined PSF star images discussed in Section~\ref{subsec.psf}. To derive the magnitudes, the total flux in the flux calibrated star cube is summed in the wavelength range defined for calibration of KMOS data\footnote{https://www.eso.org/sci/facilities/paranal/instruments/kmos/doc/VLT-MAN-KMO-146606-002\_P100.pdf} and corrected for flux loss outside of the IFU using a Moffat model. The fluxes measured from the combined star data agree within $\sigma\lesssim$19\% with the fluxes derived from known magnitudes, with minor offsets with mean and standard deviations of $\magYJmedian\pm\magYJstddev$, $\magHmedian\pm\magHstddev$,$\magKmedian\pm\magKstddev$, in $YJ$, $H$, and $K$ bands respectively.  

%%%%%%%%%%%%%%%%%%%%%%%%%%%%
\subsection{Background subtraction}
%%%%%%%%%%%%%%%%%%%%%%%%%%%%
The steps described in Section 4.1 (detector-level corrections) provide an initial correction for the detector-level background in individual exposures.  In many cases an additional correction is required to remove residual spatial non-uniformity driven by channel-to-channel variation, as well as correct non-zero background levels, which otherwise limit our ability to push to low (continuum) surface brightness levels. 

We model the background in each reconstructed data cube as the combination of individual (detector) output channels and a spatially- and spectrally-uniform background component.  Because individual dispersed spectra are tilted with respect to the KMOS detectors, the relative contribution of output channels at a given spatial position in the reconstructed cubes varies as a function of wavelength.  We model this variation using "channel cubes" which provide a mapping between pixels in the reconstructed IFU data (in $x$,$y$,$\lambda$) and their corresponding detector output channels.  On average 4 distinct output channels contribute to any given IFU.  The median background correction derived per detector is small, $\sim$a few $\times10^{-21}~\mathrm{erg~s^{-1}~cm^{-2}}$ \AA$^{-1}$, with interquartile values ranging between $\pm 5\times10^{-20}~\mathrm{erg~s^{-1}~cm^{-2}}$ \AA$^{-1}$.  Channel-to-channel variations are typically an order of magnitude smaller.  For comparison, the median surface brightness at $r_\mathrm{e}$ of \kmostdns{} sources implies flux densities of order $10^{-20}~\mathrm{erg~s^{-1}~cm^{-2}}$ \AA$^{-1}$ such that, if left uncorrected, this background variability severely limits the detection of faint sources.

In the case of bright sources, where the object continuum is detected in an individual 300 second exposure, the fitting procedure outlined above can systematically overestimate the true background level.  The magnitude of this overestimation is small, $\lesssim$10 \% of the variation in background level between frames, but is systematic and can otherwise bias flux measurements for bright objects.  Correcting for this effect requires a comparison with the surface brightness profile derived from HST imaging, and is described in the next Section.

%%%%%%%%%%%%%%%%%%%%%%%%%%%%
\subsection{Combined cubes and astrometric alignment}
\label{sec.combines}
%%%%%%%%%%%%%%%%%%%%%%%%%%%%
%COMBINING
%{\color{Gray}Combines:}\\
Finally, the reduced science frames are combined for each galaxy with the standard KMOS pipeline using 3-$\sigma$ clipping and then taking the average to create a single datacube for each observed object (`ksigma' combine method). To produce the final combined datacubes the following steps are taken. 
A small fraction, \nzrotangle\% of frames were observed at a non-zero rotator angle. As a result,  \nzrotanglenum~galaxies have data observed at multiple rotator angles. To resolve the angle mis-match all individual frames are first de-rotated to 0 degrees. 

% BAD FRAMES
Individual exposures are inspected for any bad or failed data or reductions. This can include failed sky-subtraction or telluric correction, spectral fringing, bad seeing, clouds, bright background from twilight, odd continuum shapes, and or a failed reference star fit. Possible bad frames and nights with poor conditions are automatically flagged and then manually checked. For some exceptional cases (e.g.\ humidity $>$ 60\%) individual IFUs are flagged as bad within a science frame. Flagged frames or IFUs are not included in the final combination. In total 554 frames, or 6.6\% of frames were thus rejected. 

Astrometric shifts between exposures are computed using the average measured offsets from the three stars included in the same pointing. The measured astrometric shifts are the combined effect of dithering and gradual drift. For data taken over the same night or series of nights this method provides an improvement with respect to using the shifts recorded in the header keywords. However, instrument interventions and longer timescale variations in the instrument and telescope over the 5 years of the survey can result in larger spatial shifts that are not well accounted for by the standard correction applied from the PSF stars. These offsets of $>1-2$ pixels can lead to spurious `double images' of the observed galaxy. To correct for these spatial shifts partial combined datacubes are created for each galaxy, which are the sum of all useful data taken for a given galaxy, within a given KARMA setup (based on jumps in the arm telemetry). Each galaxy may have $1-6$ partial combined frames depending on the number of observations throughout the survey. 

In brief, model KMOS images are generated by convolving HST postage stamp images in the closest available band to the KMOS PSF, cropping to the KMOS FOV, and binning in KMOS 0.2" pixels. This is compared to the KMOS data cube, optimally collapsed to form a continuum image. The centroid is allowed to vary and the best fit is the one which minimizes chi-squared. This provides a visually good solution in every case where the source is clearly visible in continuum which includes ~98.6\% (730/\totalobs) accepted after visual inspection. Shifts in four of the remaining 10 cubes are confirmed via the \halpha image: the remaining 6 objects are all fainter than $Ks=22.3$ The resulting shifts are applied to the cube headers, with a median shift of $\sim1$ KMOS pixel and with $\sim10$\% of cubes exceeding shifts of 2 pixels.

The partial combined frames are each registered to the HST imaging following a procedure that is described in detail by \citet{Wilman19}.
In brief, model KMOS { continuum} images are generated by convolving HST images in the closest available band to the KMOS PSF, cropping to the KMOS FOV, and binning to KMOS $0\farcs 2$ pixels. This is compared to a KMOS continuum image obtained by taking a weighted average of the data cube in wavelength (masking contamination by sky emission). The centroid is allowed to vary and the best fit is determined through a non-linear least squares fitting algorithm.
The resulting astrometric shifts are visually inspected. Corrections for relative shifts between partial combined frames are then applied to all the individual exposures to generate the final total combined datacube. Of 355 objects with multiple setups, the median residual shift is {$\sim1.33$} KMOS pixels {($0\farcs 267$)} with {$\sim27$\%} of shifts above 2 KMOS pixels ($0\farcs 4$), ranging as high as 4.35 pixels ($0\farcs 87$). Not accounting for such shifts can artificially blur the combined cube by an average of $\sim2$ kpc and up to $\sim7$ kpc.

The final datacubes are astrometrically registered to the HST imaging by repeating the same procedure now using the fully combined cubes for each galaxy.
This provides a visually good solution in every case where the source is clearly visible in continuum, which includes 98.6\% (730/\totalobs) of the targeted sources. Shifts in four of the other ten cubes are confirmed via the \halpha image.  The remaining six objects are all fainter than $Ks=22.3~{\rm mag}$.  The resulting shifts are applied to the cube headers, with a median shift of $\sim 1$ KMOS pixel and with $\sim 10$\% of cubes exceeding shifts of 2 pixels.

%%%%%%%%%%%%%%%%%%%%%%%%%%%%
\subsection{Associated PSF images}
\label{subsec.psf}
%%%%%%%%%%%%%%%%%%%%%%%%%%%%
%PSF IMAGES
%{\color{Gray}PSF images:}\\
For each science frame the PSF stars are collapsed in the wavelength range used for the flux calibration to produce PSF images. The PSF images are fit with a Moffat profile to extract the centroid and the total flux of the star. Then each image is normalised to a total flux value equal to unity allowing different stars to be combined, as the same PSF stars may not have been observed across the multiple pointings. Different stars are combined to produce a PSF image representative of the observing conditions for galaxies observed in multiple pointings.  In this case, the stars for the final PSF image are preferentially selected to be from the same detector as the galaxy observations.  Once all the relevant individual PSF images are selected they are shifted and combined using Swarp \citep{2002ASPC..281..228B} on a $21\times21$ pixel grid. This process resamples and co-adds the input frames onto a final common grid. This process also generates a noise frame from the standard deviation of the input frames. We then obtain a PSF image which reflects the specific observing conditions for each object. The natural sampled images ($\rm 0\farcs 2~pixel^{-1}$) are re-fit with a Moffat and Gaussian function separately using custom IDL routines to characterise the PSF and photometric conditions as shown in Figure~\ref{fig.psffwhm}. The units are normalized flux units such that the total flux in the PSF model is unity.

%%%%%%%%%%%%%%%%%%%%%%%%%%%%
\subsection{Spectral resolution}
\label{subsec.res}
%%%%%%%%%%%%%%%%%%%%%%%%%%%%
%RESOLUTION
%{\color{Gray}Resolution:}\\
The spectral resolution of KMOS varies from IFU to IFU as well as in both the wavelength direction and across spaxels \citep{2013AA...558A..56D}. It is sensitive to the focus of the optical elements in the KMOS instrument and as a result can change after an instrument intervention (when KMOS is warmed up for maintenance). We account for all but the spatial variations by measuring the resolution as a function of wavelength for each galaxy data cube. Combined arc lamp cubes are assembled from the arc frames for each IFU. Sky cubes for each target are created from the same science frames as the science cubes prior to the sky-subtraction and heliocentric correction being applied. Therefore the combined sky cube and galaxy cube were created from the same raw data. They are combined following the same procedure as the final science cubes and thus take into account changes in resolution for targets observed in multiple IFUs.  
Then we fit Gaussian profiles to the arclines in each spaxel and we average the spectral resolution values obtained for a given line. Finally, we fit a 4th order polynomial to the spectral resolution values as a function of wavelength. However, the spectral resolution obtained with this method is likely to be inaccurate since the science data has been passed through different steps of the reduction procedure (e.g. shifting of the wavelength solution to match the OH line spectrum).  To overcome this issue we fit $\sim 10$ bright and isolated night sky lines in the non sky subtracted cubes and we adjust the zeroth order term of the polynomial to fit these values instrumental resolution, while preserving the overall shape of the polynomial.
The resolution at a given wavelength can then be recovered such that

\begin{figure}
\begin{center}
\includegraphics[ scale=0.38,angle=90, trim=0cm 0cm 0cm 0.5cm, clip]{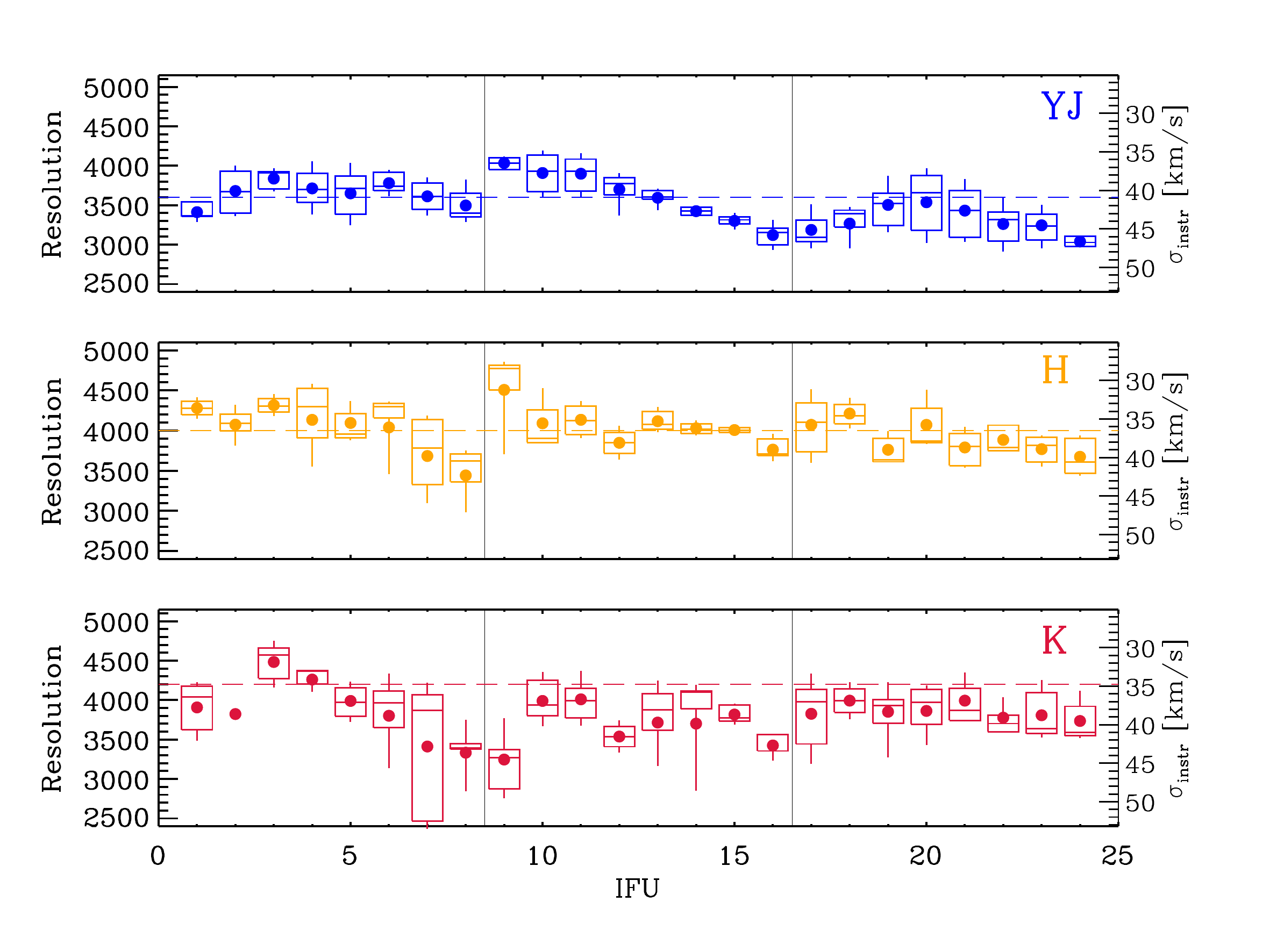}
\caption{Spectral resolution at \halpha as a function of KMOS IFU number for detected \kmostd galaxies observed in a single IFU using the $YJ$ (top, blue), $H$ (middle, orange), and $K$ (bottom, red) gratings. For each IFU the mean (circle), median (horizontal line), central 50\% (box), and central 90\% (vertical line) of the distribution is shown. The expected spectral resolution for each waveband is shown as a dashed horizontal line. Black vertical lines identify the division between the three KMOS detectors. The right hand y-axis shows the spectral resolution in km s$^{-1}$.}
\label{fig.res}
\end{center}
\end{figure}
%
%resolution equation
\begin{multline}
R = {\scriptstyle RES~COEFF0} + {\scriptstyle RES~COEFF1}\times\lambda_\mathrm{obs} + {\scriptstyle RES~COEFF2}\times\lambda_\mathrm{obs}^2  \\
+ {\scriptstyle RES~COEFF3}\times\lambda_\mathrm{obs}^3 + {\scriptstyle RES~COEFF4}\times\lambda_\mathrm{obs}^4
\end{multline}
where $\lambda_\mathrm{obs}$ is the wavelength of the observed line and RES~COEFF1 to RES~COEFF4 are the
coefficients of the polynomial fit. There are a few cases where the polynomial solutions extend towards very low or high values at the edges of the spectrum. A minimum and maximum spectral resolution are given for these cases, RES~MIN, RES~MAX. The minimum and maximum spectral resolutions are derived separately in each band from the upper and lower $3\sigma$ limits of the resolution distribution (excluding data within $\sim200$ \AA~of the ends of the wavelength range).
 
The average effective spectral resolutions at the location of \halpha detections in \kmostd are \resYJ, \resH, and \resK~in the $YJ$, $H$, and $K$ bands respectively. The variation of spectral resolution across waveband and IFU is shown in Figure~\ref{fig.res}. The measured spectral resolution obtained is close to the nominal KMOS resolution in each band with the exception of $K$-band, which yielded lower resolution by $\Delta R\sim200$. In each waveband there is variation between IFUs up to $\Delta R=1000$. We therefore stress the importance of using the wavelength- and IFU-dependent effective spectral resolution (provided with the data release) for scientific analysis, in particular to determine accurate velocity dispersions.

%%%%%%%%%%%%%%%%%%%%%%%%%%%%
\subsection{Bootstrap cubes}
%%%%%%%%%%%%%%%%%%%%%%%%%%%%
We generated 100 bootstrap realizations of each final combined datacube by selecting random exposures to combine with replacement. The KMOS pipeline produces a noise cube corresponding to the rms of all pixels contributing to a $x$, $y$, $\lambda$ position in the final combined cube. The bootstrap cubes complement this default noise cube and provide more realistic noise estimates, which are typically $\sim2-3\times$ larger. 

%----------------------------------------------------------------------
\section{Integrated \halpha properties}
\label{sec.results}
\begin{figure*}[htbp]
\begin{center}
\includegraphics[ scale=1.0, trim=0cm 0.7cm 0cm 0cm, clip]{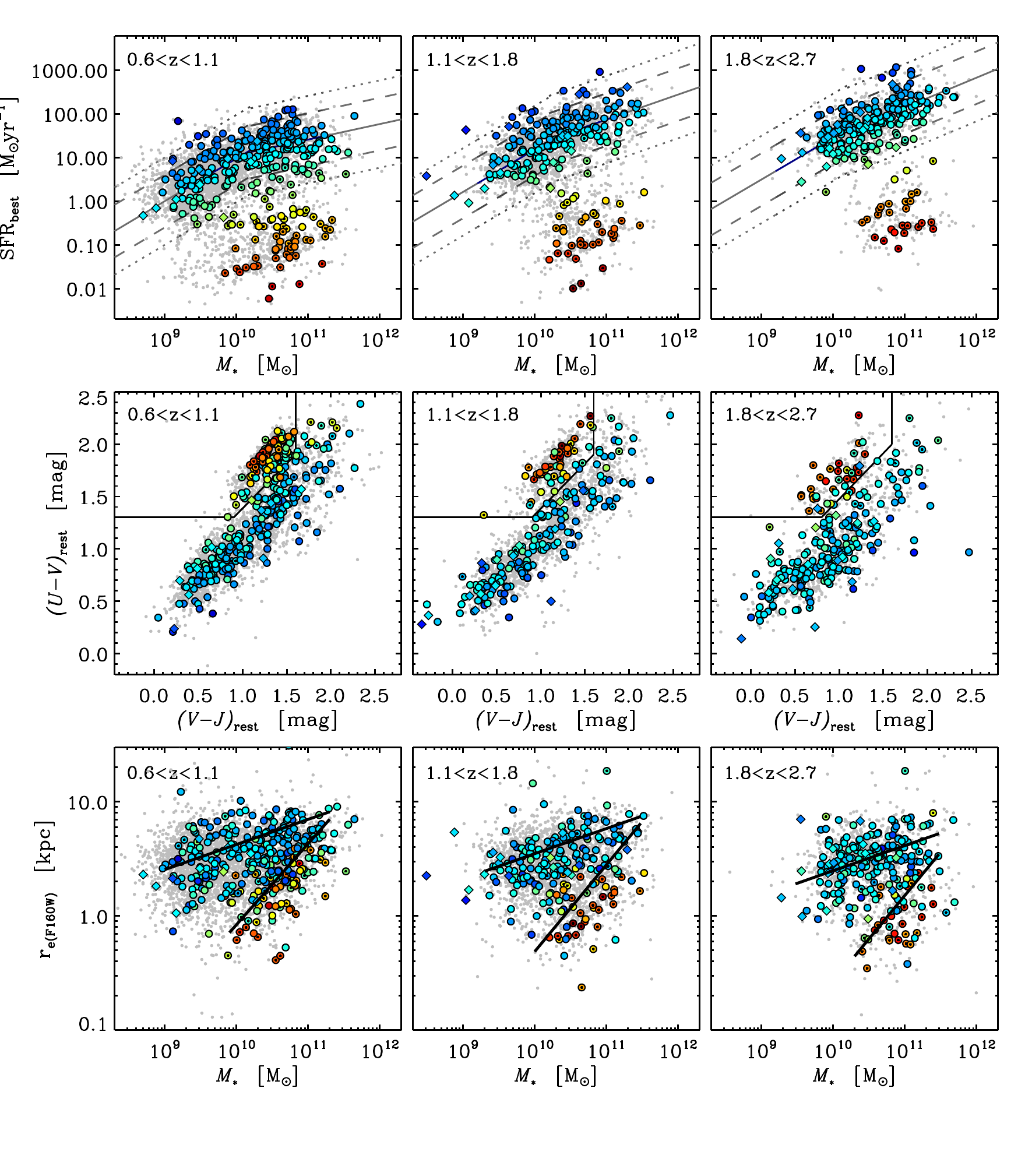} %kmos3d_sum_plots_3pan_v2.pro
\caption{ Properties of the observed \kmostd sample spanning three redshift bins in the SFR$-M_*$ plane \textit{(top)}, ($U-V$)$_\mathrm{rest}-$($V-J$)$_\mathrm{rest}$ plane \textit{(middle)}, and $r_e$(F160W)$-M_*$ plane \textit{(bottom)}. Small grey points show the parent 3D-HST sample without the magnitude and OH contamination selection criterion imposed (Section~\ref{sec.selection}). Large symbols represent galaxies observed as part of \kmostdns. Symbols are color-coded by offset from the main sequence for each individual galaxy as seen in the top panels. Non-detections of \halpha are shown with black dots within the colored circles. Diamonds represent serendipitous galaxies detected within the IFUs of the targeted galaxies. SFRs in the top panels are derived from a Herschel calibrated ladder of SFR indicators \citep{2011ApJ...738..106W} and $M_*$ are derived from SED fits. In the top panels, the broken power-law parameterization, valid between $\log(M_{\star}/M_{\odot})=9.2-11.2$, is shown by the solid lines, as defined using 3D-HST data in all CANDELS fields from $0.5<z<2.5$ using UV+IR SFRs \citep{2014ApJ...795..104W}. Power law coefficients for redshifts between the bins given in \cite{2014ApJ...795..104W} are obtained through interpolation of the coefficients as a function of stellar mass. Dashed lines and dotted lines show $4\times$ and $10\times$ above and below the canonical MS respectively. Lines defining the UVJ passive region in the middle panels are defined by \cite{2009ApJ...691.1879W}. Lines denoting the star-forming and passive galaxy loci on the size-mass plane in the bottom panels are defined by \cite{2014ApJ...788...28V}. }
\label{fig.msuvj}
\end{center}
\end{figure*}
%----------------------------------------------------------------------

% DEFINE NUMBERS
\def\percentgrismz{64} %update after DW &NMFS agree on final galaxy numbers to release
\def\uniquekarma{X}

%GENERAL DETECTIONS
\def\detectedgals{581}
\def\observedgals{739}
\def\detectfn{79}

\def\detectedgalsYJ{245}
\def\observedgalsYJ{319}
\def\detectfnYJ{77}
\def\minzkmosYJ{0.602}
\def\maxzkmosYJ{1.039}

\def\detectedgalsH{159}
\def\observedgalsH{201}
\def\detectfnH{79}
\def\minzkmosH{1.275}
\def\maxzkmosH{1.924}%1.768}

\def\detectedgalsK{177}
\def\observedgalsK{219}
\def\detectfnK{81}
\def\minzkmosK{1.996}
\def\maxzkmosK{2.675}

%ON MS DETECTIONS
\def\MSdetectedgals{541}
\def\MSobservedgals{592}
\def\MSdetectfn{91}

\def\MSdetectedgalsYJ{219}
\def\MSobservedgalsYJ{243}
\def\MSdetectfnYJ{90}

\def\MSdetectedgalsH{148}
\def\MSobservedgalsH{160}
\def\MSdetectfnH{93}

\def\MSdetectedgalsK{174}
\def\MSobservedgalsK{189}
\def\MSdetectfnK{92}

%BELOW MS DETECTIONS
\def\bMSdetectedgals{40}
\def\bMSobservedgals{147}
\def\bMSdetectfn{27}

\def\bMSdetectedgalsYJ{26}
\def\bMSobservedgalsYJ{76}
\def\bMSdetectfnYJ{34}

\def\bMSdetectedgalsH{11}
\def\bMSobservedgalsH{41}
\def\bMSdetectfnH{27}

\def\bMSdetectedgalsK{3}
\def\bMSobservedgalsK{30}
\def\bMSdetectfnK{10}

\def\Nzqone{60}
\def\Nserendip{46}
\def\Nserendipktd{16}
\def\serendipMlo{8.4}
\def\serendipMhi{10.9}
\def\serendipzlo{0.4}
\def\serendipzhi{2.6}

\def\sigmaNMADall{463}
\def\sigmaNMADgrism{1020}
\def\sigmaNMADspec{155}
\def\sigmaNMADgrismbMS{1546}

From the completed \kmostd observations, \detectedgals~of the targeted galaxies have \halpha emission line detections, translating into a 79\% detection rate across the full survey. This is a greater than $3\times$ increase in the number of \halpha detections presented in \citetalias{2015ApJ...799..209W}. 
In the last years of the \kmostd survey we pushed to lower masses and lower SFRs, as well as added observations of \observedgalsH~galaxies in a new redshift slice at $z\sim1.5$.
Figure~\ref{fig.msuvj} shows the location of all observed \kmostd galaxies on the SFR$-M_*$, $UVJ$, and $r_e-M_*$ diagrams color coded by offset from the MS ($\rm \Delta MS$). Undetected galaxies are indicated with a black dot within the circle. Detected \kmostd galaxies cover the mass ranges of $9.00<\log(M_{\star}/M_{\odot})<11.43$, $9.44<\log(M_{\star}/M_{\odot})<11.45$, and $9.79<\log(M_{\star}/M_{\odot})<11.68$ in the $z\sim1$, $z\sim1.5$ and $z\sim2$ redshift slices respectively. 
The location of the MS is shown at each redshift with a solid gray line.
It is defined by the broken power-law MS parametrization from \cite{2014ApJ...795..104W}, valid between $\log(M_{\star}/M_{\odot})=9.2-11.2$. The power law coefficients for redshifts between the bins given in \cite{2014ApJ...795..104W} are obtained through interpolation of the coefficients as a function of stellar mass.

\begin{figure}
\begin{center}
\includegraphics[ scale=0.75,angle=90, trim=10cm 0cm 0cm 0.5cm, clip]{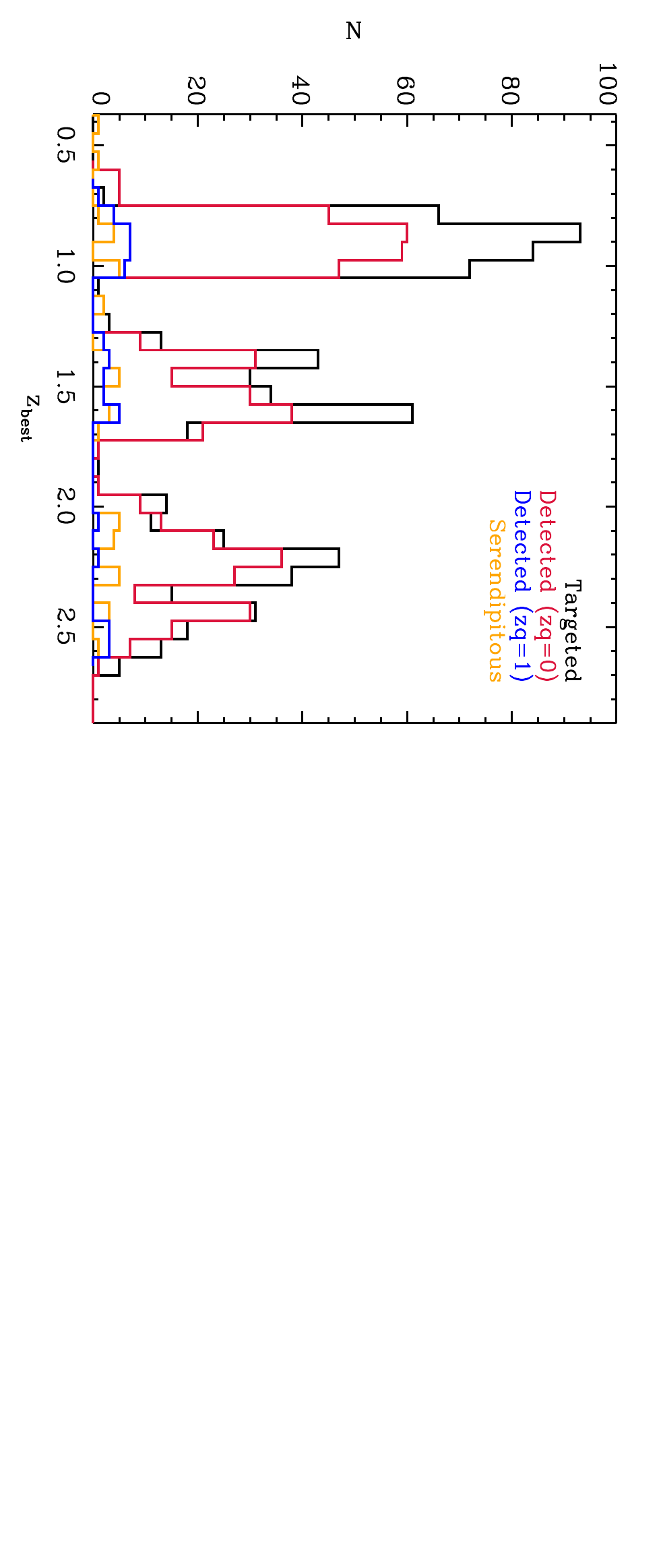}
\caption{ Redshift distribution of all targeted galaxies across the three targeted redshift slices, $z\sim1$, $z\sim1.5$, $z\sim2$, using targeted redshift, $z_\mathrm{best}$ (black) and detected galaxies (red; zq=0; blue: zq=1) using \kmostd derived redshifts. Serendipitous galaxies with redshifts in the range $0.3<z_\mathrm{KMOS}<2.7$ are shown by the orange histogram.}
\label{fig.zhist}
\end{center}
\end{figure}

%----------------------------------------------------------------------
\subsection{Detection fractions}
\label{sec.detectionfns}
%----------------------------------------------------------------------
%REDSHIFTS
%
% spectroscopic vs. grism
In this Section we characterise the \kmostd detection fractions on and off the MS and as a function of color.
We adopt the separation between star-forming and passive galaxies in terms of rest-frame colors following the $UVJ$ criteria of \cite{2009ApJ...691.1879W}, and in terms of SFR using a threshold of $\rm \Delta MS = -0.85~dex$.
A high spectroscopic redshift success rate was a key factor in the design and, ultimately, in the success of the survey.
\halpha and \NII~emission was searched for manually in each reduced cube around the 2D continuum center, using the spectroscopic or grism redshift as a prior. The global detection fraction of \detectfn\% splits between subsets as follows. Among the 36\% of targeted galaxies that had a previous spectroscopic redshift (from \citealt{2005AA...437..883M,2008AA...478...83V,2009AA...494..443P,2012MNRAS.425.2116C,2013AA...549A..63K,2013ApJ...778..114T,2015ApJS..218...15K}; see also \citealt{2014ApJS..214...24S}), 84\% are detected in H$\alpha$.  Among the other \percentgrismz\% with only a grism redshift at the time of observation, the detection fraction is 76\%.

%by redshift
Figure~\ref{fig.zhist} shows the $z_\mathrm{best, orig}$~redshift distribution of targeted galaxies, where $z_\mathrm{best, orig}$~is defined as the most accurate redshift available at the time of targeting, either from grism or higher resolution spectroscopy.  We detect \detectedgalsYJ, \detectedgalsH, and \detectedgalsK~galaxies in the redshift ranges of $\minzkmosYJ<$\zkmos$<\maxzkmosYJ$, $\minzkmosH<$\zkmos$<\maxzkmosH$, and $\minzkmosK<$\zkmos$<\maxzkmosK$ respectively. 
We achieve a comparable detection fraction within each redshift slice of $\sim79$\% as shown in Table~\ref{tab.det}.

%Redshift accuracy
To compare $z_{\rm KMOS}$ and $z_{\rm best,orig}$, we calculate the prior redshift accuracy for detected galaxies as 
\begin{equation}
\sigma_\mathrm{NMAD} = 1.48c\times \mathrm{median}\left\lvert \frac{\Delta z - \mathrm{median}(\Delta z)}{(1+z_\mathrm{kmos})} \right\rvert
\end{equation}
where $\sigma_\mathrm{NMAD}$ is equal to the standard deviation for a Gaussian distribution following \cite{2008ApJ...686.1503B}, $\Delta z=z_\mathrm{best}-z_\mathrm{kmos}$, and $c$ is the speed of light.
The overall 3D-HST redshift accuracy for the detected \kmostd galaxies corresponds to a velocity offset of \sigmaNMADall~\kms~from the previously known redshift. For galaxies selected on a prior spectroscopic redshift and detected with KMOS, $\sigma_\mathrm{nmad}$, is \sigmaNMADspec~\kms. For galaxies selected on a grism redshift and detected with KMOS $\sigma_\mathrm{nmad}= \sigmaNMADgrism$~\kms.  Below the MS, where grism redshifts are more often constrained by the continuum rather than emission lines, the standard deviation is higher with $\sigma_\mathrm{nmad}=\sigmaNMADgrismbMS$~\kms. {Possible non-detections due to larger redshift uncertainties would decrease the quoted accuracies, discussed further in Section~\ref{sec.nondet}.} The grism redshift accuracy also decreases with increasing observed $K$-band magnitude, as demonstrated in \cite{2016ApJS..225...27M}. Within the detected sample 41 galaxies have redshifts from \kmostd deviant from the expected redshift from 3D-HST by $>10,000$ \kms~(7 galaxies at $z\sim1$, 12 galaxies at $z\sim1.5$, and 22 galaxies at $z\sim2$).

\begin{table}
\caption{Target detection fractions}
\begin{tabular}{l|lll}
\hline
 &  All & $\Delta\mathrm{SFR}>-0.85$ & $\Delta\mathrm{SFR}<-0.85$ \\
\hline
All & \detectfn\%~(\detectedgals/\observedgals ) & \MSdetectfn\%~(\MSdetectedgals/\MSobservedgals )& \bMSdetectfn\%~(\bMSdetectedgals/\bMSobservedgals )\\
$\minzkmosYJ<$z$<\maxzkmosYJ$& \detectfnYJ\%~(\detectedgalsYJ/\observedgalsYJ ) & \MSdetectfnYJ\%~(\MSdetectedgalsYJ/\MSobservedgalsYJ )& \bMSdetectfnYJ\%~(\bMSdetectedgalsYJ/\bMSobservedgalsYJ )\\
$\minzkmosH<$z$<\maxzkmosH$  & \detectfnH\%~(\detectedgalsH/\observedgalsH ) & \MSdetectfnH\%~(\MSdetectedgalsH/\MSobservedgalsH )& \bMSdetectfnH\%~(\bMSdetectedgalsH/\bMSobservedgalsH )\\
$\minzkmosK<$z$<\maxzkmosK$& \detectfnK\%~(\detectedgalsK/\observedgalsK ) & \MSdetectfnK\%~(\MSdetectedgalsK/\MSobservedgalsK )& \bMSdetectfnK\%~(\bMSdetectedgalsK/\bMSobservedgalsK )\\
\hline
\end{tabular}
\label{tab.det}
\end{table}%
\begin{figure}[htbp]
\begin{center}
\includegraphics[ scale=0.75, angle=90, trim=0cm 0cm 0cm 0.8cm, clip]{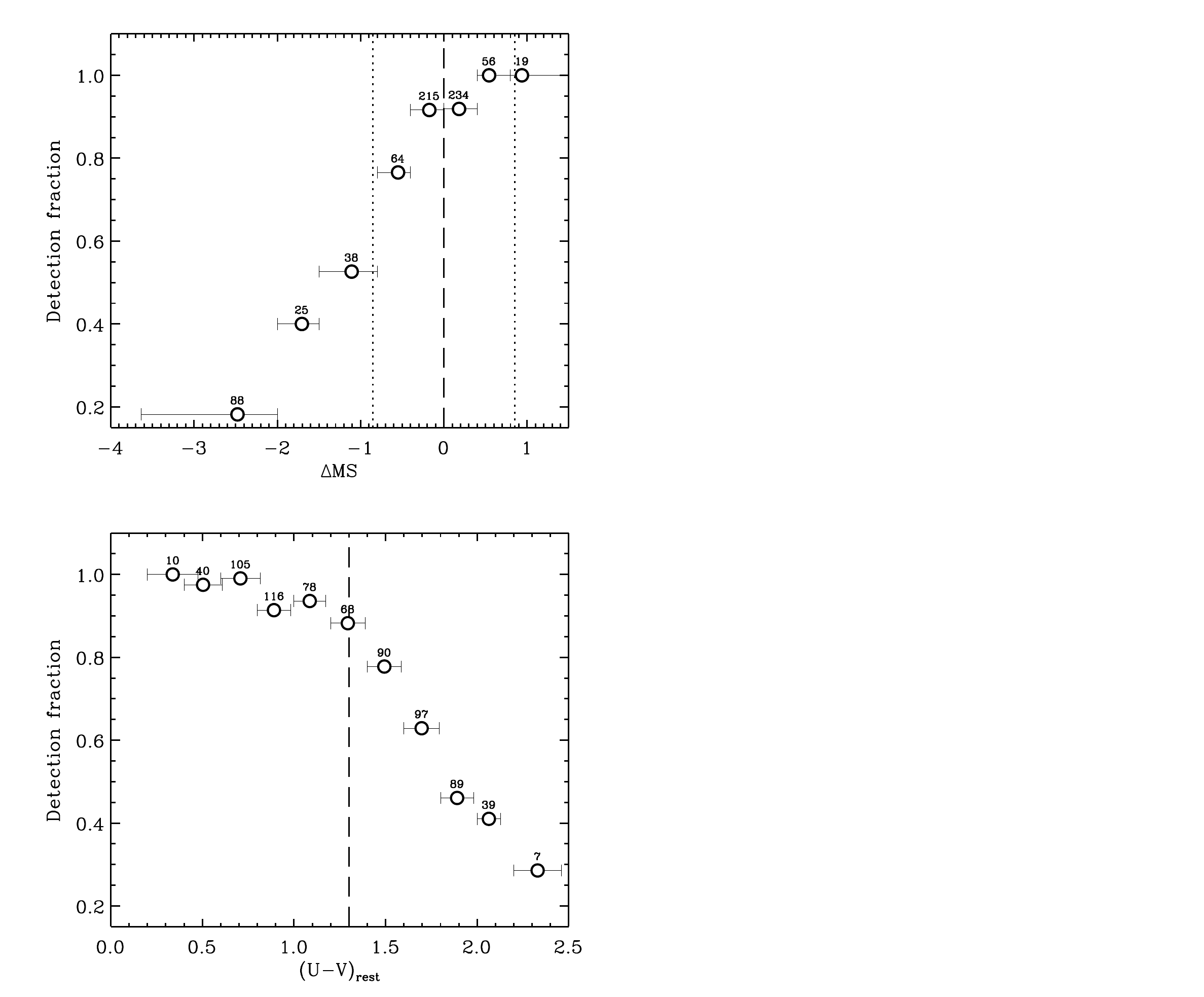}
\caption{ \halpha detection fraction as a function of MS offset (top) and ($U-V$)$_\mathrm{rest}$ color (bottom) for the full sample. In the top panel, vertical lines indicate the centre and boundaries of the defined MS. In the bottom panel, the vertical line indicates and the ($U-V$)$_\mathrm{rest}$ passive limit from \cite{2009ApJ...691.1879W}. The horizontal error bars represent the size of the bin while the numbers above the data points give the number of galaxies in each bin. }%{\color{BrickRed} should I make these cumulative distributions?}}
\label{fig.detfns}
\end{center}
\end{figure}
%on/off MS, color
There are \Nzqone~galaxies with a possible \halpha detection but S/N is low or sky features put the validity of the detection in question. These galaxies are considered detected but are plotted separately as the blue histogram in Figure~\ref{fig.zhist}. They are found primarily in the lowest redshift bin in which a higher number of $UVJ$-passive galaxies where observed. The remaining detected galaxies have a high quality detection with detection flag $z_\mathrm{q}=0$. 

%DETECTION FRACTIONS AS FUNCTION OF GALAXY PROPERTIES
Detection fractions are a strong function of MS offset and galaxy colors as shown in Figure~\ref{fig.detfns}.
Table~\ref{tab.det} gives the detection fraction and number of galaxies observed above and below $\rm \Delta MS = -0.85~dex$ in each redshift slice.
The detection fraction is as high as \MSdetectfn\% (\MSdetectedgals/\MSobservedgals), when considering only galaxies near and above the MS at $\rm \Delta MS > -0.85~dex$. It is fairly constant across redshifts, with \MSdetectfnYJ\%, \MSdetectfnH\%, and \MSdetectfnK\% detected in the redshift slices $z\sim1$, $z\sim1.5$, $z\sim2$ respectively. In contrast the detection fraction below the MS ($\rm \Delta MS < -0.85~dex$) is \bMSdetectfnYJ\%, \bMSdetectfnH\%, and \bMSdetectfnK\% respectively in the same redshift ranges.  At blue colors, $(U-V)_\mathrm{rest}<1.3$, and on/above the canonical MS ($\Delta\mathrm{MS}>-0.25$) we detect $>90$\% of all galaxies. We detect all galaxies at $\Delta\mathrm{MS}>0.6$. The detection fractions fall rapidly when moving to redder $(U-V)_\mathrm{rest}$ colors or below the MS. Galaxies with no \halpha detection below the MS do not correlate with magnitude or exposure time. They are typically small and have SFRs derived from SEDs (because undetected in the far-IR).

\begin{figure}[!t]
%\figurenum{1}
\begin{center}
\includegraphics[scale=0.7,trim=0 0.55cm 0 0, clip,angle=0]{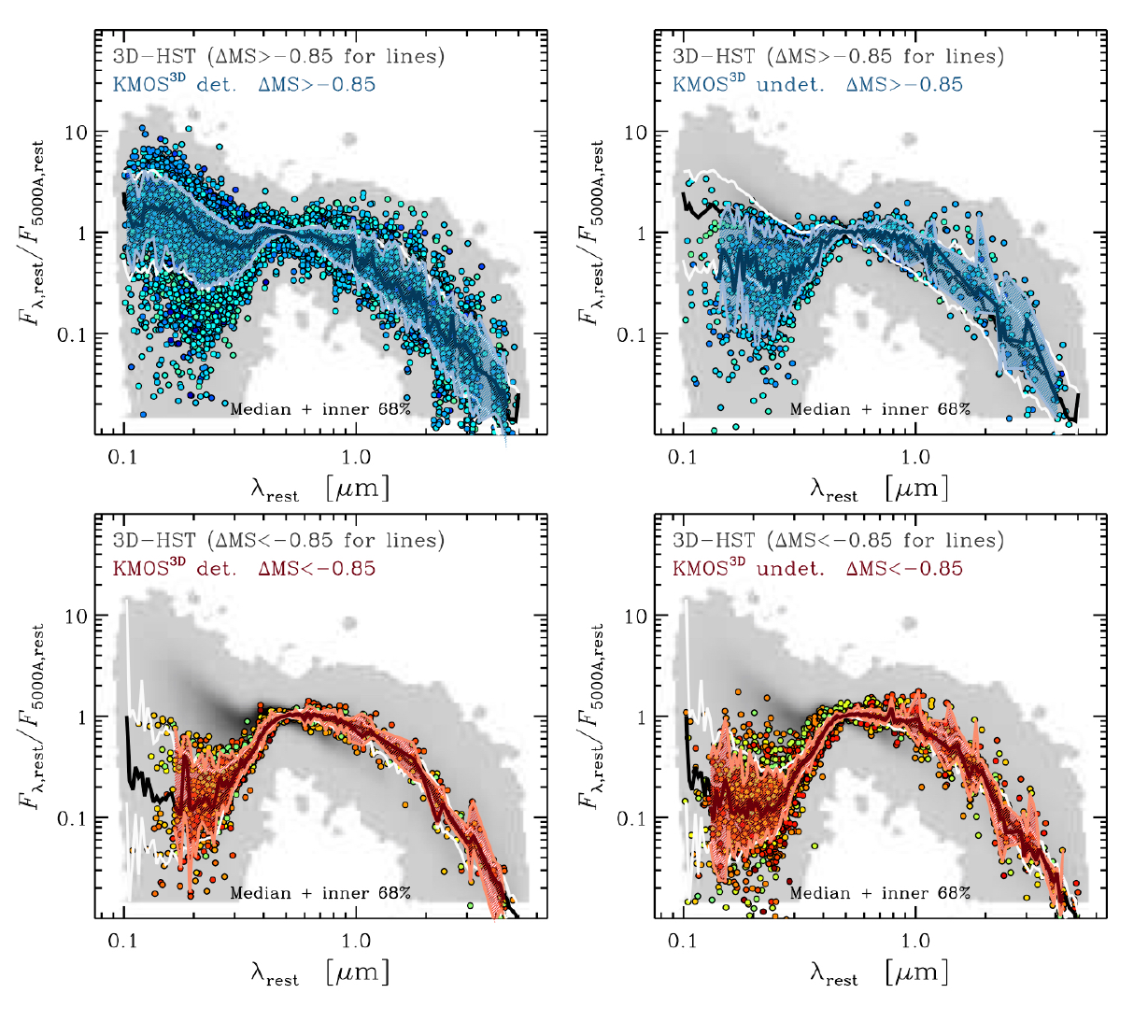}
\end{center}
\caption{
%\small
%{\color{blue}
Composite SEDs of detected and undetected \kmostd targets split by star formation activity level.
{\it Top row}: the panels show the composite SEDs of the detected \textit{(left)} and undetected \textit{(right)} subsets of galaxies with $\rm \Delta MS > -0.85~dex$.
The photometry comprising the SEDs are color-coded according to the corresponding galaxy $\rm \Delta MS$ (as shown in Fig.~\ref{fig.msuvj}). 
The individual SEDs are plotted in the rest-frame from the observed optical
to $\rm 8\,\mu m$ photometry, normalized to $\lambda_{\rm rest} = 5000\,\AA$.
The median SEDs are plotted as dark blue lines, and the central 68\% intervals
are indicated with the light-blue lines.
The grey-shaded areas in the background of all panels show the density
distributions in the respective parameter spaces of the 3D-HST parent
population (at $0.7<z<2.7$, $\rm \log(M_{\star}/M_{\odot}) > 9.0$,
$\rm F160W < 25.1~mag$, and $K < 23~{\rm mag}$), over all $\rm \Delta MS$
values.  The black and white lines correspond to the median SED and 68\% range around it of the subset of
the parent population at $\rm \Delta MS > -0.85~dex$.  
{\it Bottom row}: same as the top panels now for the \kmostd targets
at $\rm \Delta MS < -0.85~dex$. 
At $\rm \Delta MS > -0.85~dex$, the undetected \kmostd targets have typically
redder SEDs than those of the detected targets and of the star-forming subset
of the parent 3D-HST sample.  In contrast, detected and undetected objects
at low MS offsets differ little.
%}  %  End color blue
\label{fig.SEDs}
}
%\vspace{2.0ex}

\end{figure}

Figure~\ref{fig.SEDs} shows the composite rest-frame SEDs of detected and undetected galaxies for each of the subsets above and below $\rm \Delta MS = -0.85~dex$.  These SEDs were constructed from the optical-to-8$\mu$m broad- and medium-band photometry, normalizing the individual SEDs at rest-frame 5000\,\AA, and computing the running median and inner 68\% range of the distributions. For comparison, we also constructed composite SEDs in a similar manner for the 3D-HST parent population (at $0.7 < z < 2.7$, $\log(M_{\star}/M_{\odot}) > 9$, $\rm F160W < 25.1~mag$, and $K < 23~{\rm mag}$), split in the same $\rm \Delta MS$ bins.
Among the star-forming subset, the median SED of the detected KMOS targets and the 68\% range around it are nearly identical to those for the 3D-HST parent SFG population, while the undetected targets tend to be significantly redder.  This difference likely reflects higher levels of dust extinction among undetected targets at $\rm \Delta MS > -0.85~dex$, which would lead to fainter emergent H$\alpha$ fluxes. {In support of this explanation, we find that the median IR/UV ratio (as a measure of dust obscuration, see Section~\ref{sec.sfrs}) of undetected targets is systematically higher by $3.5\times$ than that of \halphans-detected targets matched in $M_{\star}$ (within +/- 0.2 dex), $z$ (same band), and observing time (+/- 1 hour) at any SFR level (UV+IR or SED SFRs).}
In contrast, there is very little difference in SEDs between detected and undetected targets at $\rm \Delta MS < -0.85~dex$, with both subsets having substantially redder SEDs than the overall galaxy population as expected, and consistent with their broad- and medium-band SEDs being dominated by older stellar populations (e.g.\, \citealt{2008ApJ...677..219K,2016ApJ...822....1F}). The H$\alpha$+[\ion{N}{2}] spectral properties of the detected objects in the quiescent regime indicate that half of them may be undergoing low-level rejuvenation events, and the other half exhibits dominant signatures of gas outflows and shocks \citep{2017ApJ...841L...6B}.
%}   % End color blue

%%FIGURE%%
\begin{figure*}[!ht]
%\figurenum{1}
\begin{center}
\includegraphics[scale=0.31,trim=50 10 50 10, clip=1,angle=0]{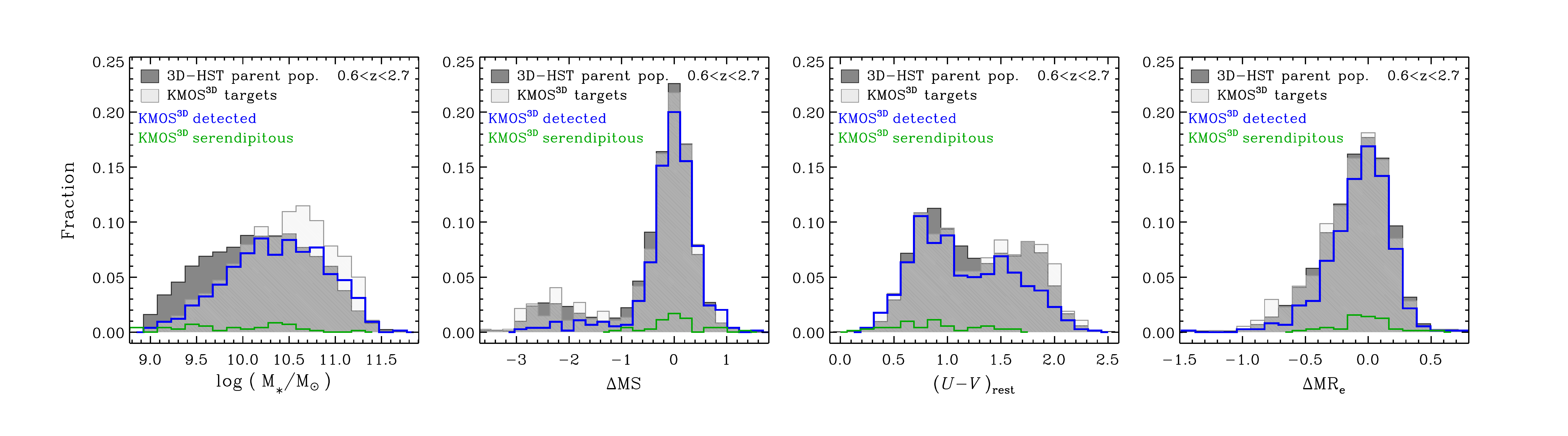}
\end{center}
\vspace{-0.9cm}
\renewcommand\baselinestretch{0.5}
\caption{
\small
%{\color{magenta}
One-dimensional distributions of the \kmostd targets in selected galaxy properties.
From left to right, the properties are stellar mass, MS offset, rest-frame $U-V$
colors, and offset from the mass-size relation of SFGs.
In each panel, the light-grey filled histogram shows the distribution of all primary targets
observed while the blue histogram shows the subset of H$\alpha$-detected objects.
The green histogram corresponds to the sources serendipitously detected through their
line emission in the KMOS IFUs.
The dark-grey filled histogram shows for comparison the distribution of the parent 3D-HST
population in the same redshift range, with the same mass and magnitude cuts as applied
in selecting the \kmostd targets.
The strategy of emphasizing a homogeneous mass and redshift coverage, with no cut on star
formation, color, or size properties, causes most of the differences in mass and color
distributions between the parent population and the selected targets; the targets are
however well representative of the parent 3D-HST distributions in \dms\ and \dmre.
The reduced success rate of H$\alpha$ detection, due to lower SFR and/or higher
dust obscuration, leads to the differences between the full target and detected
subset histograms.  Serendipitously detected sources are near-MS, typically blue
objects across the same mass range as the \kmostd targets.
%}  %  End color magenta
\label{fig.hist}
}
\vspace{2.0ex}
\end{figure*}

%----------------------------------------------------------------------
%NON DETECTION galaxies
\subsubsection{Non-detections}
\label{sec.nondet}
%----------------------------------------------------------------------
As shown in Figure~\ref{fig.msuvj} and Figure~\ref{fig.detfns} the majority of non-detections are galaxies with red colors and or low levels of star formation activity. However, a small number of galaxies with blue colors and SFRs on the MS are not detected. This is likely the result of larger uncertainties or misidentification of the target redshift. For example the \kmostd targets, {GS4\_03349}, {GS4\_42705}, {U4\_20694}, {GS4\_15735}, and {U4\_20770} have v4.1.5 $z_\mathrm{best}<1.8$ ($H$-band targets) but at the time of observation had $z_\mathrm{targeted}>1.8$ ($K$-band targets). For these galaxies it is possible that their \halpha emission falls between KMOS wavebands or in a waveband that was not targeted. Similarly, we detect \halpha for 41 galaxies $>10,000$ \kms~from the expected redshift. It is possible that there are a small number of additional galaxies with this large uncertainty that places the \halpha emission outside of the KMOS waveband observed.

%----------------------------------------------------------------------
%SERENDIPITOUS galaxies
\subsubsection{Serendipitous galaxies}
\label{sec.serendip}
%----------------------------------------------------------------------
We have robustly detected emission lines in \Nserendip~additional galaxies within the KMOS IFU of the primary targeted galaxy. These serendipitous galaxies have redshifts between $\serendipzlo<z<\serendipzhi$ and are in the mass range $\serendipMlo<\log(M_{\star}/M_{\odot})<\serendipMhi$. Of the serendipitous galaxies, \Nserendipktd~fulfil the \kmostd K-band cut and had a prior spectroscopic or grism redshift within $0.7<z<2.7$. The majority of serendipitous detections are from a single emission line. In most cases part of the serendipitous galaxy is outside of the field of view of the KMOS IFU. We therefore do not include the kinematics of these galaxies in future sections, however interacting galaxies are discussed further in Section~\ref{sec.discuss}. The redshift distribution of serendipitous galaxies is shown in yellow in Figure~\ref{fig.zhist} while their SFRs, masses, and colors are represented with diamonds in Figure~\ref{fig.msuvj}.

%----------------------------------------------------------------------
%SERENDIPITOUS galaxies
\subsubsection{Final sample distributions}
\label{sec.sampledist}
%----------------------------------------------------------------------
%{\color{magenta}

Figure~\ref{fig.hist} compares the one-dimensional distributions of \kmostd
sample galaxies in \lmstar , \dms, \uvrest, and \dmre\ to those of the parent 3D-HST
source catalog, subjected to the same cuts in redshift, mass, and magnitude
applied to select our targets. \dmre\ is defined as the logarithmic offset
in $R_{\rm e}$ from the mass-size relation for SFGs of \cite{2014ApJ...788...28V} (and shown in Figure~\ref{fig.msuvj})  at the same $\rm M_{\star}$ and $z$ as each target.
In stellar mass, the full \kmostd sample selected for observations has an excess
above $\log(M_{\star}/M_{\odot}) \sim 10.5$, and a deficit below that mass,
relative to the parent 3D-HST population.  This difference reflects our strategy
of emphasizing a more homogeneous coverage in mass and redshift compared to the
underlying galaxy population.
The \kmostd target distribution follows closely the parent population in \dms\ 
and \dmre, as well as in \uvrest\ colors although with a slight deficit in the
bluer half that is tied to the more uniform mass and redshift distribution of
our selection.
In turn, the distributions for the H$\alpha$ detected subset reflect the lower
success rate in the redder, higher-mass regime well below the MS discussed above
in Section~\ref{sec.detectionfns}.
Unsurprisingly, the distributions for the serendipitously detected line-emitting
sources show they are all SFGs with SFRs within a factor of $\sim 10$ of the MS
and half-light radii within a factor of $\sim 3$ of the mass-size relations.
They have a broad mass distribution covering the same range as the primary
targets, and are typically in the bluer part of the \uvrest\ range. 
Similar trends as just discussed are seen when considering the distributions
in the three redshift slices corresponding to the objects observed in the
$YJ$, $H$, and $K$ bands.
%}   % End color magenta

%----------------------------------------------------------------------
%FLUXES
\def\NIIdetectfn{70}
\def\SIIdetectfn{X}
\def\OIdetectfn{X}

\subsection{Integrated spectra and \halpha fluxes}
\label{sec.intspec}
%----------------------------------------------------------------------

%How spectra are extracted
A total galaxy spectrum is extracted for each galaxy. The spectrum is extracted from the data cube within a $1\farcs 5$ radius aperture centered on the continuum center. Serendipitous galaxies, as discussed in the previous section, are masked when extracting an aperture spectrum. 

Prior to extracting a galaxy spectrum, the continuum is subtracted from each pixel of the datacube. The continuum in the KMOS cubes is a combination of real galaxy continuum and possible residual background, and is not well captured by a simple polynomial fit. To more robustly estimate the shape of the continuum, we mask channels within 1000~km~s$^{-1}$ (2200~km~s$^{-1}$ for seven galaxies with particularly broad emission lines) of strong emission lines (OI~$\lambda$6300, [N~II]$\lambda$6548, H$\alpha$, [N~II]$\lambda$6584, [S~II]$\lambda$6716 or [S~II]$\lambda$6731), calculate the moving median across each spectrum in 30 pixel wide windows, and then perform linear interpolation across the line channels. After the continuum is subtracted, spikes (defined as channels more than 1000~km~s$^{-1}$ from strong emission lines with values exceeding twice the rms or three times the median value of all non-line channels in a given pixel spectrum) are masked. The galaxy spectrum is then extracted, and the continuum subtraction is performed again to remove any residual continuum. 

%how emission lines are measured measured
We fit the combined \halphans, \NII$\lambda\lambda$6548,6563 emission complex with Gaussians for all aperture spectra. For the multi-line fit, the positions of the \NII~lines are tied to the \halpha position and the width of all the lines are equal. In most cases, having constraints from the \NII~emission improves the fit to \halpha emission, particularly when atmospheric contamination is present near the emission lines. In contrast when the detection of \NII~is weak or contaminated a single Gaussian component can provide a better fit to the data. In these cases, the resulting \halpha fluxes from single and multiple Gaussian fits are compared. The spectral fits for galaxies with large difference between measurements are visually inspected and the $\chi^2$ of the two fits are considered. The flux from the better fit is adopted. For the majority of galaxies the \halpha flux measurement from the joint \halphans+\NII~fit is adopted. The resulting \halpha and \NII~fluxes, and a flag indicating which fitting method was used are given in Table~\ref{tab.kmos}.
{
Errors on the line fluxes are calculated by repeating the spectrum extraction and line fitting for each of the bootstrap cubes, and taking the standard deviation across all fit values for each of the parameters.
}

%%%
\begin{figure*}[htbp]
\begin{center}
\includegraphics[ scale=0.75,angle=90,  trim=13.5cm 0cm 0cm 0.5cm, clip]{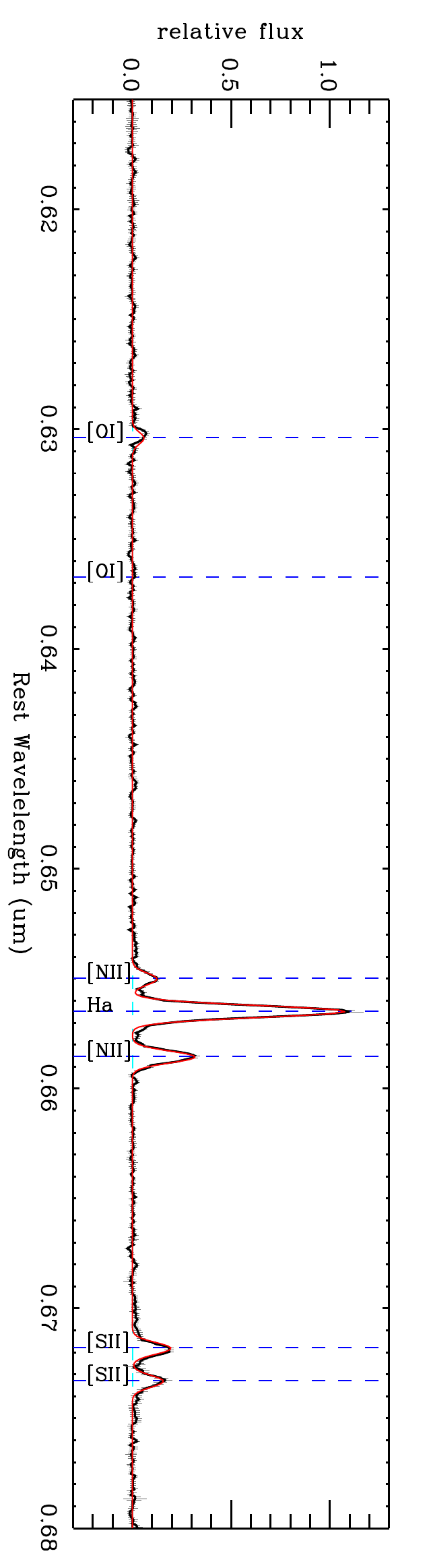}
\caption{{Median stacked spectrum of all detected \kmostd galaxies with a secure \halpha detection where the 47 galaxies hosting a broad component have been removed \citep{2019ApJ...875...21F}}.  The normalised spectrum is shown at rest wavelength with key emission lines labels and denoted with dashed vertical lines. \halphans, \NII, [S II] are detected as well as [O I]. Error bars shows the $\pm1\sigma$ uncertainties on the stacked spectra, derived using bootstrap samples. %The individual spectra are resampled at 3$\times$ the native spectral resolution before stacking.
}
\label{fig.stack}
\end{center}
\end{figure*}%%%
\begin{figure*}
\begin{center}
\includegraphics[ scale=0.71, trim=1.25cm 18.0cm 10.5cm 3.6cm, clip]{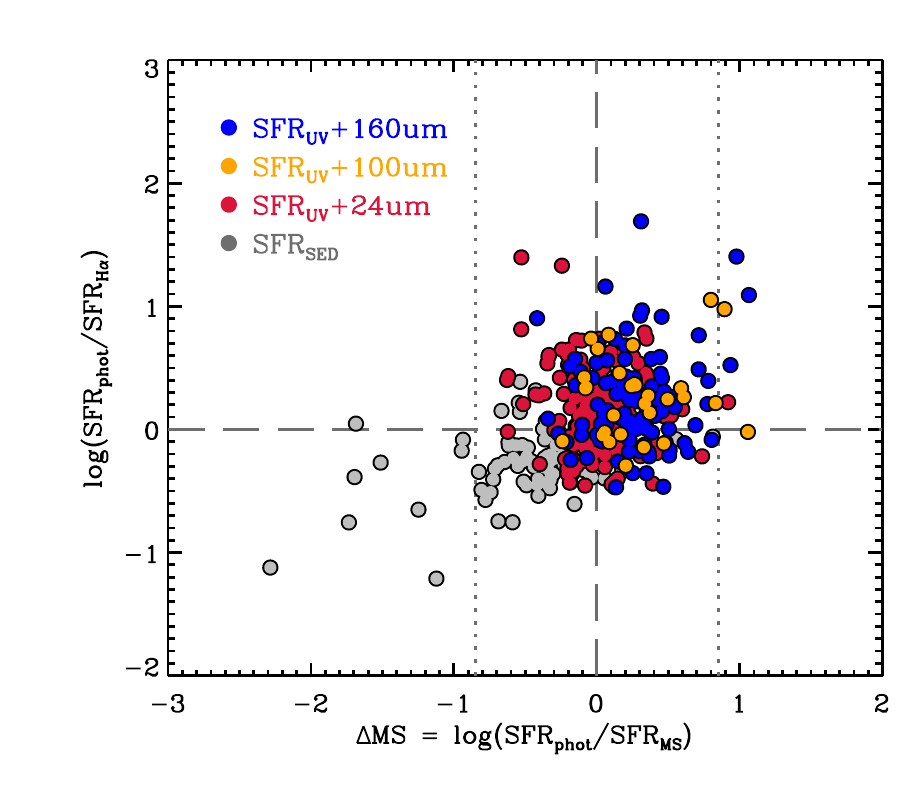}
\includegraphics[ scale=0.71, trim=12.2cm 10.27cm 1.25cm 11.7cm, clip]{k3d_sfr_comp_v415_fnlsp1b.pdf}
\includegraphics[ scale=0.71, trim=12.2cm 2.55cm 0.7cm 19.4cm, clip]{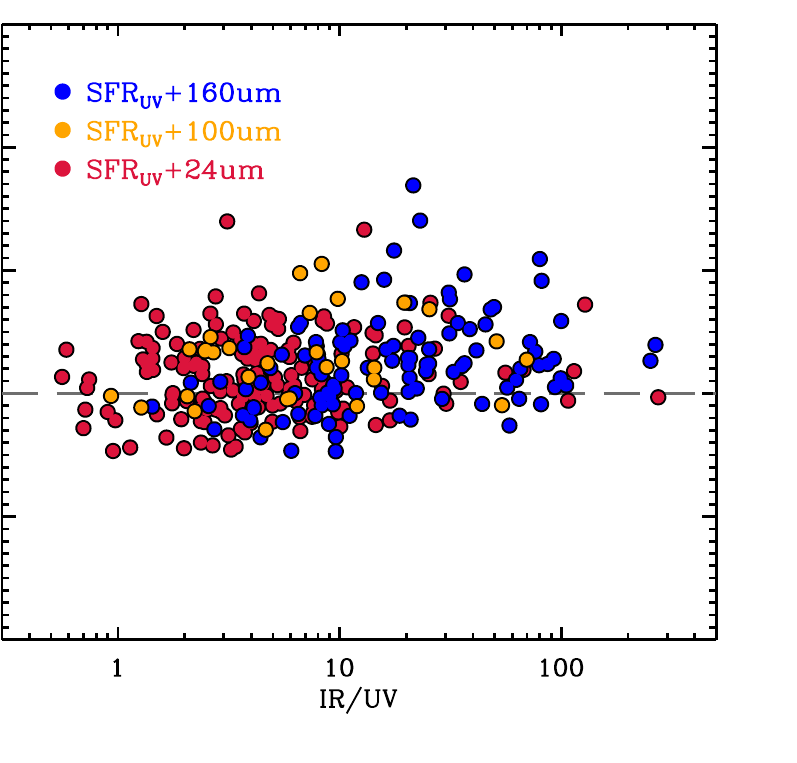}
\caption{Ratio of best SFR derived from photometry to the measured KMOS \sfrha, with dust and aperture corrections applied, as a function of offset from the main sequence, SED derived dust attenuation, A$_V$, and the IR/UV ratio. Galaxies with a broad emission line component \citep{2019ApJ...875...21F} are removed. Colors indicate which SFR indicators are used to derive the SFR$_\mathrm{phot}$ (e.g. 24$\mu$m, 100$\mu$m, 160$\mu$m).}
\label{fig.sfrcomp}
\end{center}
\end{figure*}

The choice of a single aperture for all galaxies, despite size differences provides a simple, robust, and repeatable measurement to characterise the full sample. We note that, while optimal for the majority of the sample, for many sub-MS galaxy detections the relatively large $1\farcs 5$ radius aperture reduces the S/N of detections and is not optimal for the study of compact galaxies near our detection limits.  
However, when the aperture size is reduced to $1\farcs 2$ radius for example, the median \halpha flux decreases by 23\% for the $z\sim1$ sample, 16\% for the $z\sim1.5$ sample, and 15\% for the $z~\sim2$ sample. Thus we adopt the $1\farcs 5$ radius aperture. Given the continuum half-light sizes of massive galaxies especially at $z\sim1$, the effects of beam smearing, and the size of the KMOS IFUs, it is likely that a fraction of the total \halpha flux is outside of the presented IFU observations. \citet{Wilman19} presents \halpha fluxes, addressing these issues, derived from exponential \halpha profile fitting for a subset of the \kmostd sample. By comparing our aperture flux measurements to total fluxes derived from exponential profile fitting for this subset, we find that the estimated flux losses outside the aperture shows the expected dependence on the ratio $1\farcs 5/r_e$, such that, for instance, $\sim50$\% flux is lost where $r_e=1\farcs 5$.
For the purpose of discussing H$\alpha$ SFRs in the next section, we estimated simple aperture corrections for all detected galaxies based on the results from the more detailed work by \citet{Wilman19} and using the structural parameters fitted on HST F160W images \citep{2014ApJ...788...28V}.  We created mock 2D H$\alpha$ exponential profiles with the same axis ratio and a half-light radius equal to $1.19\times r_e$(F160W), convolved with the associated KMOS PSF (Section~\ref{subsec.psf}).  The aperture correction was then taken as the ratio between the total flux and that within $r = 1\farcs 5$ of the model center.

For a small fraction of galaxies we find that a single spectral component fit to the \halpha emission
provides an inadequate description of the observed line profiles.
In most of these cases, there is excess emission in high-velocity wings that are associated
with strong gas outflows \citep{2014ApJ...796....7G,2019ApJ...875...21F}.
For these objects, two Gaussian components are used to fit the emission lines with the narrower
of the two emission lines assumed to be associated with star formation.  The emission from the
narrow component is the one used in the analysis in the
next sections.
More detail on the fitting and interpretation of the broad emission component is given by
\cite{2014ApJ...796....7G,2014ApJ...787...38F,2019ApJ...875...21F}. 

If \halpha is not detected we calculate an upper limit, using a fixed linewidth corresponding to 120 \kms, by summing the errors on the \halpha channels (defined as channels separated from the \halpha line centre by less than the FWHM of the \halpha line) in quadrature, and then multiplying by 3 to obtain the 3$\sigma$ upper limit. For galaxies with a prior spectroscopic redshift the upper limit is measured assuming \halpha is centered at the known redshift. For non-detections with only a grism redshift a weighted average upper limit is estimated using the 3D-HST redshift probability distribution. 

We detect \NII$\lambda6563$ at S/N$>3$ in \NIIdetectfn\% of galaxies. The detection fraction of \NII~is a strong function of stellar mass as expected from the mass-metallicity relation
{
 (e.g.\ \citealt{1979AA....80..155L,2004ApJ...613..898T,2006ApJ...644..813E,2016ApJ...827...74W}).
}
In the mass ranges $9.5<\log(M_{\star}/M_{\odot})<10.5$, $10.0<\log(M_{\star}/M_{\odot})<11.0$, $10.5<\log(M_{\star}/M_{\odot})<11.5$ the detection fraction of \NII~is 66\%, 74\%, and 80\%.  If the resulting \NII$\lambda6563$ flux is zero, we estimate an upper limit following the same procedure described above but assuming a linewidth fixed to the \halpha linewidth. 

% STACK & bootstrap errors
In Figure~\ref{fig.stack} we show a stacked spectrum of all galaxies with a secure \halpha detection. In the stack we detect the \SII$\lambda\lambda$6716,6731 doublet at S/N $= 20$ and \OI$\lambda6302$ at S/N $= 8$. These additional lines are detected in a subsample of \kmostd galaxies. The interpretation of the \NII/\halpha ratios are discussed further in \cite{2014ApJ...789L..40W,2016ApJ...827...74W} and that of the \SII~ratios are discussed in \cite{2019ApJ...875...21F}.

%----------------------------------------------------------------------
%SFRs
\subsection{SFR comparisons}
\label{sec.sfrs}
%----------------------------------------------------------------------

From the flux measurements described above \halphans-based SFRs, \sfrha, are calculated using \cite{1998ARAA..36..189K} adjusted to a \cite{2003PASP..115..763C} IMF such that
\begin{equation}
\mathrm{SFR}_{\mathrm{H}\alpha}=4.65\times10^{-42}L_{\mathrm{H}\alpha}10^{-0.4A_\mathrm{extra}}10^{-0.4A_\mathrm{cont}}
\end{equation}
where $A_\mathrm{extra} = ( 0.9A_\mathrm{cont} - 0.15A_\mathrm{cont}^2)$ and $A_\mathrm{cont} = 0.82A_\mathrm{v,SED}$ following \cite{2013ApJ...779..135W}. A factor of 1.7 is used to convert from a \cite{1955ApJ...121..161S} to \cite{2003PASP..115..763C} IMF. Here $A_\mathrm{cont}$ is the attenuation of the continuum light at the wavelength of \halpha following \cite{2000ApJ...533..682C} and $A_\mathrm{v,SED}$ is the attenuation at $V$-band derived from the SED fitting described in Section~\ref{sec.selection}. The derived aperture and dust corrected \halpha SFRs range between 0.2 and 319 \sfrunits, with an average \sfrha $=37$ \sfrunits, not including upper-limits. 

%describe the plot: FIGURE SFR_ha vs SFR_best vs SFR_ha
Figure~\ref{fig.sfrcomp} shows the comparison of \sfrha with the UV+IR or SED SFRs (hereafter $\rm SFR_{phot}$) described in Section~\ref{sec.selection} as a function of $\Delta$MS, $A_V$, and IR/UV; the IR to UV flux ratio for IR detected sources. Each SFR indicator, SFR$_\mathrm{UV}+160\mu$m, SFR$_\mathrm{UV}+100\mu$m, SFR$_\mathrm{UV}+24\mu$m, and \sfrsed~is shown by a different color. In general, we find good agreement between \halpha and UV+IR SFR indicators for the majority of galaxies on the MS. Below the MS, where SED based SFRs dominate (due to detection limits in the IR), \sfrsed~is $\sim0.5$ dex lower than the derived \sfrha~as shown in the left panel. This is discussed further in \cite{2017ApJ...841L...6B} and may result from the assumptions in the SED models (e.g. SFHs). In contrast, above the MS where UV+IR SFRs dominate (in particular SFR$_\mathrm{UV}+160\mu$m, SFR$_\mathrm{UV}+100\mu$m), SFR$_\mathrm{phot}$ is typically greater than the \sfrha. 

{One possible explanation for the galaxies with SFR$_\mathrm{phot}$ $\gg$\sfrha~is that the dust correction is underestimated at higher masses and/or in ``starbursts'' above the MS. In the middle and right panel of Figure~\ref{fig.sfrcomp} we look at the ratio of SFR indicators as a function of dust properties, $A_{V}$ and IR/UV.  Galaxies with high SFR$_\mathrm{phot}$ relative to \sfrha~have high IR/UV ratios indicating more dust-attenuated star-formation. This is most common among galaxies with far-IR SFR indicators, SFR$_\mathrm{UV}+160\mu$m and SFR$_\mathrm{UV}+100\mu$m. The same subset of galaxies have relatively low $A_V<1$ suggesting that the $A_V$ derived from the SED fitting does not capture the global dust attenuation. For these highly star-forming galaxies, reddening as a tracer of extinction may saturate regardless of dust geometry. In these cases the extinction may be underestimated both towards the stellar and nebular regions \citep[e.g.,][]{2011ApJ...738..106W}.}

%----------------------------------------------------------------------
\section{Resolved \halpha properties}
\label{sec.resolved}
\begin{figure*}[!tb]
\begin{center}
\includegraphics[scale=0.8,  trim=1.5cm 0cm 1.0cm 0cm, clip, angle=90]{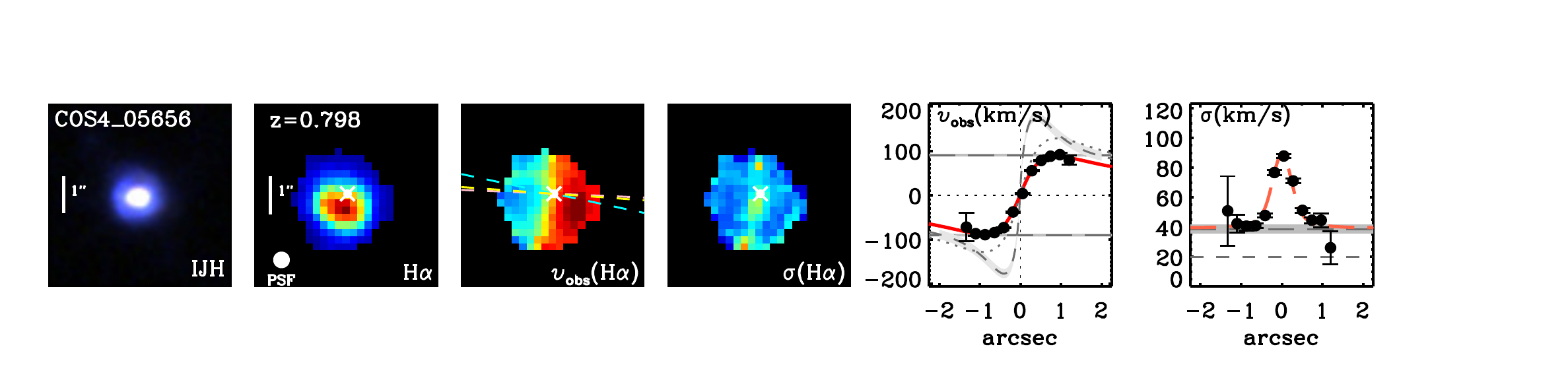}
\includegraphics[scale=0.8,  trim=1.5cm 0cm 1.0cm 0cm, clip, angle=90]{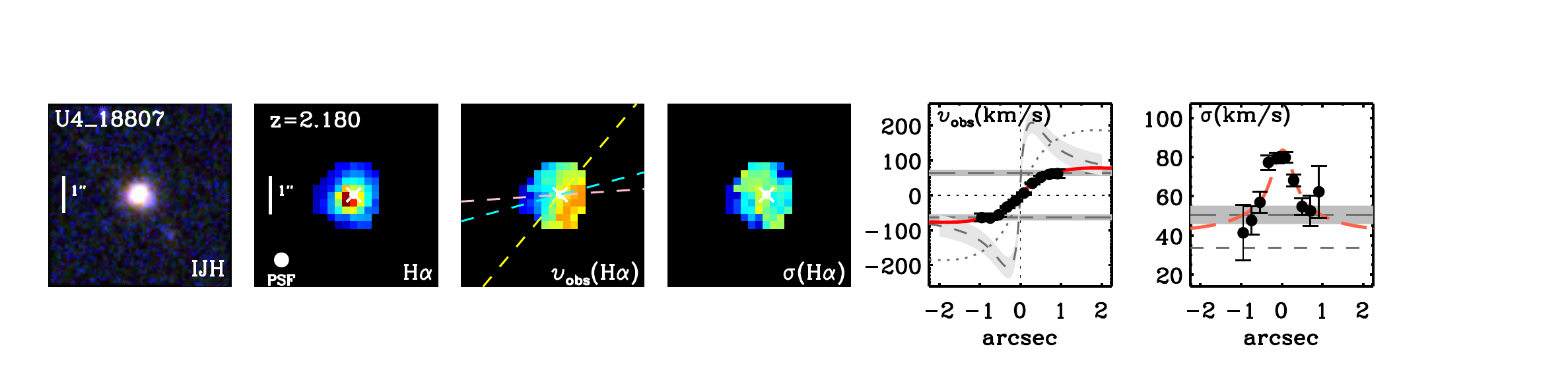}
\includegraphics[scale=0.8,  trim=0.5cm 0cm 1.0cm 0cm, clip, angle=90]{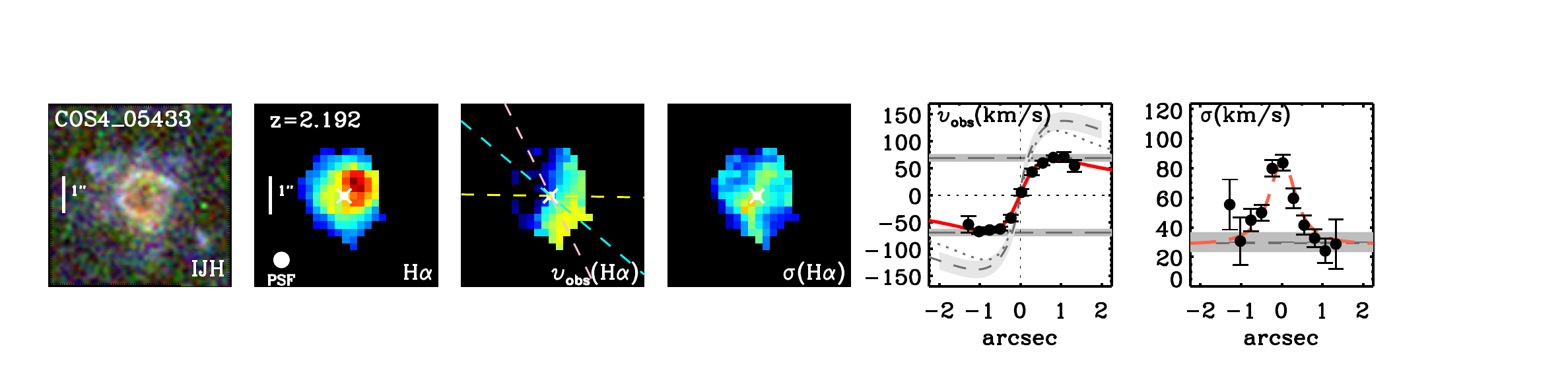}
\caption{Example $1-$ and $2-$D kinematic extractions for \kmostd galaxies. From left to right the panels for each galaxy correspond to an $IJH$ HST colour-composite image 5" on a side, \halpha image derived from KMOS, \halpha velocity field relative to the systemic redshift, \halpha velocity dispersion corrected for instrumental broadening, \halpha velocities (black points) extracted along the major kinematic axis in apertures with the PSF FWHM in diameter, and the corresponding \halpha velocity dispersions (black points) extracted along the major kinematic axis. Scale bars are shown for reference on the HST and \halpha imaging. The HST images are centred on the object, the KMOS images are centred on the centre of the KMOS cube. A circle with the diameter of the FWHM of the PSF is shown in the \halpha image.  The axis profiles are extracted along the kinematic PA as denoted by the light blue line over plotted on the velocity map. The photometric PAs, as determined by F160W and the F814W HST images are shown by the pink and yellow lines respectively. In the fifth panels, the red line shows a best fit exponential disk model, the dotted gray velocity curves show the best-ft exponential disk model with the inclination correction applied. The dashed gray velocity curve shows the intrinsic rotation curve. The associated shaded region shows the error on the rotational velocity, $v_\mathrm{rot,corr}$, corrected for both inclination and beam-smearing effects. The horizontal gray dashed lines and horizontal bars correspond to the \vobs and $\sigma_0$ derived kinematic values and errors respectively. The short dashed horizontal line in the last panels correspond to the beam smearing corrected value of $\sigma_0$.}
\label{fig.examplekin}
\end{center}
\end{figure*}

%----------------------------------------------------------------------
In the following section we discuss the resolved \halpha kinematics of galaxies and resulting disk fractions. In this paper we consider only values measured directly from the data and then corrected for beam smearing effects, which can be derived for all galaxies.  In other works \citep{2016ApJ...831..149W, 2019ApJ...880...48U} we investigated more detailed forward-modeled kinematic measurements on subsets of \kmostd disks with high S/N data.

\subsection{Spatial \halpha fitting}
\label{sec.linefit}
% Description of LINEFIT to get maps, how masks are made
% PLOTS: 1: example single object figure with all plots
% PLOTS: 2: SFR-M* plots from NMFS
The \halpha emission is fit in every spatial pixel in each galaxy using the IDL emission line fitting code LINEFIT (\citealt{2011ApJ...741...69D}; see further descriptions in \citealt{2009ApJ...706.1364F,2018ApJS..238...21F};\citetalias{2015ApJ...799..209W}). In short, the code fits an intrinsic Gaussian convolved with a line profile representing the spectral resolution, thereby implicitly taking into account instrumental broadening. The uncertainties of the fit are determined by 100 Monte Carlo simulations, where the spectrum is perturbed according to a Gaussian distribution from the associated noise spectrum.

The peak, position, and width of the fitted Gaussians are used to create the \halpha flux, velocity, and velocity dispersion maps respectively. A mask is created automatically including pixels with S/N$>2$ and velocities and velocity dispersions with errors $<100$ \kms. Single spaxels detached from the central source that are erroneously included due to fitted sky-lines or noise spikes are removed. Visual inspection of the Gaussian fits in spaxels around the edges of each galaxy is performed to include or remove spaxels as appropriate. For example, in the low S/N regime velocity dispersions can be artificially inflated due to the surrounding noise. A simple S/N cut cannot unequivocally determine the robustness of a fit, especially when the emission line falls
 on or near a sky residual. Example spatially resolved maps are shown in Figure~\ref{fig.examplekin}.

\subsection{Kinematic parameters}
\begin{figure*}%![thb]
\begin{center}
\includegraphics[ scale=0.7,angle=90, trim=10cm 0cm 0cm 0cm, clip]{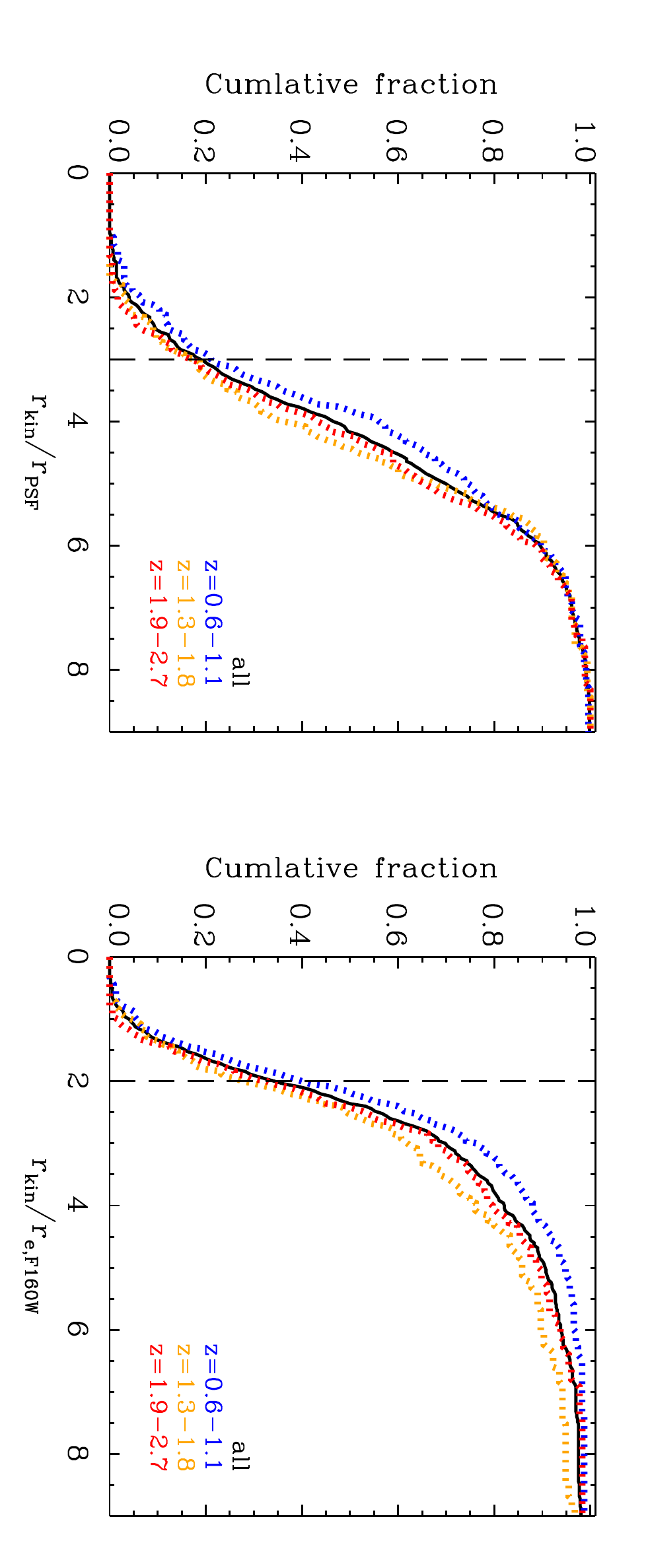}
\caption{The ratio of the maximum radius of kinematic extraction to the Moffat PSF $\sigma$ (left) and the ratio of the maximum radius of kinematic extraction to the $H$-band effective radius (right). Of the detected galaxies, 81\% are are resolved with $r_\mathrm{kin}/r_\mathrm{PSF}>3$ (vertical line), and 60\% reach 2$r_e$ and 30\% reach 3$r_e$, beyond where the turn-over is expected for a self-gravitating exponential disk (vertical line). 
 }
\label{fig.resolved}
\end{center}
\end{figure*}
% what is resolved, how kinematic axis is extracted, velocity, and dispersion, BS (update sini, axis mis-alignment plots?)
% PLOTS: 1: histogram of r_maxkin/r_psf
% PLOTS: 2: possibly axis mis-alignment plot (if something new/interesting)

We define a kinematic axis by identifying the highest and lowest 5\% of spaxels in the velocity map for larger \halpha detections ($\geq50$ high S/N spaxels) or the highest and lowest 5 spaxels for smaller galaxies ($<50$ high S/N spaxels). The positive and negative nodes of the velocity map are determined by taking the weighted average of the selected spaxels. The kinematic axis is defined as the angle between the North-South vertical axis and the line created by the positive and negative nodes. The kinematic center is defined as the half-way position between the two nodes. All kinematic axes and centroids are then inspected by team members to assess the success of this method. The kinematic axes and centroids are well defined by this process. However, for {a small number of} galaxies real (e.g.\ differential dust extinction) or data-driven (e.g.\ sky-line contamination) artefacts can bias the automated method. In these cases, the kinematic axes and centroids are adjusted by hand.

Along the kinematic axis a velocity and velocity dispersion profile are extracted within apertures with diameters equivalent to the FWHM of the PSF.  Emission line fitting for each aperture spectrum is made with LINEFIT as described in Section~\ref{sec.linefit} with associated errors from a Monte Carlo analysis. Each spectral fit along the kinematic axis is inspected. Resulting fits are included based on satisfying the following criteria: S/N$>2$, the difference between two successive velocity points is less than 150 \kms, the error on velocity is $\delta V_{xy}\mathrm{[km~s}^{-1}]<25$, and the error on velocity dispersion is $\delta \sigma_{xy}\mathrm{[km~s}^{-1}]<100$, where $\delta V_{xy}$ and $\delta \sigma_{xy}$ are the error on velocity offset and velocity dispersion from an aperture spectrum at point, $x,y$ in the galaxy. The maximum radius of kinematic extraction, $r_\mathrm{kin}$, is defined as the largest distance from the kinematic center that an emission line can be fit satisfying these criteria. Kinematic extractions are shown in Figure~\ref{fig.examplekin} for a few example galaxies. More 1D kinematic extractions from the $z\sim1$ and $z\sim2$ data sets are given in the Appendix of \citetalias{2015ApJ...799..209W} and in Figure 3 of \cite{2016ApJ...831..149W}. 

We resolve \halpha emission at and beyond three resolution elements, $r_\mathrm{kin}>3r_\mathrm{PSF}$, in 81\% of the detected galaxies. Figure~\ref{fig.resolved} shows the ratio of the maximum radius of kinematic extraction to the model Moffat PSF, $ r_\mathrm{kin}/r_\mathrm{PSF}$, where $r_\mathrm{PSF}=\mathrm{FWHM}/2$. {As seen in Figure~\ref{fig.resolved}, the resolved fraction is comparable in each of the observing bands, YJ:79\%, H:82\%, K:83\%, as expected given the increased observing time enabling detection of extended low surface brightness out to the highest redshift bin. The marginally lower fraction of resolved galaxies at $z\sim1$ is due to the larger number of below-MS sources.} Of the resolved galaxies 50\% have $>4.25$ resolution elements across the galaxies. We measure out to $\sim2r_\mathrm{e}$ in 60\% of detected galaxies and $\sim3r_\mathrm{e}$ in 30\% of detected galaxies, as shown on the right of Figure~\ref{fig.resolved}, beyond where a change in slope, or flattening, of the velocity curve is expected for a self-gravitating exponential disk.

For resolved galaxies we measure an observed velocity, \vobs, and velocity dispersion, $\sigma_0$. The observed velocity is calculated as the average of the absolute value of the minimum and maximum velocity measured along the kinematic axis. Inclination corrected rotational velocity is defined as \vrot = \vobs / $\sin{i}$, where $i$ is the inclination measured from F160W HST axis ratios assuming an intrinsic disk thickness, $q_0=0.25$. The measured velocity dispersion is calculated by taking the weighted mean of all data points from the 1D velocity dispersion profile at $>0.75\times\lvert r_\mathrm{kin}\rvert$. This methodology is designed to measure the line of sight velocity dispersion in disk galaxies where beam smearing from large velocity gradients does not inflate the dispersion by spreading galaxy-wide rotational motions across multiple wavelength channels. We apply this method to all galaxies regardless of kinematic type. {We stress that since one of our goals is to quantify the disk fraction among the full sample, the approach followed here is appropriate as it can be applied to non-disk systems as well.} All resulting \vobs~and $\sigma_0$ measurements from these automated methods are checked against the corresponding kinematic maps and 1D profiles. In a small number of cases adjustments are made when the automated method fails to capture the dispersion in the outer regions or is biased by an individual data point. The derived \vobs, $\sigma_0$ and associated errors are represented by the horizontal lines and gray bands respectively in the 1D kinematic profiles (last column) for the example galaxies shown in Figure~\ref{fig.examplekin}. 

In rotating galaxies where a flattening of velocity is not detected (i.e. the velocity curve is a simple linear gradient) the effects of beam smearing are large, inflating the measured $\sigma_0$ and reducing the measured \vobs. Even galaxies mapped with the most resolution elements may be mildly affected by beam smearing \citep{2011ApJ...741...69D}. In Appendix A.2.4 of \cite{2016ApJ...826..214B} we derived corrections for beam smearing based on the intrinsic size of the galaxy, stellar mass, inclination, and observed PSF size. This approach has the advantage that is easy to apply for all galaxies consistently (regardless of S/N), is not time intensive, and is based on galaxy models. On the other hand, it is based on relatively simple models and the interpolation between a fixed set of galaxy parameters. The alternative is time intensive 2D kinematic models simultaneously taking into account the measured velocities and dispersions (e.g.\ \citealt{2015arXiv150106586B,2015MNRAS.451.3021D,2018ApJ...854L..24U}) that may only converge for a subset of the highest S/N data.  {For the substantial subset of disk galaxies that are sufficiently well resolved and have sufficiently high S/N to be modeled, the simpler approach described here yields \vrot~(and $\sigma_0$) measurements that agree with results of full forward modeling (e.g., \citealt{2016ApJ...831..149W,2019ApJ...880...48U}) within 8 \kms~for velocities (and 10 \kms~for dispersions) on average.}
For the present paper we adopt model-independent measured values corrected for beam smearing using the methods presented in \cite{2016ApJ...826..214B}.

The beam smearing corrections are valid only for galaxies that are well described by a disk model. As shown in \citetalias{2015ApJ...799..209W} and in Section~\ref{sec.discuss} the majority of resolved galaxies in \kmostd fulfil this criteria.  The median velocity beam smearing correction factor for the rotation-dominated galaxies identified in Section~\ref{sec.discuss} is 1.36, with a range from $1.08-1.97$. However, inferred intrinsic \halpha sizes can be $1-4\times$ greater than the $H$-band sizes used to derive the corrections \citep{2013ApJ...763L..16N,2018ApJ...855...97W,Wilman19}. In these cases the beam-smearing corrections may be over-estimated when using the $H$-band size as has been done for this work. {In Figure~\ref{fig.examplekin} the intrinsic non-beam-smeared rotation curve assuming the exponential disk scalelength is equal to $r_\mathrm{e}[{\rm F160W}]/1.68$ is shown for the rotation-dominated galaxies by the dashed line.}  The grey band surrounding the dashed line reflects the errors on the observed velocity, inclination correction, and beam-smearing corrections. 

Errors on the beam smearing corrections are estimated from Monte Carlo simulations of the galaxy parameters that enter into the beam smearing calculations. For the velocity beam smearing, only $r_\mathrm{e}$ is varied as the correction depends very little on other parameters. The resulting 16 and 84 percentile errors on the velocity beam smearing correction are small, typically a few percent.  Beam smearing corrections to $\sigma_0$ are dependent on $M_*$, $i$, and $r_\mathrm{e}$ as detailed in Appendix A.2.4 of \cite{2016ApJ...826..214B}. Multiplicative corrections range from $0.13-0.98$ for the full sample with a median of 0.53.  The 16 and 84 percentile errors on the dispersion beam smearing correction are larger, typically 25\%.

%%%%%%%%%%%%%%%%%%%%%%%%%%%%%%%%%%%%%%%%%%%%%%%%
\section{Analysis} 
\label{sec.discuss}
%%%%%%%%%%%%%%%%%%%%%%%%%%%%%%%%%%%%%%%%%%%%%%%%

%%%%%%%%%%%%%%%%%%%%%%%%%%%%%%%%%%%%%%%%%%%%%%%%
\subsection{Evolution of disk fractions}
\label{sec.diskcrit}
% Criteria for disk fractions, results, comparison to literature?
% PLOTS: 1: histograms/cum. distributions of each criteria? 
% PLOTS: 2: disk fractions vs z, disk fractions vs. mass (discuss simmons +17)
% PLOTS: 3: comparison with literature (though no way of bringing this into alignment!)
%%%%%%%%%%%%%%%%%%%%%%%%%%%%%%%%%%%%%%%%%%%%%%%%

In \citetalias{2015ApJ...799..209W} we presented the fraction of `rotation-dominated' and `disk-like galaxies' of 83\% and 71\% respectively in our combined $z\sim1$ and $z\sim2$ samples using five morpho-kinematic criteria. The high-incidence of rotationally dominated kinematics in star-forming galaxies was more than $\sim2\times$ what had been previously reported (e.g.\ \citealt{2009ApJ...706.1364F, 2009ApJ...697.2057L}). 
In the largest deep, high-resolution AO-assisted IFU survey at
  $1.5 \la z \la 2.5$,
  including 35 galaxies probing massive MS SFGs, as many as $\sim 70\%$ of the
  objects are kinematically classified as rotation-dominated disks
\citep{2018ApJS..238...21F}.
  It has become increasingly clear from the literature that large samples and high S/N data are crucial to accurately characterize disk fractions at any redshift. Recently, it has been proposed that the rotation-dominated fraction among SFGs is monotonically increasing over cosmic time, with fractions as low as 35\% at $z\sim3.5$ \citep{2012ApJ...758..106K,2016MNRAS.457.1888S,2017MNRAS.471.1280T,2017ApJ...843...46S}.  Although there was a hint of evolution between the $z\sim2$ and $z\sim1$ galaxy samples with our first year of data, at that time it was not clear if this was a result of the evolving galaxy sizes and morphologies, making it increasingly difficult to resolve higher-redshift samples and fairly apply the same criteria. 

Here we present disk fractions for galaxies on the MS, $\rm -0.85<\Delta MS<0.85$, across three redshift slices for the full \kmostd sample at a greater depth and larger stellar mass range than our previous results. Following \citetalias{2015ApJ...799..209W} we use the same five disk criteria outlined below with minor adjustments. However, we note the known caveats relevant to this selection as outlined in \cite{2018ApJS..238...21F} and discuss the validity of this kinematic selection in Section~\ref{sec.env}.\\

\begin{figure*}
\begin{center}
\includegraphics[scale=0.9,  trim=1.8cm 0cm 1.4cm 0.5cm, clip, angle=90]{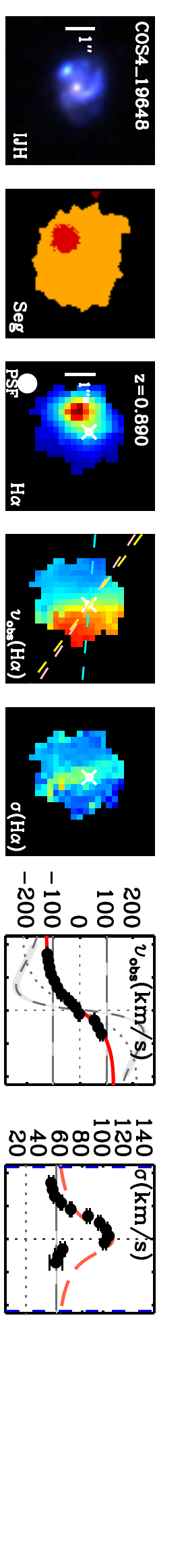}
\includegraphics[scale=0.9,  trim=1.0cm 0cm 1.4cm 0.5cm, clip, angle=90]{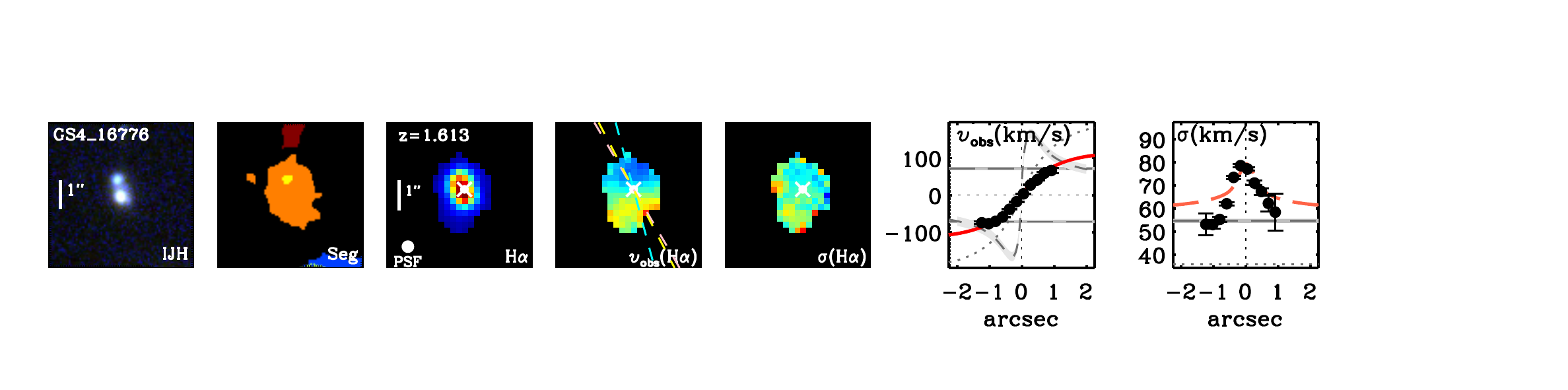}
\caption{Example $1-$ and $2-$D kinematic extractions for \kmostd galaxies with close bright kinematic components at different inclinations. The panels are the same as in Figure~\ref{fig.examplekin} with the additional panel showing the associated segmentation map from 3D-HST. In the segmentation map different colors represent different unique IDs from the 3D-HST catalog. 
} 
\label{fig.exampleclumpy}
\end{center}
\end{figure*}

The criteria applied are as follows:\\

1.) The \halpha velocity map exhibits a continuous velocity gradient along a single axis. In larger systems this is synonymous with the detection of a `spider' diagram (\citealt{1978ARAA..16..103V}, {third} column in Figure~\ref{fig.examplekin}).\\

2.) \vrot/$\sigma_0 > \sqrt{3.36}$, where \vrot~is the rotational velocity corrected for inclination, $i$, by \vrot = \vobs/$\sin{i}$, and both kinematic parameters are corrected for beam smearing \citep{2018ApJS..238...21F}.\\

3.) The position of the steepest velocity gradient, as defined by the midpoint between the velocity extrema along the kinematic axis, is coincident within the uncertainties ($\sim$1.6 pixels) with the peak of the velocity dispersion map.\\%; Figure~\ref{}

4.) For inclined galaxies ($q < 0.6$) the photometric and kinematic axes are in agreement ($< 30$ degrees).\\

5.) The position of the steepest velocity gradient is coincident, within the uncertainties, with the centroid of the KMOS continuum center (a proxy for the center of the potential, i.e. in the higher mass galaxies this is usually a bulge).\\

{For the first criterion, careful visual inspection is performed on the 2D maps, 3D cubes, as well as from the extracted major axis profiles and compared to simple disk models (e.g.\ as plotted in Figure~\ref{fig.examplekin}). To satisfy this criterion the galaxy must exhibit a unique kinematic axis and be monotonically increasing/decreasing as expected for disk velocity curves.}

For a subset of observations, discussed in Section~\ref{sec.serendip}, multiple objects are detected within a single IFU. This includes cases where the multiple galaxies detected and segmented in 3D-HST and the CANDELS catalogs are within $\sim500$ \kms~and likely merging. If the galaxies in these cases are clearly spatially segmented the kinematic parameters discussed in the previous section and used for the disk criteria are derived for the primary galaxy while the serendipitous galaxy is masked. Some examples include GS4\_29773, COS4\_21030, GS4\_19676. When there is smaller physical separation, e.g.\ $<5$ kpc, between bright knots in the imaging it is ambiguous if the imaging reflects multiple galaxies in a merger or multiple clumps of SF within a galaxy (e.g.\ COS4\_19648, GS4\_16776, U4\_25808, U4\_36568) and the image segmentation is variable. In edge-on systems it can be especially difficult to identify clumps within a galaxy, in comparison to face-on systems as shown in Figure~\ref{fig.exampleclumpy}. In these cases the KMOS data can help disentangle the nature of the system. Similar conclusions are reached using simulations of galaxies and the kinemetry analysis of \cite{2008ApJ...682..231S}.

\begin{table}
%\begin{minipage}{\textwidth}
\caption{\% of galaxies satisfying disk criteria}
\begin{tabular*}{\linewidth}{@{\extracolsep{\fill}}lc|ccc}
\hline
{Criteria:} &  { 1,2} & {1,2,3} & { 1,2,3,4} & { 1,2,3,4,5}\\
\hline
$10.0<\log(M_{\star}/M_{\odot})<11.75$\\
\hline
Full Sample   & 79\% & 65\% & 64\% & 59\%  \\
$z\sim1.0$       & 91\% & 73\% & 70\% & 66\% \\
$z\sim1.5$       & 79\% & 68\% & 68\% & 65\% \\
$z\sim2.0$       & 70\% & 56\% & 56\% & 49\% \\
\hline
\label{tab.diskfractions_w15}
\end{tabular*}%\end{minipage}
\end{table}

\begin{table}[!h]
%\begin{minipage}{\textwidth}
\caption{\% of galaxies satisfying disk criteria}
\begin{tabular*}{\linewidth}{@{\extracolsep{\fill}}lc|ccc}
\hline
{Criteria:} &  { 1,2} & {1,2,3} & { 1,2,3,4} & { 1,2,3,4,5}\\
\hline
$9.0<\log(M_{\star}/M_{\odot})<11.75$\\
\hline
Full Sample   & 77\% & 61\% & 60\% & 55\%  \\
$z\sim1.0$       & 87\% & 67\% & 65\% & 61\% \\
$z\sim1.5$       & 72\% & 57\% & 57\% & 54\% \\
$z\sim2.0$       & 69\% & 56\% & 55\% & 48\% \\
\hline
\hline
\hline
$10.5<\log(M_{\star}/M_{\odot})<11.75$\\
\hline
Full Sample   & 85\% & 72\% & 71\% & 66\%  \\
$z\sim1.0$       & 88\% & 75\% & 74\% & 70\% \\
$z\sim1.5$       & 86\% & 75\% & 75\% & 72\% \\
$z\sim2.0$       & 81\% & 66\% & 66\% & 58\% \\
\hline
$9.5<\log(M_{\star}/M_{\odot})<10.5$\\
\hline
Full Sample   & 73\% & 54\% & 52\% & 49\%  \\
$z\sim1.0$       & 91\% & 66\% & 62\% & 60\% \\
$z\sim1.5$       & 66\% & 49\% & 49\% & 46\% \\
$z\sim2.0$       & 58\% & 46\% & 45\% & 39\% \\
\hline
$9.0<\log(M_{\star}/M_{\odot})<9.5$\\
\hline
Full Sample   & \ldots & \ldots & \ldots & \ldots \\
$z\sim1.0$       & 63\% & 47\% & 47\% & 42\% \\
$z\sim1.5$       & \ldots & \ldots & \ldots & \ldots \\
$z\sim2.0$       & \ldots & \ldots & \ldots & \ldots \\
\hline
\label{tab.diskfractions}
\end{tabular*}%\end{minipage}
\end{table}

We first update the fraction of rotation-dominated systems (criteria 1 \& 2) for the most massive galaxies, $\log(M_{\star}/M_{\odot})>10$, for comparison with \citetalias{2015ApJ...799..209W} as shown in Table~\ref{tab.diskfractions_w15}. We find excellent agreement with the results from our first year data presented in Table~1 of \citetalias{2015ApJ...799..209W}. With the larger sample presented here we find that the main evolution between $z\sim2.3$ to $z\sim0.9$ is a result of the evolving \vrot/$\sigma_0$, or criteria 2, primarily driven by the evolution of $\sigma_0$ \citep{2019ApJ...880...48U}.

\begin{figure*}
\includegraphics[ scale=0.73, trim=10.7cm 12cm 0cm 0cm, clip, angle=90]{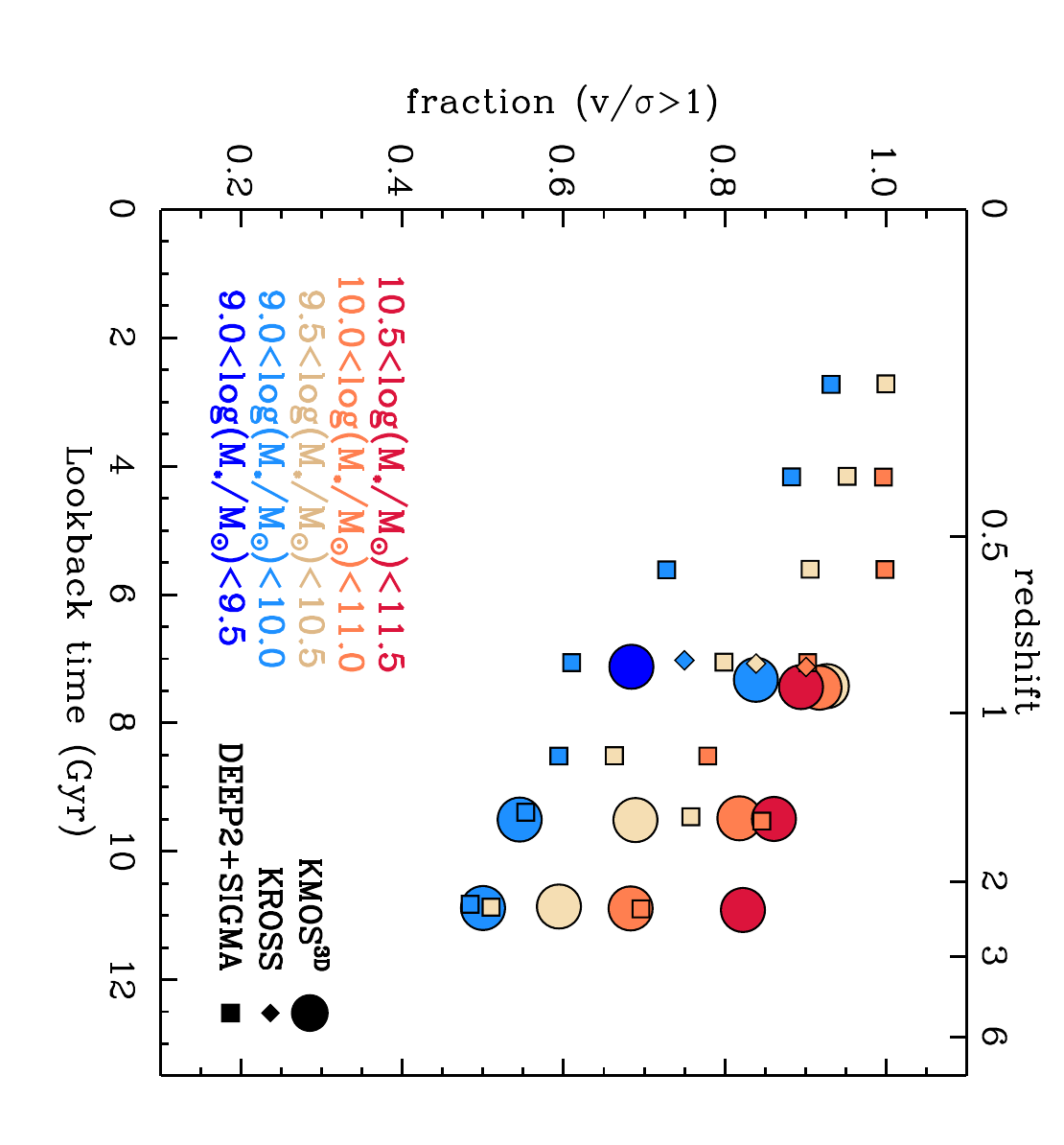}
\includegraphics[ scale=0.73, trim=10.7cm 12cm 0cm 0.5cm, clip, angle=90]{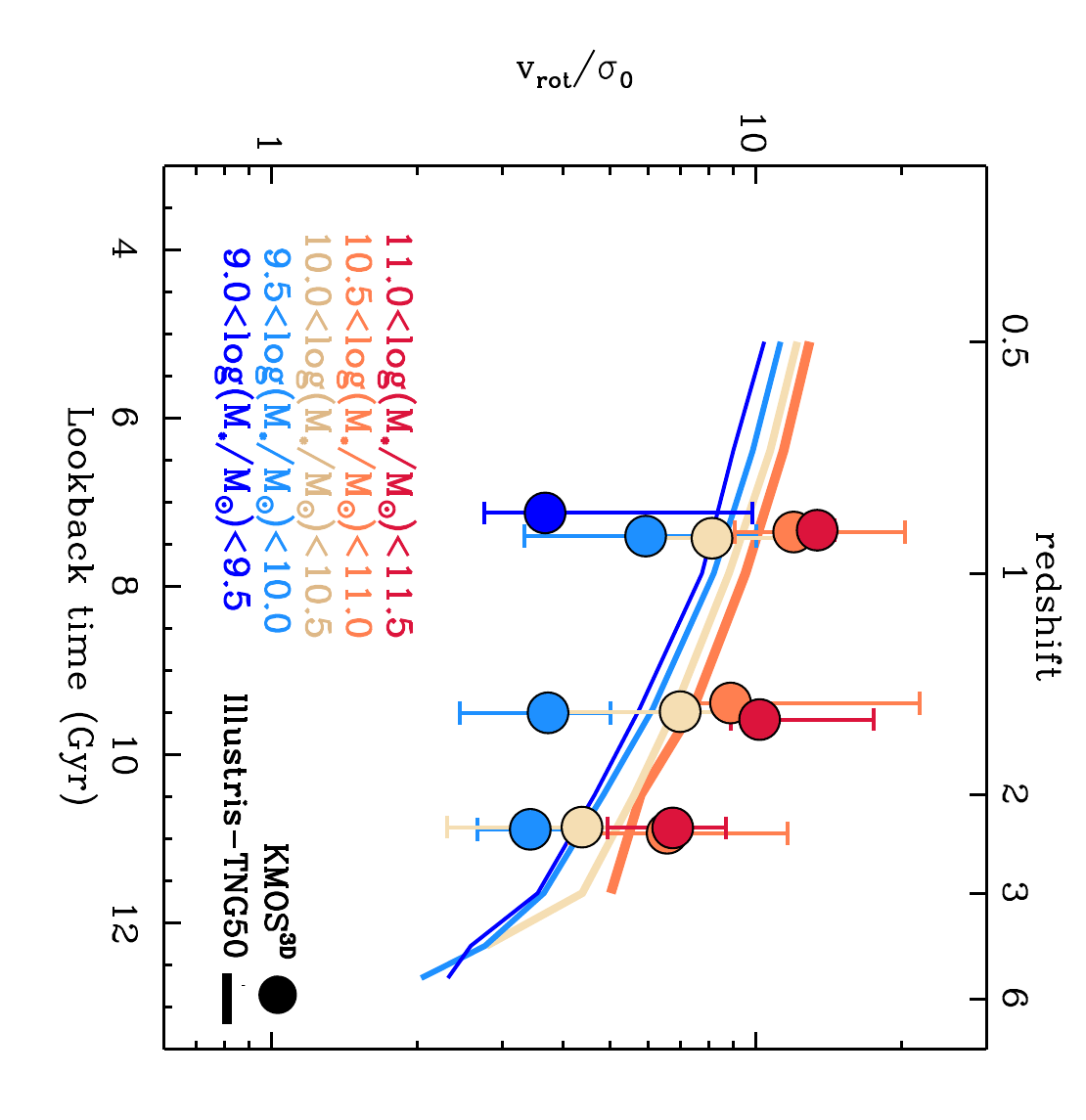}
\caption{Evolution of $v/\sigma>1$ as a function of stellar mass for resolved MS galaxies. Stellar mass bins from $\log{M_*}=9-11.5$ are shown with different colors. In the left panel, the ratio of  $v/\sigma>1$ threshold is used for ease of comparison to the literature. Squares show the results from the DEEP2 and SIGMA slit surveys \citep{2017ApJ...843...46S}. Diamonds show the results from the KMOS KROSS survey at $z\sim1$ \citep{2016MNRAS.457.1888S}. {In the right panel, the ratio $v_\mathrm{rot}/\sigma_0$ as measured from the \kmostd survey is compared to a similarly derived ratio for SFGs in the Illustris-TNG50 simulations \citep{2019arXiv190205553P}.}}
\label{fig.rotdom}
\end{figure*}

With the complete \kmostd survey we have a large enough sample to investigate rotation-dominated (criteria 1 \& 2) and disk (criteria 1-5) fractions as a function of stellar mass as shown in Table~\ref{tab.diskfractions} and Figure~\ref{fig.rotdom}. For the two highest redshift bins we split the sample into two mass bins, $9.5<\log(M_{\star}/M_{\odot})<10.5$ and $10.5<\log(M_{\star}/M_{\odot})<11.5$. For the lowest redshift bin we are able to include an additional low mass bin of $9.0<\log(M_{\star}/M_{\odot})<9.5$. The percentage of galaxies satisfying each criterion is given in Table~\ref{tab.diskfractions}. The fraction of rotation-dominated galaxies  (criteria 1 \& 2) depends on both mass and redshift with galaxies in the mass bin, $9.5<\log(M_{\star}/M_{\odot})<10.5$, and lowest redshift bin, $z\sim1$, having the highest fraction of rotation-dominated galaxies, 93\%.  The largest evolution is seen in the $9.5<\log(M_{\star}/M_{\odot})<10.5$ mass range, evolving from 58\% at $z\sim2$ to 93\% at $z\sim1$. In contrast, the highest mass bin shows a significantly shallower evolution from 82\% to 89\% respectively. 

%{
In Figure~\ref{fig.rotdom} we show the dependence of the fraction of ``rotation-dominated'' galaxies on stellar mass and cosmic time. To facilitate comparison to literature results we adopt a $v/\sigma>1$ threshold to define the `rotation-dominated' fraction of galaxies rather than a threshold of $\sqrt(3.36)$ as adopted in criterion 2.
%} 
The galaxy samples are split into overlapping stellar mass bins of 1.0 dex. Our results are in general agreement with \cite{2017ApJ...843...46S} and \cite{2016MNRAS.457.1888S}. With the \kmostd survey we are able to add an additional high mass bin with $\log(M_{\star}/M_{\odot})=10.5-11.5$ which shows higher factions of $v/\sigma>1$ than at $\log(M_{\star}/M_{\odot})=10.0-11.0$ in the $z\sim1.5$ and $z\sim2.0$ samples. In contrast, at $z\sim1.0$ the rotation-dominated fraction as a function of mass from the \kmostd survey are higher than the literature data. The apparent disagreement between surveys at $z\sim1$ in the mass bin $\log(M_{\star}/M_{\odot})=9.0-10.0$ may be due to the large bin size, and distribution of masses in each bin. When a smaller bin size of 0.5 dex is used we have a large enough sample to explore the \kmostd data at $z\sim1$ between $\log(M_{\star}/M_{\odot})=9-9.5$, shown by the dark blue point. This subset of the low mass data is in better agreement with the literature for low-mass galaxies.

In addition to the caveats already discussed, we also note additional uncertainties in comparison between samples. First, the depth of the \kmostd data may help identify more galaxies with $v/\sigma>1$ as the \halpha emission is probed to larger radii. For example, at $z\sim1$ the \kmostd data are typically twice the depth of the KROSS data. Methods to extract rotational velocity, velocity dispersion, and beam smearing corrections also differ across publications. Finally, slit-based analyses can lead to higher velocity dispersions than IFU-based analyses due to slit misalignment (for discussion see \citealt{2016ApJ...819...80P}).

The right panel of Figure~\ref{fig.rotdom} compares the kinematic ratio, $v_\mathrm{rot}/\sigma_0$, measured for SFGs in the \kmostd sample and simulated SFGs in the Illustris-TNG50 sample \citep{2019arXiv190205553P}. {While the kinematics of the simulations are extracted with comparable methods as the KMOS data, the effects of noise, beam smearing, and inclination are not included in the Illustris simulation data.} Each data point shows the median value of $v_\mathrm{rot}/\sigma_0$ from \kmostd at a given redshift and mass, however, we note that a wide range in values is present in each bin. This variation is reflected in the error bars, which represent the interquartile range. 
Interestingly, the data and simulated galaxies show qualitative agreement both in trends with mass and cosmic time despite the idealised extraction of kinematic values from the simulations. We note that due to both resolution limits of the data and simulations as well as completeness limits with the target selection the comparison of results below $10^{10}$ \Msun~is more uncertain. 

Ordered rotation yields clear observational signatures that lead to the criteria defined in this section (see also: \citetalias{2015ApJ...799..209W}; \citealt{2018ApJS..238...21F}). A small fraction of galaxies dominated by rotation, as captured by our disk fractions, may be in some stage of an instability or interaction sequence as it is not always a binary distinction (e.g.\ \citealt{2017MNRAS.465.1157R}). In contrast, some accretion events, internal processes, and interactions are capable of perturbing rotational motions $-$ producing a variety of kinematic signatures ranging from subtle to extreme (dependent on e.g.\ time scale, halo mass, orientation, gas fraction; \citealt{2003ApJ...597..893N,2008ApJ...685L..27R,Shapiro:2009sj,2017PASA...34...50F}). Here we report high rotation fractions providing constraints on the duty cycle of such processes for the gas-rich star-forming galaxies.
The high fraction of star-forming galaxies on/near the MS characterized
by rotational motions, along with the typically disk-like distributions
of H$\alpha$ and of stellar light and mass \citep{2011ApJ...742...96W,2013ApJ...763L..16N,2016ApJ...828...27N,2014ApJ...788...11L,2014ApJ...792L...6V}, point to major mergers playing a minor role
in setting galactic structure observed at $z \sim 1 - 3$, or to disks
being largely preserved or regrown on short timescales as indicated by
numerical simulations of gas-rich systems (e.g.\ \citealt{2008ApJ...685L..27R,2009ApJ...691.1168H,2017PASA...34...50F,2017MNRAS.470.3946S,2018MNRAS.480.2266M}).
Taking advantage of the large \kmostd sample size and the extensive
characterization of galaxies in the 3D-HST/CANDELS fields, we explore
further the role of interactions at various stages on the kinematics,
as well as metallicity and star formation, through a
statistical analysis of neighbouring galaxies (Mendel et al. in prep.).

%%%%%%%%%%%%%%%%%%%%%%%%%%%%%%%%%%%%%%%%%%%%%%%%
\subsection{Environmental effects on axis misalignment}
\label{sec.env}
\begin{figure}
\begin{center}
\includegraphics[ scale=0.45, trim=0cm 0cm 0cm 1.2cm, clip, angle=90]{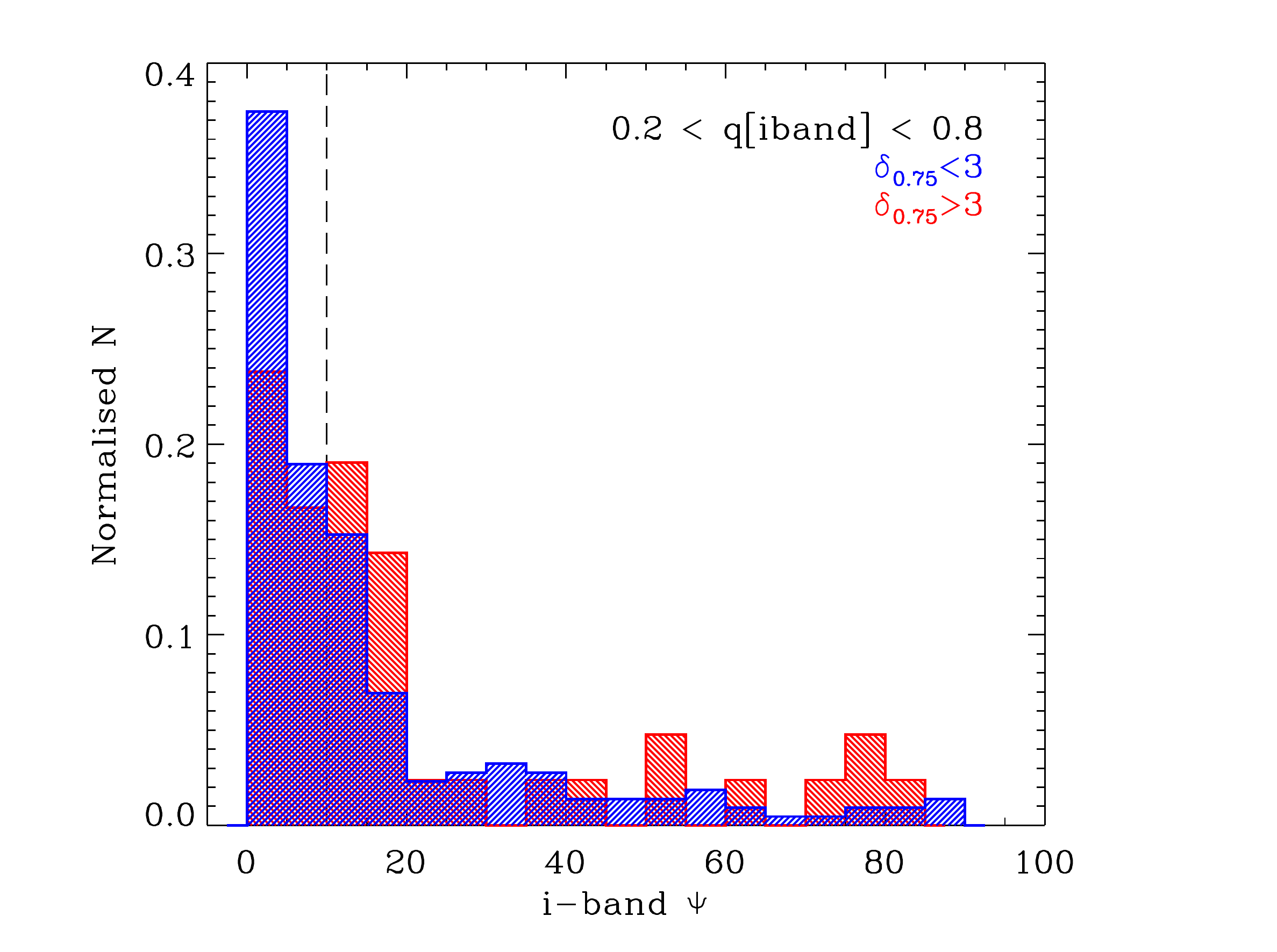}
\caption{Distribution of kinematic misalignment from $i$-band photometric major axis, $\Psi$, split by environmental metric $\delta_{0.75}$, the relative over-density of galaxies in an aperture of 0.75 Mpc radius  \citep{2017ApJ...835..153F}. Galaxies in dense environments are shown in red ($\delta_{0.75}\geq3$) and galaxies in less dense environments are shown in blue  ($\delta_{0.75}<3$). Due to difficulties in measuring misalignments at both high and low inclination the sample is restricted to axis ratios, $0.2<q<0.8$. Histograms are normalised to the same area.}
\label{fig.psienv}
\end{center}
\end{figure}
%%%%%%%%%%%%%%%%%%%%%%%%%%%%%%%%%%%%%%%%%%%%%%%%
{In this and previous works the alignment of the kinematic and photometric axis is used to assess whether a galaxy is perturbed from a recent or ongoing interaction \citep{2006AA...455..107F,2012AA...539A..92E,2015ApJ...799..209W,2017MNRAS.465.1157R,2018ApJS..238...21F} under the assumption that interactions can cause changes in the distribution of angular momentum \citep{2015MNRAS.451.3269V}. We test if it is possible to recover these differences using our observables (\halpha kinematics representing the gaseous component and $i$-band or $H$-band imaging representing the stellar component) with the recently published environment catalog, \cite{2017ApJ...835..153F}, utilising high-quality grism redshifts from 3D-HST and all available spectroscopic redshifts in COSMOS, GOODS-S, UDS, GOODS-N, and AEGIS.  
{Position angles from both $i$-band or $H$-band are used to assess misalignment as they may be sensitive to different mass components of the galaxy (e.g. \citealt{2017MNRAS.465.1157R,2018ApJS..238...21F}).}

In Figure~\ref{fig.psienv} we show the distribution of photometric to kinematic axis misalignment, $\Psi$, where $\Psi=\arcsin(|\sin{(PA_\mathrm{phot}-PA_\mathrm{kin})}|)$. For this analysis only resolved galaxies with $i$-band axis ratios (van der Wel, \textit{private communication}), $0.2<q_\mathrm{F814W}<0.8$, are included as the ability to measure accurate photometric PAs is reduced at high axis ratios when galaxies appear mostly spherically symmetric or face-on and the ability to recover axis misalignments in edge-on systems is reduced due to projection effects. A quarter of galaxies, 24\%, have a kinematic axis misaligned from the $i$- or $H$-band photometric axes by $>30$ degrees. The majority of galaxies with large axes misalignments have axis ratios $>0.5$ where determination of the photometric PA becomes more uncertain (\citetalias{2015ApJ...799..209W}).  

In Figure~\ref{fig.psienv} the \kmostd sample is separated by the relative over-density of galaxies in an aperture of 0.75 Mpc radius, $\delta_{0.75}$ \citep{2017ApJ...835..153F}. Galaxies in a more over-dense environment ($\delta_{0.75}\geq3$) are shown with red ($N=43$) compared to galaxies in a less over-dense environment ($\delta_{0.75}<3$) shown in blue ($N=214$).The possible signal of misalignments on order $\sim10-20$ degrees, shown in In Figure~\ref{fig.psienv}, to more commonly be found in dense environments is comparable to the errors on the photometric and kinematic PA.  
{A two-sided Kolmogorov-Smirnov test using the $i-$band PAs gives a 10\% probability of the two populations being drawn from the same distribution. In contrast, using the $H-$band PAs there is a 60\% chance of the populations being drawn from the same distribution.} While mitigated by having a cut on axis ratio, issues remain in determining photometric parameters such as PA and axis ratio in face-on galaxies and galaxies with strongly light-weighted features such as clumpy star formation, spiral arms, bulges, etc.. Furthermore, due to the typically high stellar masses of the resolved sample, \kmostd galaxies are also more likely to be classified as "central" galaxies in massive halos as defined by the \cite{2017ApJ...835..153F} environment catalog. This test would therefore miss any signatures of misalignment present in "satellite" galaxies. A more robust measurement of kinematic misalignment would result from comparing stellar \textit{kinematics} and gas kinematics \citep{2014AA...568A..70B,2015AA...582A..21B}. Furthermore, SFGs at this epoch have high gas fractions (e.g.\ \citealt{2010Natur.463..781T,2013ApJ...768...74T,2010ApJ...713..686D}) making gas and star misalignments either unlikely or very short lived \citep{2015MNRAS.451.3269V}}.

%%%%%%%%%%%%%%%%
\section{Data Release}
\label{sec.datarelease}
%%%%%%%%%%%%%%%%
With this paper we include a final data release of all \totalobs~targeted galaxies\footnote{\url{http://www.mpe.mpg.de/ir/KMOS3D}} of the full \kmostd Survey. This release includes fully reduced data cubes and their associated noise, exposure, PSF data, and bootstrap cubes.  A list of key galaxy properties including redshifts, $M_{\star}$, SFRs, magnitudes, and sizes is also provided. The data were reduced with a combination of custom routines and the standard SPARK reduction package for KMOS $-$ designed to decrease the sky noise and push to low surface brightness levels.
In particular, we implemented a PCA approach to background subtraction for the first time in a large near-IR dataset \citep[based on the ZAP software described by][with modifications for KMOS data]{2016MNRAS.458.3210S}, leading to a reduction in background noise by a factor of $\sim2$ over SPARK.  The full data reduction procedures are described in Section~\ref{sec.datareduction} and the relevant header keywords are given in Table~\ref{tab.keywords}.  We achieved a 20\% flux calibration accuracy and release $1\farcs 5$ radius aperture \halpha fluxes for \detectedgals~detected primary galaxies.

The data release document available on the release web page describes the data products as well as the properties and flags included in the accompanying catalog. The data cubes are provided in FITS format, including four extensions corresponding to the combined science data cube, noise cube, exposure map, and PSF image. The header keywords are listed in Table~\ref{tab.keywords}.  The catalog is provided as FITS binary data table containing the properties listed in Table~\ref{tab.table}. In addition, the details of the galaxy observations, and photometrically and spectroscopically derived properties, are given in Tables~\ref{tab.obs} to \ref{tab.kmos}.
Table~\ref{tab.obs} lists the observational properties of the galaxies including the coordinates, the original target redshift, the $K$ band magnitude, the KMOS observing band and total exposure time, and the spatial and spectral resolution described in Sections~\ref{sec.obs} and \ref{sec.datareduction}.
Table~\ref{tab.phot} gives the $M_{\star}$, SFR, rest-frame absolute $U$, $V$, and $J$ band magnitudes, the visual extinction, and the HST F160W-based effective radius and axis ratio described in Section~\ref{sec.selection}.
Table~\ref{tab.kmos} lists the emission line properties derived from the integrated KMOS spectra including the redshift, H$\alpha$ and \ion{N}{2} fluxes, and line width described in Section~\ref{sec.results}.
We stress that the flux measurements used a fixed aperture size and fitted a single (or double in some cases) Gaussian profile to the
spectrum, to serve as simple reference characterizing the emission of all detected galaxies.
Measurements can be extracted from the reduced data cubes using other methods tailored to specific requirements of science goals.

%%%%%%%%%%%%%%%%
\section{Summary}
\label{sec.summary}
%%%%%%%%%%%%%%%%
This paper presents the completed \kmostd survey and accompanying data release of \observedgals~galaxies observed with the near-IR multi-IFU KMOS at the VLT in $2013-2018$.
\kmostd mapped the ionized gas distribution and kinematics of galaxies on and off the star-forming MS through the \halphans, \NII, and \SII~emission lines.
Deep observations were obtained, with median on-source times of \obstimeYJ, \obstimeH, and \obstimeK~hours for $z\sim1$, $z\sim1.5$, and $z\sim2$ targets, respectively, under excellent typical near-IR seeing conditions of $0\farcs 5$.
The targets were drawn from the 3D-HST survey at $0.7 < z < 2.7$, $\log(M_{\star}/M_{\odot}) > 9$ and $K < 23~{\rm mag}$, with the requirement of having a sufficiently accurate redshift (either grism or spectroscopic) and the lines of interest falling away from telluric emission lines and low transmission spectral regions.  No explicit criterion involving SFR, colors, or AGN activity was applied.  The survey was designed to provide a population-wide census of spatially-resolved kinematics, star formation, outflows, and nebular gas conditions and has delivered on these goals through a number of publications (\citealt{2014ApJ...796....7G,2017Natur.543..397G,2014ApJ...789L..40W,2016ApJ...827...74W, 2015ApJ...799..209W,2018ApJ...855...97W,2016ApJ...826..214B,2016ApJ...831..149W,2017ApJ...840...92L,2017ApJ...841L...6B,2017ApJ...842..121U,2019ApJ...880...48U,2019ApJ...875...21F,Wilman19}). 

Among the sample of \observedgals~targeted galaxies, \detectedgals~are detected in H$\alpha$, for a global fraction of 79\%; these galaxies span $0.6<z<2.7$ and $9.0<\log(M_{\star}/M_{\odot})<11.7$. 
At $\rm \Delta MS>-0.25~dex$ and $(U-V)_\mathrm{rest}<1.3$, 90\% of the targets are detected. Unsurprisingly, the H$\alpha$ detection fraction is a strong function of both color and MS offset. With the strategy emphasizing depth, line emission was nonetheless detected in $\sim 25\%$ of galaxies classified as quiescent based on their $\rm \Delta MS<-0.85~dex$ or their $UVJ$ colors --- a regime poorly explored in previous near-IR IFU studies.  The sensitivity of the data also probes extended faint line emission, contributing to the high {resolved fraction of 81\%} for detected galaxies with $\geq3$ resolution elements along the major kinematic axis. 

%SFRS
Our spatially- and spectrally-resolved KMOS IFU data allow measurements of H$\alpha$ fluxes over most or all of the emission regions of the galaxies, with no contamination by the neighbouring [\ion{N}{2}] lines, an accuracy of better than 20\%, and over four orders of magnitudes in derived \halpha SFR.
From the comparison of dust-corrected \halpha SFRs to UV+IR and SED SFRs, we find a general good agreement.  We confirm that extra extinction towards \halpha is required to closely match \halpha and UV+IR derived star-formation rates \citep{2013ApJ...779..135W} but find that SED derived $A_V$ values may be underestimated for galaxies with high IR/UV ratios. 

%Disk fractions
Confirming our first year results (\citetalias{2015ApJ...799..209W}) we find that the majority, 78\%, of galaxy kinematics on the MS are dominated by rotational motions, with $v/\sigma>\sqrt{3.36}$ (satisfying criteria 1 \& 2 of Section~\ref{sec.diskcrit}). The fraction of rotation-dominated galaxies increases with both mass and redshift. The largest evolution is seen at moderate stellar masses ($9.0<\log(M_{\star}/M_{\odot})<10.5$), evolving from 58\% at $z\sim2$ to 93\% at $z\sim1$, while in the highest mass bin ($10.5<\log(M_{\star}/M_{\odot})<11.75$) a significantly shallower evolution is measured, 82\% to 89\% respectively. While five criteria are used to identify disk-like structure in the \halpha kinematics, it is the ratio of velocity to random motions that dominates the evolution of disks over cosmic time. Within the $\Lambda$CDM paradigm, the high measured disk fractions among SFGs indicates that gas is able to resettle into a semi-equilibrium state of a rotating disk quickly from events that may cause morphological and kinematic disruptions such as accretion, strong outflows, and interactions. 

Given the large sample and broad selection presented here, a number of galaxies have close companions and close kinematic alignment and may minimally inflate our rotation dominated fractions (e.g.\ \citealt{2008ApJ...682..231S}). Using stricter criteria motivated by these concerns we define a purer disk selection, taking into account photometric data. This selection technique identifies 56\% of galaxies in our sample as well described by an exponential disk model. 
Recent work, applying the criteria in Section~\ref{sec.diskcrit} to the Illustris \citep{2014MNRAS.444.1518V} simulations, suggests that the observed disk fractions accurately capture rotationally-dominated systems (not in the state of merging) at a 5\% and 15\% level for galaxies with $\log(M_{\star}/M_{\odot})>10$ and $\log(M_{\star}/M_{\odot})=9-10$ respectively \citep{2019ApJ...874...59S}.

KMOS has filled the literature with many rich datasets including \kmostd \citep[e.g.,][]{2014ApJ...796....7G,2015ApJ...799..209W,2015ApJ...804L...4M,2016MNRAS.457.1888S,2016MNRAS.460..103T,2016MNRAS.456.1195H,2017MNRAS.467.1965H,2017ApJ...846..120B,2017ApJ...850..203P,2017MNRAS.471.1280T,2017ApJ...838...14M,2018AA...613A..72G}. With deep observations, the seeing-limited nature of KMOS is a strength $-$ allowing ionised gas kinematics to be mapped beyond $2r_\mathrm{e}$ for hundreds of galaxies. Further progress in the evolution of kinematic properties will require similarly large investments of time with new instruments. Near infrared IFS studies of $z\gtrsim1$ galaxies have typically been limited to medium resolution ($R\sim2000-4000$) and narrow wavelength ranges covering only single emission line complexes (e.g.\ \halphans -\NII, \hbetans-\OIII). New capabilities on future IFS instruments such as higher spectral and spatial resolution (e.g.\ ERIS/VLT, GIRMOS/Gemini; \citealt{2018arXiv180705089D,2018arXiv180703797S}) and broader wavelength coverage (e.g.\ NIRSPEC/JWST, MIRI/JWST, GMTIFS/GMT; \citealt{2008SPIE.7010E..11C,2016SPIE.9908E..1YS}) will enable surveys to provide insight into small-scale motions, 10 \kms, of the ionised gas and to map the spatially varying ISM conditions, currently only possible from local IFS studies. Matched to similar resolution molecular gas maps of the same galaxies (a synergy that has been realized only for a handful of galaxies) we can study where and how star formation occurs. A full census of the physical mechanisms driving the early growth and lifecycle of galaxies is necessary to piece together the formation of stellar structure in the oldest components of today's massive galaxies.

%%%%%%%%%%%%%%%%%%%%%%%%%%%%%%%
\acknowledgments{
We acknowledge the whole 3D-HST team for a productive collaboration and access to early data for selection. We wish to thank the ESO staff, and in particular the staff at Paranal Observatory, for their helpful and enthusiastic support during the many observing runs over which the KMOS GTO were carried out. We thank the entire KMOS instrument and Commissioning team for their hard work, which allowed our observational program to be carried out so successfully. We also thank the software development team of \spark for all their work with us to get the most out of the data. This paper and the \kmostd survey have benefitted from many constructive, insightful, and enthusiastic discussions with many colleagues whom we are very grateful to, especially A. van der Wel, A. Renzini, A.E. Shapley, M. Franx.  ESW and JTM acknowledge the support of the Australian Research Council Centre of Excellence for All Sky Astrophysics in 3 Dimensions (ASTRO 3D), through project number CE170100013. DJW and MFossati acknowledge the support of the Deutsche Forschungsgemeinschaft via Project ID WI 3871/1-1, and WI 3871/1-2. PL and MFossati acknowledge funding from the European Research Council (ERC) under the European Union's Horizon 2020 research and innovation programme (grant agreement Nos.\ 694343 and 757535 respectively). }

%%%%%%%%%%%%%%%%%%%%%%%%%%%%%%%%%%%%%%%%%%%%%%%%
\bibliographystyle{apj}

\clearpage
\clearpage
%\vspace*{15cm}
%%%%%%%%%%%%%%%%%%%%%%%%%%%%%%%%%%%%%%%%%%%%%%%%

%%% @@@ HERE
\begin{table*}
%\begin{minipage}{\textwidth}
\caption{KMOS-specific FITS header keywords for the released data cubes}
\begin{tabular*}{\linewidth}{@{\extracolsep{\fill}}ll}
\hline
{Keyword} & {Description} \\
\hline
\multicolumn{2}{c}{Primary header} \\
\hline
OBJECT    & Object ID in 3D-HST v4 catalog \\
OBJ\_TARG  & Object ID in 3D-HST at time of KMOS observations (v2 or v4 catalog) \\
OBSBAND   & Observing band \\
EXPTIME   & Total exposure time (minutes) \\
NEXP      & Number of combined exposures \\
VERSION   & \kmostd release version \\
INSTRUME  & KMOS for all cubes \\
EXT1      & Information contained in FITS extension 1 \\
EXT2      & Information contained in FITS extension 2 \\
EXT3      & Information contained in FITS extension 3 \\
EXT4      & Information contained in FITS extension 4 \\
HIERARCH ESO K3D RES ORDER   &  Order of polynomial for spectral resolution \\
HIERARCH ESO K3D RES COEFF0  & Constant polynomial coefficient \\
HIERARCH ESO K3D RES COEFF1  & 1st-order polynomial coefficient \\
HIERARCH ESO K3D RES COEFF2  & 2nd-order polynomial coefficient \\
HIERARCH ESO K3D RES COEFF3  & 3rd-order polynomial coefficient \\
HIERARCH ESO K3D RES COEFF4  & 4th-order polynomial coefficient \\
HIERARCH ESO K3D RES COEFF5  & 5th-order polynomial coefficient \\
HIERARCH ESO K3D RES MIN     & Floor for spectral resolution across band \\
HIERARCH ESO K3D RES MAX     & Ceiling for spectral resolution across band \\
\hline
\multicolumn{2}{c}{Fourth extension header for PSF} \\
\hline
HIERARCH ESO K3D PSF MOFFAT INTFLUX    & Total PSF model flux from Moffat fit \\
HIERARCH ESO K3D PSF MOFFAT FRACFLUX   & Flux fraction in image from Moffat fit \\
HIERARCH ESO K3D PSF MOFFAT AMPL       & Amplitude from Moffat fit \\
HIERARCH ESO K3D PSF MOFFAT BETA       & Moffat fit beta parameter \\
HIERARCH ESO K3D PSF MOFFAT FWHM\_MIN   & Minor axis FWHM from Moffat fit \\
HIERARCH ESO K3D PSF MOFFAT FWHM\_MAJ   & Major axis FWHM from Moffat fit \\
HIERARCH ESO K3D PSF MOFFAT AXRAT      & Axis ratio from Moffat fit \\
HIERARCH ESO K3D PSF MOFFAT PA         & Position angle from Moffat fit  \\
HIERARCH ESO K3D PSF MOFFAT TOTABSRES  & Total fit residual with Moffat model \\
HIERARCH ESO K3D PSF MOFFAT CHISQ      & Best fit chi squared with Moffat model \\
HIERARCH ESO K3D PSF GAUSS INTFLUX     & Total model flux from Gaussian fit \\
HIERARCH ESO K3D PSF GAUSS FRACFLUX    & Flux fraction in image from Gaussian fit \\
HIERARCH ESO K3D PSF GAUSS AMPL        & Amplitude from Gaussian fit \\
HIERARCH ESO K3D PSF GAUSS FWHM\_MIN    & Minor axis FWHM from Gaussian fit \\
HIERARCH ESO K3D PSF GAUSS FWHM\_MAJ    & Major axis FWHM from Gaussian fit \\
HIERARCH ESO K3D PSF GAUSS AXRAT       & Axis ratio from Gaussian fit \\
HIERARCH ESO K3D PSF GAUSS PA          & Position angle from Gaussian fit  \\
HIERARCH ESO K3D PSF GAUSS TOTABSRES   & Total fit residual with Gaussian model \\
HIERARCH ESO K3D PSF GAUSS CHISQ       & Best fit chi squared with Gaussian model \\
HIERARCH ESO K3D PSF CONST             & Data background level \\
HIERARCH ESO K3D PSF AMPL              & Data peak flux \\
HIERARCH ESO K3D PSF FWHM\_MIN          & Data minor axis FWHM \\
HIERARCH ESO K3D PSF FWHM\_MAX          & Data major axis FWHM \\
\hline
\label{tab.keywords}
\end{tabular*}%\end{minipage}
\end{table*}

\clearpage

\begin{table*}
%\begin{minipage}{\textwidth}
\caption{Keywords for the released data table}
\begin{tabular*}{\linewidth}{@{\extracolsep{\fill}}ll}
\hline
{Keyword} & {Description} \\
\hline
ID    & KMOS3D ID with field and 3D-HST v4 catalog object ID \\
FIELD   & Field identifier; COS=COSMOS, GS=GOODS-SOUTH, U=UDS \\
ID\_SKELTON  & Object ID in 3D-HST v4 catalog \citep{2014ApJS..214...24S} \\
ID\_TARGETED  & KMOS3D ID when targeted, with field and 3D-HST (v2 or v4 catalog) object ID \\
FILE   & Associated datacube in fits format \\
FLAG\_PRIMARYTARG      & $1=$ targeted as a primary KMOS3D target, \\
& $0=$ serendipitous galaxy detection within IFU of a primary target  \\
FLAG\_ADDGALDET   & $1=$ additional galaxy detected in the IFU of the primary target, \\
& $0=$ no additional galaxy detected \\
FLAG\_SEGMENTATION  & $1=$ possible issues with photometry and derived parameters resulting from over or under segmentation, \\
& $0=$ no issues identified with segmentation map \\
FLAG\_ZQUALITY      & $1=$ redshift/detection is uncertain, \\
& $0=$ redshift is secure	 \\
& {$-1=$ Non-detection}	 \\
RA      & Right ascension \\
DEC      & Declination \\
Z\_TARGETED      & Best known redshift at time of observations \\
OBSBAND  &  Observing band \\
EXPTIME & Total exposure time (minutes)\\
PSF\_FWHM & FWHM of PSF using Moffat model, minor axis (arcsec)\\
Z  & Measured redshift from KMOS3D observations, -9999. if not detected \\
SPEC\_RES & Estimated spectral resolution from arc and OH sky lines as described in Section~\ref{subsec.res}\\
M\_KS & Apparent Ks magnitude (AB) \\
RF\_U  & Rest frame absolute U-band magnitude (AB) \\
RF\_V  & Rest frame absolute V-band magnitude (AB) \\
RF\_J   & Rest frame absolute J-band magnitude (AB)  \\
SFR    & SFR from ladder of SFR indicators in \sfrunits~assuming a \cite{2003PASP..115..763C} IMF (see \cite{2011ApJ...738..106W,2011ApJ...742...96W} - Section 2.2.3)\\
SFR\_TYPE    & SFR indicator of SFR\\
& $5=$ SFR\_UV+160um; \\
& $4=$ SFR\_UV+100um; \\
& $3=$ SFR\_UV+70um; \\
& $2=$ SFR\_UV+24um; \\
& $1=$ SFR\_SED\\
LMSTAR   & Stellar mass derived from SED modeling following \cite{2011ApJ...738..106W}, using the FAST \citep{2009ApJ...700..221K} fitting code, \\
 & \cite{2003MNRAS.344.1000B}; Chabrier IMF; solar metallicity; Exponentially declining SFH with tau > 300 Myr; \\
 & 0 < Av < 4; 50 Myr < age\_since\_onset\_SF < age\_universe \\
SED\_AV       & Dust attenuation towards V-band derived from SED modeling \\
RHALF       & CANDELS H-band major axis effective radius (arcsec) \\
RHALFERR   & error on CANDELS H-band major axis effective radius (arcsec)\\
Q   & CANDELS H-band axis ratio\\
QERR     & error on CANDELS H-band axis ratio \\
FLAG\_HSOURCE         & Source of Rhalf, Rhalferr, Q, Qerr: \\
& $1=$ H-band fit from \cite{2012ApJS..203...24V};\\
&  $2=$ H-band fit from \cite{2014ApJ...788...11L} \\
\hline
\label{tab.table}
\end{tabular*}%\end{minipage}
\end{table*}

\clearpage

\begin{table*}
%\begin{minipage}{\textwidth}
\caption{\kmostd galaxy target and observing properties}
\begin{tabular*}{\linewidth}{@{\extracolsep{\fill}}lcccccccc}
\hline
{ID} &  {R.A.} & {Decl.} & { z$_\mathrm{best,orig}^a$} & {$K_\mathrm{AB}$} & {Band} & {Exposure time$^b$} & {PSF FWHM$^c$} & {$R^d$}\\
{} &  {} & {} & { } & {(mag)} & {} & {(min)} & {(arc sec)} & {}\\
\hline
COS4\_00779   &   150.10114   &  2.1906323  &   0.92133   &   19.77 & $YJ$  &   230  &   0.585    &    3682\\
COS4\_00937   &   150.12886   &  2.1932354  &   0.87830  &    19.07 & $YJ$  &   230  &   0.585    &    3160\\
COS4\_00970   &   150.14334   &  2.1926434   &  0.79940   &   19.23 & $YJ$  &   285  &   0.585    &    3430\\
COS4\_01351   &   150.14261   &  2.1969705   &  0.85380  &    19.72 & $YJ$  &   285  &   0.585     &   3604\\
COS4\_01598   &   150.11681   &  2.1967461  & 1.02223  &    20.91 & $YJ$   &  290  &    0.585 &     \ldots \\
\hline
\label{tab.obs}
\end{tabular*}%\end{minipage}
\tablecomments{Table~\ref{tab.obs} is published in its entirety in the machine-readable format. A portion is shown here for guidance regarding its form and content.\\
$^a$ Best available redshift at the time of target selection, including grism redshifts from 3D-HST and spectroscopic redshifts from the literature.\\
$^b$ Total on-source integration time in minutes of the combined data sets used for the analysis, excluding low-quality exposures for some sources (e.g., taken under poorer
observing conditions or having poor sky subtraction).
\\
$^c$ The PSF FWHM corresponds to the effective resolution of the combined observations for a given object. It is estimated from the combined data of stars observed simultaneously with different KMOS arms, by fitting a 2D Moffat profile (discussed in Section~\ref{subsec.psf}).\\
$^d$ Spectral resolution at the location of \halpha as determined from polynomial fits to arc lines and sky emission (described in Section~\ref{subsec.res}).
}
\end{table*}

\begin{table*}
%\begin{minipage}{\textwidth}
\caption{\kmostd photometrically derived properties}
\begin{tabular*}{\linewidth}{@{\extracolsep{\fill}}lccccccccccc}
\hline
{ID} &  {$\log(M_*)$}  & {SFR$_\mathrm{phot}^a$} & {SFR type$^b$} & {$U_\mathrm{rest}$} & {$V_\mathrm{rest}$} & {$J_\mathrm{rest}$} & {$A_v^d$} & {$r_e$[F160W]$^e$} & {$q^f$} & {flag$^g$}\\
{} &  {(\Msun)}  & {(\sfrunits)} & {} & {(mag)} & {(mag)} & {(mag)} & {} & {(arc sec)} & {} & {}\\
\hline
COS4\_00779  &  10.85  &    16.46 &  2 & -20.57  &  -22.26  &  -23.52  &  0.3 & 1.053 &  0.772  &     2   \\
COS4\_00937  &  11.17  &    14.59 &  5 & -20.80  &  -22.62  &  -24.10  &  0.7 & 0.640 &  0.835  &     2   \\
COS4\_00970  &  11.04  &     0.87 &  1 & -20.33  &  -22.34  &  -23.77  &  0.6 & 0.385 &  0.497  &     2   \\
COS4\_01351  &  10.73  &    57.40 &  5 & -20.28  &  -21.82  &  -23.40  &  1.7 & 1.119 &  0.256  &     2   \\
COS4\_01598  &  10.50  &     0.74 &  1 & -17.97  &  -20.18  &  -22.20  &  1.5 & 0.478 &  0.192  &     2   \\
\hline
\label{tab.phot}
\end{tabular*}%\end{minipage}
\tablecomments{Table~\ref{tab.phot} is published in its entirety in the machine-readable format. A portion is shown here for guidance regarding its form and content.\\
$^a$  Stellar mass derived from SED modeling following \cite{2011ApJ...738..106W}, using the FAST \citep{2009ApJ...700..221K} fitting code,  \cite{2003MNRAS.344.1000B}; Chabrier IMF; solar metallicity; Exponentially declining SFH with tau > 300 Myr;  0 < Av < 4; 50 Myr < age\_since\_onset\_SF < age\_universe \\
$^b$ SFR from ladder of SFR indicators in \sfrunits~assuming a \cite{2003PASP..115..763C} IMF (see \cite{2011ApJ...738..106W,2011ApJ...742...96W} - Section 2.2.3)\\
$^c$ $5=$ SFR\_UV+160um;  $4=$ SFR\_UV+100um;  $3=$ SFR\_UV+70um;  $2=$ SFR\_UV+24um;  $1=$ SFR\_SED\\
$^d$ Dust attenuation towards $V$-band derived from SED modeling \\
$^e$ CANDELS $H$-band major axis effective radius (arcsec) \\
$^f$ CANDELS $H$-band axis ratio\\
$^g$ Source of $r_e$[F160W], q and associated erros: $1=$ H-band fit from \cite{2012ApJS..203...24V}; $2=$ H-band fit from \cite{2014ApJ...788...11L}.} 
\end{table*}

\begin{table*}
%\begin{minipage}{\textwidth}
\caption{\kmostd KMOS derived properties}
\begin{tabular*}{\linewidth}{@{\extracolsep{\fill}}lcccccc}
\hline
{ID} &  {$z_\mathrm{kmos}^a$} & {$z_\mathrm{q}^b$} & {$f_{\mathrm{H}\alpha}^c$} & {apperture} & {$\sigma_\mathrm{int}$} & {Serendipitous flag$^d$}\\
{} &  {} & {} & {($10^{17}$ erg s$^{-1}$ cm$^{2}$)}& {correction} & {(km s$^{-1}$)} & {}\\
\hline
COS4\_00779  &  0.92430  &  1  &  0$\pm$0  & 0$\pm$0 & 0$\pm$0 &      0\\
COS4\_00937  &  0.87789  &  0  &  0$\pm$0  & 0$\pm$0 & 0$\pm$0 &      0\\
COS4\_00970  &  0.82045  &  0  &  0$\pm$0  & 0$\pm$0 & 0$\pm$0 &      0\\
COS4\_01351  &  0.85345  &  0  &  0$\pm$0  & 0$\pm$0 & 0$\pm$0 &      0\\
COS4\_01598  &  \ldots   &  0  &  0$\pm$0  & 0$\pm$0 & 0$\pm$0 &      0\\
\hline
\label{tab.kmos}
\end{tabular*}%\end{minipage}
\tablecomments{Table~\ref{tab.kmos} is published in its entirety in the machine-readable format. A portion is shown here for guidance regarding its form and content.\\
$^a$ Redshift (vacuum) from the \halpha line fits to the spectrum of each galaxy integrated in the circular aperture.\\
$^b$ Redshift quality: $0=$ redshift is secure; $1=$ redshift/detection is uncertain.\\
$^c$ Properties from Gaussian line profile fits to the spatially-integrated spectrum of each galaxy extracted in a circular aperture of $1.5''$ radius. The total \halpha flux and velocity dispersion, $\sigma_\mathrm{int}$, are given. The velocity dispersion is corrected for the instrumental resolution at \halpha. The uncertainties are derived using the bootstrap cubes. $3\sigma$ upper limits are given when \halpha emission line is undetected.\\ 
$^d$ Flag indicating if a serendipitous galaxy is detected in the observations: $0=$ no additional galaxy detected, $1=$ additional galaxy detected in the IFU of the primary target.
\\}
\end{table*}

\label{lastpage}

\end{document}